\documentclass{aa}

\usepackage{graphicx}
\usepackage{amsmath}
\usepackage{txfonts}
\usepackage{natbib}
\usepackage{hyperref}
\usepackage{caption}
\usepackage{amssymb}
\usepackage{soul}

\hypersetup{pagebackref=true, colorlinks=true, linkcolor=red, citecolor=blue, urlcolor=blue, bookmarks=true} 

\def\gsimeq{\hbox{\raise0.5ex\hbox{$>\lower1.06ex\hbox{$\kern-1.07em{\sim}$}$}}} 
\def\lsimeq{\hbox{\raise0.5ex\hbox{$<\lower1.06ex\hbox{$\kern-1.07em{\sim}$}$}}} 

\bibpunct{(}{)}{;}{a}{}{,} % to follow the A&A style

\begin{document}

\title{A super-linear ``radio-AGN main sequence'' links mean radio-AGN power and galaxy stellar mass since $z$$\sim$3}

\author{I.~Delvecchio\inst{1}\thanks{email: ivan.delvecchio@inaf.it}
\and E.~Daddi\inst{2}
\and M. T. Sargent\inst{3,4}
\and J. Aird\inst{5,6}
\and J. R. Mullaney\inst{7}
\and B. Magnelli\inst{2}
\and D. Elbaz\inst{2}
\and L. Bisigello\inst{8,9,10}
\and \\L. Ceraj\inst{11}
\and S. Jin\inst{12,13}
\and B. S. Kalita\inst{2}
\and D. Liu\inst{14}
\and M. Novak\inst{15}
\and I. Prandoni\inst{16}
\and J. F. Radcliffe\inst{17,18,19}
\and C. Spingola\inst{16}
\and \\G. Zamorani\inst{10}
\and V. Allevato\inst{20}
\and G. Rodighiero\inst{8,9}
\and V. Smol{\v{c}}i{\'c}\inst{21}
}

\institute{
INAF - Osservatorio Astronomico di Brera, via Brera 28, I-20121, Milano, Italy \& via Bianchi 46, I-23807, Merate, Italy
\and Universit\'e Paris-Saclay, Universit\'e Paris Cit\`e, CEA, CNRS, AIM, 91191, Gif-sur-Yvette, France
\and International Space Science Institute (ISSI), Hallerstrasse 6, CH-3012 Bern, Switzerland
\and Astronomy Centre, Department of Physics \& Astronomy, University of Sussex, Brighton, BN1 9QH, England
\and Institute for Astronomy, University of Edinburgh, Royal Observatory, Edinburgh, EH9 3HJ, UK
\and School of Physics \& Astronomy, University of Leicester, University Road, Leicester LE1 7RJ, UK
\and Department of Physics and Astronomy, University of Sheffield, Sheffield S3 7RH, UK
\and Department of Physics and Astronomy, Università degli Studi di Padova, Vicolo dell'Osservatorio 3, I-35122, Padova, Italy
\and INAF - Osservatorio Astronomico di Padova, Vicolo dell’Osservatorio 5, I-35122, Padova, Italy
\and INAF - Osservatorio Astronomico di Bologna, via P. Gobetti 93/3, 40129 Bologna, Italy
\and Ru{\dj}er Bo\v{s}kovi\'c Institute, Bijeni\v{c}ka cesta 54, 10000 Zagreb, Croatia
\and Cosmic Dawn Center (DAWN)
\and DTU-Space, Technical University of Denmark, Elektrovej 327, DK-2800 Kgs. Lyngby, Denmark
\and Max-Planck-Institut f\"{u}r extraterrestrische Physik (MPE), Giessenbachstrasse 1, D-85748 Garching, Germany
\and Independent Researcher
\and INAF - Istituto di Radioastronomia, Via P. Gobetti 101, 40129 Bologna, Italy 
\and Department of Physics, University of Pretoria, Lynnwood Road, Hatfield, Pretoria 0083, South Africa
\and Jodrell Bank Centre for Astrophysics, University of Manchester, Oxford Road, Manchester M13 9PL, UK 
\and National Institute for Theoretical and Computational Sciences (NITheCS) South Africa
\and INAF-Osservatorio astronomico di Capodimonte, Via Moiariello 16, I-30131 Naples, Italy
\and Department of Physics,  University of Zagreb,  Bijeni{\v{c}}ka cesta 32, 10002 Zagreb, Croatia
}

   \date{Received }

\abstract{
Mapping the average AGN luminosity across galaxy populations and over time encapsulates important clues on the interplay between supermassive black hole (SMBH) and galaxy growth. This paper presents the demography, mean power and cosmic evolution of radio AGN across star-forming galaxies (SFGs) of different stellar masses ($\mathcal{M_{*}}$). We exploit deep VLA-COSMOS 3~GHz data to build the rest-frame 1.4~GHz AGN luminosity functions at 0.1$\leq$$z$$\leq$4.5 hosted in SFGs. Splitting the AGN luminosity function into different $\mathcal{M_{*}}$ bins reveals that, at all redshifts, radio AGN are both more frequent and more luminous in higher $\mathcal{M_*}$ than in lower $\mathcal{M_*}$ galaxies. The cumulative kinetic luminosity density exerted by radio AGN in SFGs peaks at $z$$\sim$2, and it is mostly driven by galaxies with 10.5$\leq$$\log$($\mathcal{M_{*}}$/$\mathcal{M_{\odot}}$)$<$11. Averaging the cumulative radio AGN activity across \textit{all} SFGs at each ($\mathcal{M_{*}}$,$z$) results in a ``radio-AGN main sequence'' that links the \textit{time-averaged} radio-AGN power $\langle$$L_{1.4}^{\mathrm{AGN}}$$\rangle$ and galaxy stellar mass, in the form: $\log$$\langle$[$L_{1.4}^{\mathrm{AGN}}$/~W~Hz$^{-1}]\rangle$ = (20.97$\pm$0.16) + (2.51$\pm$0.34)$\cdot$$\log$(1+$z$) + (1.41$\pm$0.09)$\cdot$($\log$[$\mathcal{M_{*}}$/$\mathcal{M_{\odot}}$] -- 10). The \textit{super-linear} dependence on $\mathcal{M_{*}}$, at fixed redshift, suggests enhanced radio-AGN activity in more massive SFGs, as compared to star formation. We ascribe this enhancement to both a higher radio AGN duty cycle \textit{and} a brighter radio-AGN phase in more massive SFGs. A remarkably consistent $\mathcal{M_{*}}$ dependence is seen for the evolving X-ray AGN population in SFGs. This similarity is interpreted as possibly driven by secular cold gas accretion fueling both radio and X-ray AGN activity in a similar fashion over the galaxy's lifetime. 
}

\titlerunning{A radio-AGN main sequence}
\authorrunning{I. Delvecchio et al.}

  \maketitle

%________________________________________________________________

\section{Introduction} \label{intro}

Understanding what drives the interplay between supermassive black holes (SMBHs) and their host galaxies is among the most debated topics in extragalactic astrophysics. Multi-wavelength surveys in the last decade have enabled us to reconstruct the cosmic history of SMBH accretion and star formation \citep{Madau+14} finding a seemingly consistent decline since the ``cosmic noon'' ($z$$\sim$1--3, e.g., \citealt{Forster+20}). Fossil remnants of such interplay, the empirical BH-galaxy scaling relations seen at $z$$\sim$0 might suggest an intertwined evolution \citep{Kormendy+13}, which is likely self-regulated by the galaxy's baryon cycle of feeding and feedback mechanisms (e.g., \citealt{Harrison17}). 

To explain BH-galaxy scaling relations, cosmological simulations advocate a two-fold phase of Active Galactic Nuclei (AGN) feedback, characterized by high radiative (``quasar mode'') and high kinetic (``jet mode'') power, whose cumulative effect are able to regulate star formation in massive galaxies (stellar mass $\mathcal{M_{*}}$$>$10$^{10}$~$\mathcal{M_{\odot}}$), preventing the runaway galaxy mass growth (\citealt{DiMatteo+05}; \citealt{Croton+06}; \citealt{Zubovas+12}). While the scenario of ``expulsive'' AGN feedback is not fully backed-up by observations (e.g., \citealt{Sanders+22}), alternative manifestations of AGN feedback have been recently gaining consensus. For instance, compact ($<$1~kpc), low-power radio jets have been widely observed in local (``radio-quiet'') Seyferts (e.g., \citealt{Jarvis+20}; \citealt{Venturi+21}; \citealt{Girdhar+22}), powering ionized-gas outflows that inject heat and turbolence in the interstellar medium, thus possibly reducing the host' SF efficiency. Nevertheless, the \textit{long-term} impact (or lack thereof) of AGN feedback on galaxy growth is still highly controversial (e.g., \citealt{Harrison17}).

In the local Universe ($z$$<$0.3), a number of studies have reported an increasing incidence of jet-driven AGN activity (often visible at radio frequencies) in more massive galaxies \citep{Heckman+14}, which reaches close to 100\% at $\mathcal{M_{*}}$$>$10$^{11.5}$~$\mathcal{M_{\odot}}$ \citep{Sabater+19}, especially if above low-frequency radio luminosities of $\log$[$L_{\mathrm{150~MHz}}$ / W~Hz$^{-1}$]$\geq$21.7. Given the known BH-galaxy mass scaling relations \citep{Kormendy+13}, this trend might reflect the increasing ability of more massive BHs to launch more powerful jets (e.g., \citealt{Best+05}).

However, capturing the \textit{instantaneous} effect of AGN feedback has proven to be difficult. The distribution of black hole accretion rate (BHAR) normalized by galaxy $\mathcal{M_{*}}$ (or ``specific BHAR'', sBHAR$\propto$$L_X$/$\mathcal{M_{*}}$\footnote{If not otherwise specified, $L_X$ is the absorption-corrected rest-frame [2--10]~keV luminosity from AGN.}; \citealt{Aird+12}) is notably broad ($>$1~dex; e.g., \citealt{Mendez+13}; \citealt{Azadi+15}; \citealt{Aird+18}) and is subject to short-term variability relative to the stellar emission from the host (0.1--1~Myr vs 100 Myr; \citealt{Novak+11}; \citealt{Mullaney+12}; \citealt{Aird+13}; \citealt{Hickox+14}; \citealt{Schawinski+15}). Moreover, the competing host-galaxy light from star-forming processes or the effect of circum-nuclear gas and/or dust obscuration may easily wash out AGN emission in individual objects. Thus, a growing practice is now to measure the \textit{average} AGN power imprinted on large and homogeneous galaxy samples. These techniques often include either image stacking (e.g., \citealt{Mullaney+12}; \citealt{Chen+13}; \citealt{Rodighiero+15}; \citealt{Yang+17}; \citealt{Carraro+20}; \citealt{Ito+22}), phenomenological modeling of the AGN luminosity function (\citealt{Caplar+15}, \citeyear{Caplar+18}; \citealt{Weigel+17}; \citealt{Jones+19}; \citealt{Bernhard+19}; \citealt{Delvecchio+20}), Bayesian modeling of sBHAR distributions with detections and non-detections (e.g., \citealt{Aird+18}; \citeyear{Aird+19}; \citealt{Grimmett+19}), or N-body simulations via continuity equations and ``abundance matching'' (e.g., \citealt{Behroozi+13}; \citealt{Grylls+19}; \citealt{Shankar+19}; \citealt{Allevato+21}).

The emerging consensus from deep X-ray observations at 0$\lesssim$$z$$<$3 is that radiative (X-ray) AGN activity appears to be prevalent in more massive and distant star-forming galaxies (SFGs; e.g., \citealt{Yang+18}; \citealt{Aird+19}; \citealt{Carraro+20}; \citealt{Delvecchio+20}). Specifically, \citet{Aird+19} found that X-ray AGN at a given redshift are more frequently triggered in more massive galaxies, while - at fixed $\mathcal{M_{*}}$ - the typical sBHAR (or $L_{X}$) increases with redshift, possibly induced by the increasing cold gas fractions (\citealt{Tacconi+18}; \citeyear{Tacconi+20}; \citealt{Liu+19}).

When averaging over all X-ray luminosities, the linear \textit{mean} $\langle$$L_{X}$$\rangle$ is found to strongly correlate with galaxy $\mathcal{M_{*}}$, featuring a \textit{super-linear} "X-ray AGN main sequence" (gradient $\sim$ 1.5; e.g., \citealt{Aird+19}). This behaviour suggests enhanced radiative AGN activity in massive galaxies relative to star formation, which by contrast \textit{sub-linearly} increases with $\mathcal{M_{*}}$ along the ``star-forming main sequence'' (MS, e.g., \citealt{Noeske+07}; \citealt{Elbaz+11}; \citealt{Schreiber+15}; \citealt{Lee+15}; \citealt{Rinaldi+22}). Therefore, the shape and evolution of $\langle$$L_{X}$$\rangle$ encapsulates important clues on the interplay between SMBH and galaxy growth.
\smallskip

Building upon the above studies, our main goal is to investigate the relationship between radio-AGN activity and galaxy star formation within SFGs, at various stellar masses and redshifts. Specifically, this study aims to elucidate the role of the host galaxy in triggering and sustaining long-term radio-AGN activity over cosmic time. Similarly to what was done for X-ray AGN, we thus explore the existence of a possible ``radio-AGN main sequence'' that links mean rest-frame 1.4~GHz AGN luminosity ($\langle$$L_{1.4}^{\mathrm{AGN}}$$\rangle$) and galaxy $\mathcal{M_{*}}$, at different redshifts (0.1$\leq$$z$$\leq$4.5). Measurements of the typical radio AGN power across $\mathcal{M_*}$-selected SFGs are currently inferred only in the local Universe ($z$$<$0.3; e.g., \citealt{Sabater+19}). At higher redshift, a common roadblock for the calculation of $\langle$$L_{1.4}^{\mathrm{AGN}}$$\rangle$ is how to quantify contamination from SF-driven synchrotron emission, which not only varies across the galaxy population, but is notoriously dominant over AGN emission at 1.4~GHz flux densities below 100~$\mu$Jy (e.g., \citealt{Prandoni+15}; \citealt{Smolcic+17a}; \citealt{Novak+18}; \citealt{Kono+22}), where the bulk of radio SFGs and radio-faint AGN lies (\citealt{Padovani16}).

Radio-synchrotron emission at rest-frame 1--10~GHz offers an independent baseline to estimate dust-unbiased star formation rates (SFRs; \citealt{Condon92}; \citealt{Murphy+11}; \citealt{Prandoni+15}; but see \citealt{Algera+21} for free-free emission at higher frequencies), and thereby decompose the radio emission into SF and AGN contributions. Local SFGs follow a strikingly tight ($\sigma$$<$0.2 dex; \citealt{Molnar+21}) correlation between total IR luminosity (rest-frame 8-1000$~\mu$m, $L_{IR}$) and 1.4~GHz luminosity $L_{1.4}^{\mathrm{SF}}$ arising from star formation. This so-called "infrared-radio correlation" (IRRC) is often expressed in terms of $q_{\mathrm{IR}}$$\approx$$\log$($L_{IR}$/$L_{1.4}^{\mathrm{SF}}$) (e.g., \citealt{Helou+85}; \citealt{Condon92}; \citealt{Bell03} and references therein). At higher redshift, flux-limited IR/radio samples yield a significant decline of $q_{\mathrm{IR}}$ with redshift (e.g., \citealt{Magnelli+15}; \citealt{Delhaize+17}). However, when accounting for various selection effects, \citealt{Delvecchio+21} (hereafter {D21}) found that the median SF-driven $q_{\mathrm{IR}}$ (or $q_{\mathrm{IRRC}}$) shows a primary dependence on $\mathcal{M_*}$, and a much less significant redshift evolution. Their relation can be re-written in $\log$-space as:
% % % % 
\begin{equation}
 q_{\mathrm IRRC} (\mathcal{M_{*}},z) = A + B \cdot \log(1+z) + (C) \cdot \left(\log  \frac{\mathcal{M_*}}{\mathcal{M_{\odot}}} - 10 \right) ~,
   \label{eq:bestq}
\end{equation}
% % % 
where $A$=(2.646$\pm$0.024), $B$=(--0.137$\pm$0.048), $C$=(0.148$\pm$0.013). This was obtained by exploiting $>$400,000 $NUVrJ$-selected SFGs in the COSMOS field and stacking their ancillary infrared (\textit{Herschel}, SCUBA, AzTEC; \citealt{Jin+18}) and radio (VLA-3~GHz; \citealt{Smolcic+17a}, as well as depth-matched MeerKAT-1.3GHz; \citealt{Jarvis+16}; \citealt{Heywood+22}) images across an unprecedented $\mathcal{M_{*}}$-$z$ range, removing radio AGN contamination through a recursive approach. These findings argue that more massive SFGs are radio brighter, at fixed $L_{IR}$, than lower-$\mathcal{M_{*}}$ analogues. A broadly similar $\mathcal{M_{*}}$--dependence has been independently confirmed by deep LOFAR 150-MHz data at $z$$\lesssim$1 over a 10$\times$ larger area than COSMOS (\citealt{Smith+21}; \citealt{Bonato+21}), as well as by local derivations (\citealt{Molnar+21}; \citealt{Matthews+21}; \citealt{Heesen+22}).

The implications of these new $\mathcal{M_{*}}$-dependent recipes are two-fold. Firstly, they enable us to reliably convert radio emission into SFR (from $L_{IR}$; e.g., \citealt{Kennicutt+12}, or from SED-based SFRs, e.g., \citealt{Smith+21}). For instance, the prescription from D21 has proved useful to reproduce the evolution of SFR density (SFRD) at $z$$>$3 \citep{vanderVlugt+22}, reaching the current best agreement with the dust-corrected UV estimate by \citet{Madau+14} as compared to pure $z$-declining IRRCs. Secondly, these $\mathcal{M_{*}}$-dependent prescriptions are instrumental for identifying ``radio-excess'' AGN, i.e., that display excess radio emission (i.e., lower $q_{\mathrm{IR}}$ or higher radio/$\mathrm{SFR}$ ratio) relative to that expected from star formation alone (e.g., \citealt{Donley+05}; \citealt{DelMoro+13}; \citealt{Delvecchio+17}), across a wide range of $\mathcal{M_{*}}$ and redshift.

In the present study, we leverage the above-mentioned ($\mathcal{M_{*}}$,$z$)-dependent IRRC to quantify $\langle$$L_{1.4}^{\mathrm{AGN}}$$\rangle$ as a function of $\mathcal{M_{*}}$ and redshift across the global population of $\mathcal{M_{*}}$-selected SFGs. From known radio-excess AGN placed at $>$2$\sigma$ from the IRRC (with $\sigma$$\sim$0.22~dex, D21), we adopt a novel approach to factor in the statistical contribution of radio AGN within the IRRC, that is critical to compute a representative sample-averaged AGN power. Firstly, we construct the 1.4~GHz luminosity function of radio-excess AGN in SFGs (``AGN RLF'' hereafter) at different redshifts, following previous studies (e.g., \citealt{Smolcic+17c}; \citealt{Novak+18}; \citealt{Ceraj+18}; \citealt{Butler+19}; \citealt{Kono+22}). Secondly, to quantify the mean radio-AGN power at each $\mathcal{M_{*}}$, we split the AGN RLF across different $\mathcal{M_{*}}$ \textit{and} redshift bins, fitting and integrating each luminosity function down to the exact same $L_{1.4}^{\mathrm{SF}}$ set by the IRRC at that ($\mathcal{M_{*}}$,$z$). This self-consistent approach allows us to assess the \textit{cumulative} power exerted by radio AGN in SFGs at various $\mathcal{M_{*}}$, including the elusive contribution of radio-faint AGN within the IRRC. 

The layout of this paper is as follows. Sect.~\ref{sample} describes the sample selection. Sect.~\ref{class} presents the $\mathcal{M_{*}}$-dependent radio source classification. The 1.4~GHz AGN RLF split in different $\mathcal{M_{*}}$ and redshift bins is discussed in Sect.~\ref{rlf}. Sect.~\ref{results} illustrates the integrated and mean power of radio AGN across the entire SFG population, including our first derivation of the ``radio-AGN main sequence''. In Sect.~\ref{discussion}, its shape and evolution are compared to those of X-ray AGN and star formation, discussing the broad implications for AGN feedback in SFGs over cosmic time. We report our main conclusions in Sect.~\ref{summary}.

Throughout this paper, magnitudes are given in the AB system \citep{Oke74}. We assume a \citet{Chabrier03} initial mass function (IMF) and a $\Lambda$CDM  cosmology with $\Omega_{\rm m}$ = 0.30, $\Omega_{\rm \Lambda}$ = 0.70, and H$\rm _0$ = 70 km s$^{-1}$ Mpc$^{-1}$ \citep{Spergel+03}.

%________________________________________________________________

% % % -------------------------------------------------------------------------------
\section{Sample selection} \label{sample}
% % % -------------------------------------------------------------------------------

\subsection{Selection criteria and final sample}   \label{sample_selection}

\begin{figure}
\centering
     \includegraphics[width=\linewidth]{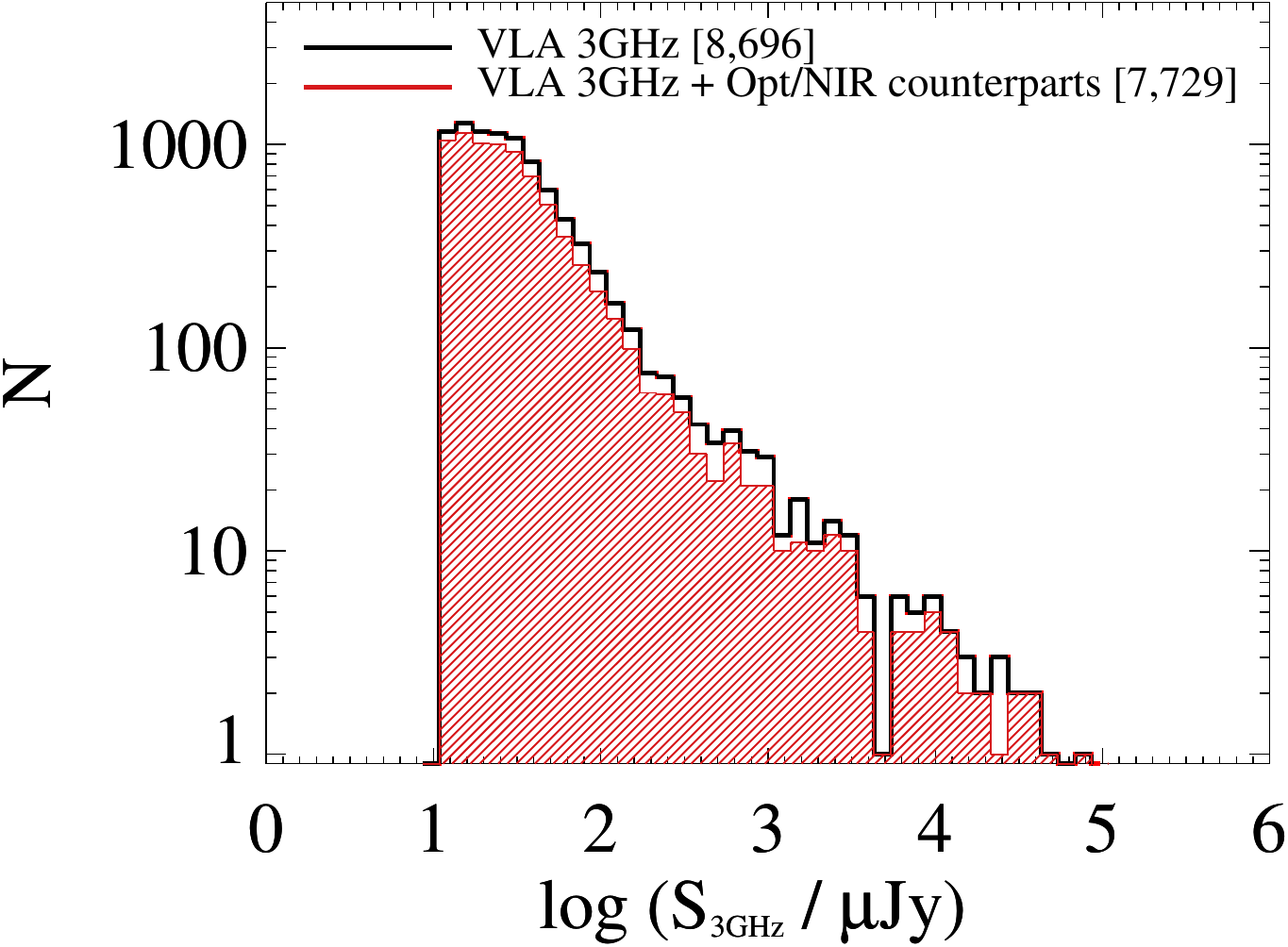}
 \caption{\small Flux distribution of VLA-3~GHz sources detected at $S/N$$\geq$5 (8,696, black) over the COSMOS 1.77~deg$^2$ field \citep{Smolcic+17a}. The subset having an optical/NIR counterpart in the COSMOS2015 catalogue \citep{Laigle+16} is also highlighted (7,729 sources, red dashed).
 }
   \label{fig:ref1}
\end{figure}

For this analysis we exploit deep radio-continuum data from the VLA-COSMOS 3 GHz Large Project \citep{Smolcic+17a}, one of the most sensitive radio surveys ever conducted across a medium sky area like COSMOS (rms=2.3~$\mu$Jy~beam$^{-1}$). With an angular resolution of 0.75'', the total number of S/N$>$5 detections reaches 10,830 over an area of 2.6~deg$^2$. As detailed in \citet{Smolcic+17b}, 8,696 detections (Fig.~\ref{fig:ref1}, black histogram) are contained within 1.77~deg$^2$ with optical and near-infrared (NIR) coverage. Of these, 7,729 (Fig.~\ref{fig:ref1}, red histogram) were assigned an optical/NIR counterpart from the COSMOS2015 catalog \citep{Laigle+16} via a maximum likelihood approach. From these 7,729 counterparts, we further restrict ourselves to the central 1.5~deg$^2$ area of the COSMOS field with deeper Ultra-VISTA coverage (6,742 sources), in order to also exploit available de-blended far-IR/sub-mm photometry \citep{Jin+18} extracted on $K_s$+MIPS~24~$\mu$m+VLA positional priors (see also \citealt{Liu+18}). This dataset was used in D21 to retrieve de-blended $L_{IR}$ estimates for individual FIR/sub-mm detections, as well as to obtain median-stacked $L_{IR}$ from non-detections in different $\mathcal{M_{*}}$ and redshift bins. 

%%%placing figure  ---------------------------
%%%-------------------------------------

Furthermore, within our 6,742 VLA-3~GHz detections, we select only radio sources within the ``blue'' wedge of the [NUV-r]/[r-J] diagram (about 85\%), using dust-corrected rest-frame magnitudes from \citet{Laigle+16}. This latter criterion restricts our final sample to 5,658 3~GHz detected \textit{star-forming} galaxies (see e.g., \citealt{Davidzon+17}). Adopting the ($\mathcal{M_{*}}$,$z$)-dependent IRRC from D21, calibrated on SFGs, enables us to self-consistently identify radio-excess AGN within our sample. We acknowledge that radio-detected galaxies are biased towards higher stellar masses, at fixed redshift, than a $\mathcal{M_{*}}$-selected sample (as in D21). Interpreting the properties of radio-detected AGN from the viewpoint of $\mathcal{M_{*}}$-selected SFGs will be addressed in Sect.~\ref{rams}.

We motivate our choice of \textit{not} including $NUVrJ$-selected passive galaxies in our analysis in Sect.~\ref{passive}. Finally, for consistency with the $\mathcal{M_{*}}-z$ space over which the IRRC was calibrated (D21), we consider sources within the same range: 0.1$\leq$$z$$\leq$4.5 and 9$\leq$$\log(\mathcal{M_{*}}$/$\mathcal{M_{\odot}}$)$\leq$12. These cuts yield our final sample of 5,658 radio-detected (S/N$>$5 at 3~GHz) star-forming galaxies across 1.5~deg$^2$. Table \ref{tab:sample} outlines the various steps towards the final sample.

%%Placing Table
\begin{table}
\centering
   \caption{Main numbers that lead to our final sample of 5,658 radio-detected galaxies. Note that the subset of radio AGN used throughout this work will be introduced in Section \ref{class}.}
\begin{tabular}{l c }
\hline
\hline
Definition      &       $\#$    \\
 \hline
VLA-3~GHz (1.77~deg$^2$)                                                            & 8,696 \\
VLA-3~GHz + Opt/NIR counterparts (1.77~deg$^2$)                                     & 7,729 \\
- UltraVISTA area (1.5~deg$^2$)                                                     & 6,742 \\
- $NUVrJ$ star-forming galaxies                                                     & 5,734 \\
- final sample:     &   \\
0.1$\le$$z$$\le$4.5 and 9$\le$$\log(\mathcal{M_{*}}$/$\mathcal{M_{\odot}}$)$\le$12    & 5,658 \\
\hline
\end{tabular}
\label{tab:sample}
\end{table}
%%%------------------------------------- 

\subsection{Parameter estimates}  \label{parameters}

For consistency with the publicly-available VLA catalogue \citep{Smolcic+17b}, we adopt the same 3~GHz radio flux densities. These are scaled to 1.4~GHz assuming $S_{\nu}$$\propto$$\nu^{\gamma}$, with the 1.4--3~GHz spectral index ($\gamma$) being inferred from the observed flux densities at both frequencies whenever available ($\approx$25\% of the sample). Otherwise, we assume $\gamma$=--0.75$\pm$0.1 (\citealt{Condon92}; \citealt{Ibar+09}, \citeyear{Ibar+10}; \citealt{Magnelli+15}), which is close to the median spectral index of the 3~GHz population using survival analysis (see Section 4 in \citealt{Smolcic+17a}). We verify that taking $\gamma$=--0.75$\pm$0.1 for all galaxies would not affect the conclusions of this work (see Appendix \ref{Appendix_B}). 

Estimates of photometric redshifts and $\mathcal{M_{*}}$ are taken from the COSMOS2015 catalogue \citep{Laigle+16} via spectral energy distribution (SED) fitting of the optical-MIR photometry, reading the median value of the likelihood distribution for each source. The typical photometric redshift accuracy is $\left \langle |\Delta z/(1 + z)| \right \rangle =$ 0.007 at $z$$<$3, and 0.021 at 3$<$$z$$<$6 \citep{Laigle+16}, which increases up to only 0.057 for the faintest galaxies (25$<$\textit{i$^{{+}}$}$<$26). The ``super-deblended'' catalogue presented by \citet{Jin+18} also contains publicly available spectroscopic redshifts ($\approx$38\% of the sample, courtesy of M. Salvato), which were prioritized over photometric measurements if deemed reliable ($z$$_{s}$ quality flag $>$3), or allowing for redshift variations within $\pm$10\% for sources with only a photometric redshift (see \citealt{Jin+18}). The same redshifts were used to compute rest-frame 1.4~GHz spectral luminosities ($L_{1.4}$). We note that $\mathcal{M_*}$ estimates were all based on photometric redshifts from COSMOS2015, after verifying a good consistency also for spec-$z$ sources \citep{Jin+18}.

Uncertainties on $L_{1.4}$ are obtained by propagating the errors on flux density and spectral index. As mentioned in Section \ref{sample_selection}, for each radio source with FIR detection (i.e., with combined S/N$>$3 over all FIR/sub-mm bands) from \citet{Jin+18}, a single $L_{IR}$ measurement is taken from their catalogue. For radio sources with no FIR detection, instead, stacked fluxes in each ($\mathcal{M_{*}}$,$z$) bin were obtained from FIR/sub-mm stacking in D21 (see their Sect. 3). These were then converted to $L_{IR}$ via SED fitting and finally re-scaled to the $\mathcal{M_{*}}$ and redshift of each source via the MS (Sect. 4.2.1 in D21) to mitigate underlying sample variance within each bin. Potential mid-IR AGN contamination is accounted for through empirical AGN templates \citep{Mullaney+11} in the SED fitting (\citealt{Liu+18}; \citeyear{Liu+21}). These $L_{IR}$ measurements are preferred to 3$\sigma$ $L_{IR}$ upper limits inferred from FIR/sub-mm SED-fitting \citep{Jin+18} as they provide more stringent constraints for non-detections.

%%%placing figure  ---------------------------
\begin{figure*}
\centering
     \includegraphics[width=\linewidth]{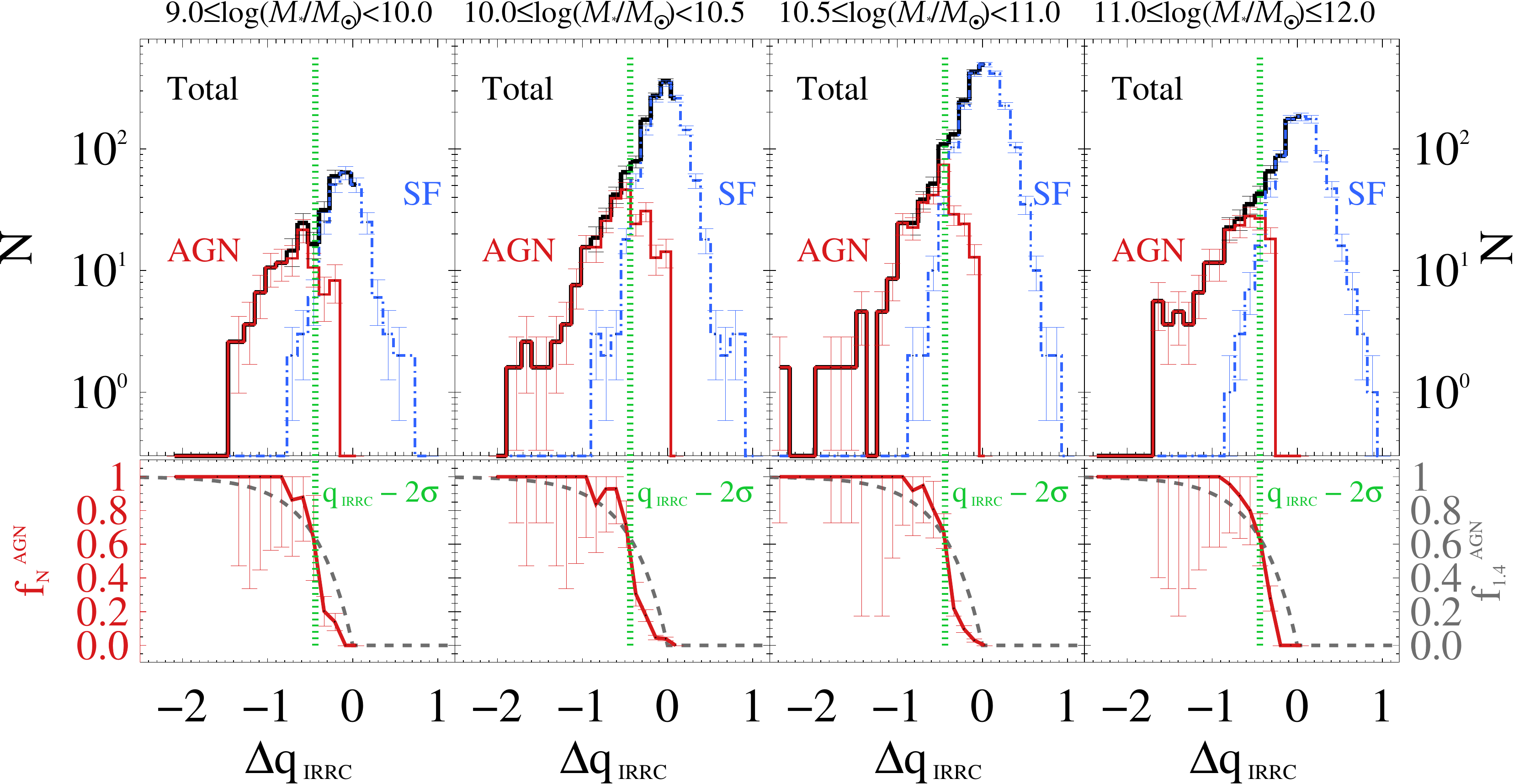}
 \caption{\small \textit{(Top panels)}: Distribution of 3~GHz detections (black) as a function of $\Delta q_{\mathrm{IRRC}}$=$q_{\mathrm{IR}} - q_{\mathrm{IRRC}}$, split in different $\mathcal{M_{*}}$ bins (increasing from left to right). The subsample of radio SFGs (blue dot-dashed) and AGN (red) are separated on statistical basis at a threshold of $\Delta q_{\mathrm{IRRC}}$=--0.44~dex (green vertical line), that is a 2$\sigma$ offset from the IRRC. \textit{(Bottom panels)}: Distribution of AGN classification purity (or $f_{N}^{\mathrm{AGN}}$, red solid line) and AGN 1.4~GHz luminosity purity ($f_{1.4}^{\mathrm{AGN}}$, grey dashed line), both shown as a function of $\Delta q_{\mathrm{IRRC}}$. At the threshold $\Delta q_{\mathrm{IRRC}}$=--0.44~dex, we get $f_{N}^{\mathrm{AGN}}$$\approx$$f_{1.4}^{\mathrm{AGN}}$$\approx$64\%. For details, see Sect.~\ref{corrections}.
 }
   \label{fig:histo_class}
\end{figure*}
%%%-------------------------------------

% % -------------------------------------------------------------------------------
\section{Radio source classification} \label{class}
% % % -------------------------------------------------------------------------------

In this section, we describe the methodology adopted in this work to identify AGN through the radio-excess criterion. This method aims to distribute galaxies into two categories depending on the physical process likely producing the radio emission, regardless of other multi-band AGN diagnostics. This can be carried out by decomposing the observed $q_{\mathrm{IR}}$ (or equivalent formalism) distribution of radio detections into a dominant (at flux densities below 100~$\mu$Jy~beam$^{-1}$, e.g., \citealt{Prandoni+15}) population of SFGs and a smaller, skewed low-$q_{\mathrm{IR}}$ tail, ascribed to radio AGN. The dividing threshold is somewhat empirical, but reflects a trade-off between AGN purity and completeness. In the literature, such radio-excess thresholds have been taken as redshift-invariant (e.g., \citealt{DelMoro+13}), or mildly redshift-dependent (e.g., \citealt{Delvecchio+17} and used in \citealt{Smolcic+17a}). Contrary to these studies, we base our radio-excess criterion on the ($\mathcal{M_{*}}$,$z$)-dependent IRRC derived in D21 (Eq.~\ref{eq:bestq}), as detailed below. A discussion on the impact of using a different radio-excess criterion is given in Appendix~\ref{Appendix_B}.

\subsection{A $\mathcal{M_{*}}$-dependent radio source classification} \label{histo}

To categorize radio detections into SFGs or radio-excess AGN, we rely upon the recent ($\mathcal{M_{*}}$,$z$)-dependent IRRC from D21. For each source we calculate its $q_{\mathrm{IRRC}}$ (from Eq.~\ref{eq:bestq}), and we measure the offset from the observed $q_{\mathrm{IR}}$, namely $\Delta q_{\mathrm{IRRC}}$=$q_{\mathrm{IR}} - q_{\mathrm{IRRC}}$. As detailed in D21, this decomposition analysis in the space of $\Delta q_{\mathrm{IRRC}}$ can be used to separate the AGN and SFG populations on \textit{statistical} basis. We recall that D21 restricted this procedure to a subset of radio detections for which the observed $q_{\mathrm{IR}}$ range is accessible by a $NUVrJ$-selected SFG, at each ($\mathcal{M_{*}}$,$z$). Their requirement translated into a $\mathcal{M_{*}}$ cut (i.e., $\mathcal{M_{*}}$$>$10$^{10.5}$~$\mathcal{M_{\odot}}$). This procedure was then extended to lower $\mathcal{M_{*}}$ using stacked $q_{\mathrm{IR}}$ measurements.

On the other hand, in the present analysis we aim at constructing the AGN luminosity function of 3~GHz detections, thus we use their full observed $q_{\mathrm{IR}}$ (and $\mathcal{M_{*}}$) range. Hence, we study the $\Delta q_{\mathrm{IRRC}}$ distribution of \textit{all} radio detections in four different $\mathcal{M_{*}}$ bins (see Fig.~\ref{fig:histo_class}): 9$<$$\log(\mathcal{M_{*}}$/$\mathcal{M_{\odot}}$)$<$10; 10$<$$\log(\mathcal{M_{*}}$/$\mathcal{M_{\odot}}$)$<$10.5; 10.5$<$$\log(\mathcal{M_{*}}$/$\mathcal{M_{\odot}}$)$<$11 and 11$<$$\log(\mathcal{M_{*}}$/$\mathcal{M_{\odot}}$)$<$12. The internal redshift variations are already factored in the $q_{\mathrm{IRRC}}$ term.

Fig.~\ref{fig:histo_class} (top panels) displays the observed $\Delta q_{\mathrm{IRRC}}$ distribution of our 5,658 radio-detections in all four $\mathcal{M_{*}}$ bins (increasing from left to right). The total distribution (black) is split between SFGs (blue dot-dashed) and AGN (red) following these steps: (i) the mode of the distribution is identified and assumed to trace pure star formation; (ii) the right-hand side of the $\Delta q_{\mathrm{IRRC}}$ distribution is mirrored to the left, building a symmetric (log-normal like) function interpreted as purely driven by star formation (blue dot-dashed); (iii) the residual distribution in excess to SFGs is statistically ascribed to AGN (red). We note that the SF histogram peaks at $\Delta q_{\mathrm{IRRC}}$$\approx$0 in all $\mathcal{M_{*}}$ bins, which highlights the validity of our IRRC prescription to the entire sample. For simplicity, a fixed threshold of $\Delta q_{\mathrm{IRRC}}$=--0.44~dex (green vertical line) is set to separate SFGs (above) from radio-excess AGN (below). This threshold indicates a 2$\sigma$ offset from the IRRC, with $\sigma$ being the dispersion of the SF gaussian found at $\mathcal{M_{*}}$$>$10$^{10.5}$~$\mathcal{M_{\odot}}$ (D21). 

D21 estimated that roughly 80\% of the AGN distribution lies below this dividing threshold, assuming a log-normal function. Here we refrain from fitting the AGN distribution at each $\mathcal{M_{*}}$, since a proper characterization of the \textit{shape} of the AGN population is beyond the scope of this paper. However, we acknowledge that radio AGN within the IRRC might be under-represented from our method, as they are defined from the residual of the SFG distribution. Hence, we do not attempt at quantifying radio AGN within 2$\sigma$ from the IRRC \textit{directly} from the observed distribution. Instead, the contribution of such radio-faint AGN will be estimated \textit{indirectly} from the AGN RLF and is discussed in Sect.~\ref{rlf}.

\subsection{Statistical corrections to classification method} \label{corrections}

A caveat of our approach concerns the purity of the radio-excess AGN sample. While galaxies within $\pm$2$\sigma$ from the IRRC are likely to have radio emission mainly powered by star formation, SFGs can partially contaminate the AGN population at the threshold $\Delta q_{\mathrm{IRRC}}$$\lesssim$--0.44~dex (D21). This is because D21 prioritized a clean identification of SFGs rather than that of radio-excess AGN. Here we quantify this effect, while the corresponding correction for these mis-classified galaxies is described in Appendix~\ref{corr_purity}.  

We thus define $f_{N}^{\mathrm{AGN}}$ = $N_{\mathrm{AGN}}$/($N_{\mathrm{AGN}}$+$N_{\mathrm{SF}}$) as the number of radio AGN ($N_{\mathrm{AGN}}$, red histogram) divided by the number of all radio-detections ($N_{\mathrm{AGN}}$+$N_{\mathrm{SF}}$, black histogram), in each $\Delta q_{\mathrm{IRRC}}$ bin. This is a proxy for AGN purity (in number), that is the probability of a radio source being classified as a radio AGN in our sample, at a given $\Delta q_{\mathrm{IRRC}}$ and $\mathcal{M_{*}}$. This is marked in Fig.~\ref{fig:histo_class} (bottom panels) and plotted as red solid line. Error bars on $f_{N}^{\mathrm{AGN}}$ are propagated from the number ratio assuming Poissonian statistics if $N_{\mathrm{AGN}}$$\geq$5, otherwise we used the tabulated 1$\sigma$ confidence intervals from \citet{Gehrels86}. It is evident that $f_{N}^{\mathrm{AGN}}$$\sim$60--70\% at the threshold of $\Delta q_{\mathrm{IRRC}}$$=$--0.44~dex, hence such a correction is \textit{non negligible} in any $\mathcal{M_{*}}$ bin. We also note that, although the $f_{N}^{\mathrm{AGN}}$ functions appear self-similar at all $\mathcal{M_{*}}$, the relative fraction of radio-excess AGN \textit{apparently} increases at $\log(\mathcal{M_{*}}$/$\mathcal{M_{\odot}}$)$<$10. This is mainly a selection effect due to our sample being progressively less complete at lower $\mathcal{M_{*}}$, and thus biased towards the brightest radio detections, i.e., AGN. 

A further caveat caused by our simple thresholding method is that 100\% of radio light is implicitly \textit{assumed} to originate from the process of the corresponding class (AGN or star formation). Instead, each object likely hides a composite nature that is important to assess, particularly in radio-faint AGN close to the IRRC, where both emission processes can substantially contribute to the total (e.g., \citealt{Maini+16}; \citealt{HerreraRuiz+17}; \citealt{Radcliffe+18}). Following previous studies (e.g., \citealt{Ceraj+18}, \citeyear{Ceraj+20}), the offset from the IRRC can statistically trace the AGN fractional contribution at 1.4~GHz, which is defined as: $f_{1.4}^{\mathrm{AGN}}$ = 1--10$^{\Delta q_{\mathrm IRRC}}$. Similar to $f_{N}^{\mathrm{AGN}}$, $f_{1.4}^{\mathrm{AGN}}$ also ranges from 0 to 1, as shown in the bottom panels of Fig.~\ref{fig:histo_class} (grey dashed line). Moreover, at the threshold $\Delta q_{\mathrm{IRRC}}$=--0.44~dex we obtain $f_{1.4}^{\mathrm{AGN}}$$\approx$64\%, implying that roughly a third of the 1.4~GHz luminosity can be contaminated by star formation. Instead, the median fractions obtained over all $\Delta q_{\mathrm{IRRC}}$ bins below --0.44~dex are 92\% (for $f_{N}^{\mathrm{AGN}}$) and 99\% (for $f_{1.4}^{\mathrm{AGN}}$), suggesting overall a highly reliable AGN selection. As discussed in more detail in Appendix~\ref{corr_l14}, each $L_{1.4}$ estimate can be scaled by the corresponding AGN fraction at 1.4~GHz to isolate the statistical AGN-related contribution in our sample, thus purifying the AGN RLF. 

It is worth noting that $f_{N}^{\mathrm{AGN}}$ (red solid line) drops with $\Delta q_{\mathrm{IRRC}}$ more steeply than $f_{1.4}^{\mathrm{AGN}}$ (grey dashed line). This suggests that our AGN identification method does not perform evenly at all $\Delta q_{\mathrm{IRRC}}$, as it picks AGN more easily in the radio-excess regime than within the IRRC, after controlling for the varying AGN fraction at 1.4~GHz. However, we also acknowledge that both $f_{N}^{\mathrm{AGN}}$ and $f_{1.4}^{\mathrm{AGN}}$ rely upon the measured  $\Delta q_{\mathrm{IRRC}}$, which becomes quite uncertain close to the IRRC, since the typical error bars on $\Delta q_{\mathrm{IRRC}}$ become larger than the actual $\Delta q_{\mathrm{IRRC}}$. This latter argument, together with a poor sampling of the radio AGN distribution at $\Delta q_{\mathrm IRRC}>$--0.44~dex (by construction), motivates our choice of not trusting the corrections for AGN classification and luminosity purity close to the IRRC. As described in Sect.~\ref{kld}, we will only include radio AGN at $\Delta q_{\mathrm IRRC}<$--0.44~dex, while the contribution of radio-fainter AGN above that threshold will be statistically factored in the integrated LF.

% % % -------------------------------------------------------------------------------
\section{Luminosity function of radio-excess AGN in SFGs} \label{rlf}
% % % -------------------------------------------------------------------------------
We describe the methods used to construct the 1.4~GHz AGN RLF (Sect.~\ref{lf_obs}) for our sample of radio-excess AGN. Details on the AGN classification purity and AGN $L_{1.4}$ purity are given in Appendix~\ref{Appendix_A}. The computation of error bars in our data is detailed in Sect.~\ref{boot}. Later, we fit our data with $z$-evolving models, discussing the model parameters estimates and corresponding uncertainties (Sect.~\ref{rlf_redshift}). Finally, in Sect.~\ref{rlf_mass} we split the AGN RLF in various $\mathcal{M_{*}}$ bins, fitting the corresponding data with luminosity or density-evolving models at each $\mathcal{M_{*}}$.

  %%%placing figure  ---------------------------
\begin{figure*}
\centering
     \includegraphics[width=\linewidth]{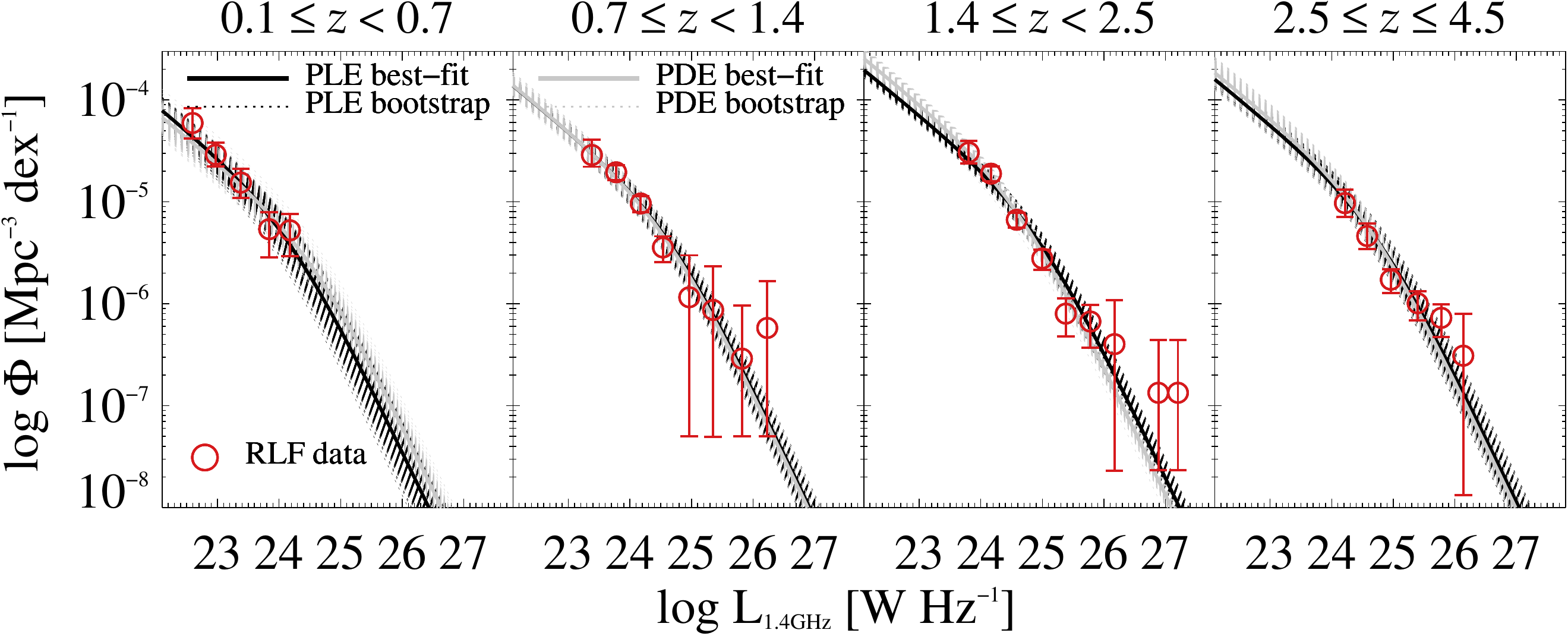}
 \caption{\small Observed AGN RLF (red circles) fitted in each redshift bin with two parametrizations: pure luminosity evolution (PLE, black lines) and pure density evolution (PDE, grey lines). The corresponding best-fitting function obtained through MCMC using the analytical form of \citet{Mauch+07} is marked with a solid line. The 1000 bootstrapped RLFs (dotted lines) delimit the $\pm$1$\sigma$ confidence interval on the fit in each redshift bin. Further details are given in Sect.~\ref{boot}, while best-fit parameters and uncertainties are listed in Table~\ref{tab:fit_z}.
 }
   \label{fig:rlf_fitting_z}
\end{figure*}
%%%-------------------------------------

\subsection{Building the radio-AGN luminosity function}   \label{lf_obs}

To calculate the luminosity function of radio-excess AGN hosted in SFGs, we follow the procedure outlined in \citet{Novak+17} (see their Sect. 3.1.). To summarize briefly, we employ the 1/V$_{\mathrm max}$ method \citep{Schmidt68}, with V$_{\mathrm max}$ being the maximum comoving volume over which a source is detectable, within the survey area and across a given redshift bin. Following \citet{Novak+17}, we further correct for a set of incompletenesses, including radio detection, heterogeneous noise, and resolution biases, as well as for missing optical/NIR counterparts (the latter being $\sim$10\%, see Table~\ref{tab:sample}).

We construct the 1.4 GHz AGN luminosity function in four redshift bins: 0.1$\leq$$z$$<$0.7; 0.7$\leq$$z$$<$1.4; 1.4$\leq$$z$$<$2.5 and 2.5$\leq$$z$$\leq$4.5. This redshift grid was chosen to be large enough to mitigate possible photometric uncertainties (i.e., sources falling into the wrong bin), while being centered on reference values of $z$ $\approx$0.5, 1, 2 and 3, respectively. We report AGN RLFs using the median $L_{1.4}$ in each luminosity bin (0.4~dex wide). The Poissonian uncertainty on the number density $\Phi$(L) in each ($L_{1.4}$,$z$) bin is calculated as in \citet{Marshall85} by weighting each galaxy by its contribution to the total (1/V$_{\mathrm max}$)$^2$. However, if there are fewer than five sources in a luminosity bin, we used the tabulated 1$\sigma$ values for small number statistics \citet{Gehrels86}. We stress that the above steps and corrections to derive $\Phi$(L) are identical to those used in previous luminosity functions based on VLA-COSMOS 3~GHz data (e.g., \citealt{Smolcic+17c}; \citealt{Novak+18}; \citealt{Ceraj+18}; \citealt{vanderVlugt+22}). 

We refer the reader to Appendix~\ref{Appendix_A} for a detailed explanation of how the AGN RLF was corrected for AGN classification purity and $L_{1.4}$ purity, following the reasoning in Sect.~\ref{corrections}.

\subsection{Error propagation via bootstrapping} \label{boot}

The AGN RLF shown in Fig.~\ref{fig:rlf_fitting_z} (red circles) represents our final dataset, whose error bars already incorporate several sources of uncertainty. Here we explain how these uncertainty propagate to the calculation of the final AGN RLF. Specifically, for each datapoint we account for errors on: 
\begin{itemize}
 \item (i) the 1.4~GHz luminosity $L_{1.4}$ of each object from the corresponding flux density and spectral index errors (see Sect.~\ref{parameters});
 \item (ii) the IR luminosity $L_{IR}$ of each object (from SED-fitting), which translates into an error on $q_{\mathrm{IR}}$;
 \item (iii) the fraction $f_{N}^{\mathrm{AGN}}$ of AGN identified from our radio-excess criterion in a given $\Delta q_{\mathrm{IRRC}}$ bin, as displayed in Fig.~\ref{fig:histo_class};
 \item (iv) the fraction $f_{1.4}^{\mathrm{AGN}}$ of AGN-related luminosity at 1.4~GHz, whose uncertainty scales directly from (i) and from the scatter of the IRRC (0.22~dex).
\end{itemize}

We bootstrap over all these uncertainties 1000 times, by assuming at each step a Gaussian shape centered on the nominal parameter value, and a standard deviation given by the 1$\sigma$ error bar on each parameter. For every realization we re-calculate the position of all datapoints (in $L_{1.4}^{\mathrm{AGN}}$ and $\Phi$(L)). Thus we end up with 1000 bootstrapped AGN RLFs, which we interpolate at the 16$^{\mathrm{th}}$ and 84$^{\mathrm{th}}$ percentiles to delimit the $\pm$1$\sigma$ confidence interval reported in Fig.~\ref{fig:rlf_fitting_z}. We note that the nominal sample size might slightly fluctuate among all AGN RLF realizations, due to some faint radio AGN crossing the threshold at $\Delta q_{\mathrm{IRRC}}$$=$--0.44~dex. 

The full baseline of AGN RLF datapoints and uncertainties is listed in Appendix~\ref{Appendix_C}.

% % %   UPDATE TABLE
\begin{table*}
\caption{Results of modelling the evolution of the radio-excess AGN population in SFGs assuming either pure luminosity evolution (PLE), or pure density evolution (PDE) models, relative to the local relation from \citet{Mauch+07}, as detailed in Sect.~\ref{rlf_redshift}. The best-fitting RLF for each model is shown in Fig.~\ref{fig:rlf_fitting_z}.}
\centering
\begin{tabular}{cccccccc}
\hline\hline
$z$-bin &  $\langle~z~\rangle$  & \multicolumn{3}{c}{PLE} & \multicolumn{3}{c}{PDE} \\
{} &    & \multicolumn{3}{l}{\phantom{}}\dotfill & \multicolumn{3}{c}{\phantom{}} \dotfill  \\
{} &    & $\log L^{\star}(z)$   & $\alpha_{\rm{L}}$ & $\chi^{2}_{\nu}$ & $\Phi^{\star}(z)\cdot 10^{-6}$ & $\alpha_{\rm{D}}$ & $\chi^{2}_{\nu}$   \\
{} &    & [$\log$(W~Hz$^{-1}$)] &                   &                  &    [Mpc$^{-3}$~dex$^{-1}$]     &    \\   
\hline
${0.1} \leq z < {0.7}$ & 0.59  & $24.16 \pm 0.07$ & $-2.15 \pm 0.51$ & 0.34 & $4.10 \pm 0.46$ & $-1.42 \pm 0.35$ & 0.22   \\
${0.7} \leq z < {1.4}$ & 1.12  & $24.64 \pm 0.04$ & $0.16 \pm 0.19$ & 0.49 & $8.46 \pm 0.66$ & $0.09 \pm 0.13$ & 0.44  \\
${1.4} \leq z < {2.5}$ & 1.97  & $24.95 \pm 0.04$ & $0.77 \pm 0.11$ & 1.26 & $15.52 \pm 1.56$ & $0.62 \pm 0.08$ & 0.83  \\
${2.5} \leq z \leq {4.5}$ & 3.11  & $24.77 \pm 0.07$ & $0.29 \pm 0.10$ & 1.79 & $11.26 \pm 1.89$ & $0.25 \pm 0.09$ & 1.78   \\
\hline
\end{tabular} \label{tab:fit_z}
\end{table*}

\subsection{Fitting the redshift evolution of the AGN RLF} \label{rlf_redshift}

The redshift evolution of a galaxy/AGN population as a function of ($L$,$z$) is generally expressed by a joint density and luminosity evolution of its local luminosity function:
% % % % 
\begin{equation}
 \Phi(L,z) = (1+z)^{\alpha_D} \cdot \Phi_{0} \left[\frac{L}{(1+z)^{\alpha_L}}\right]
   \label{eq:lf_evol}
\end{equation}
where $\alpha_D$ and $\alpha_L$ are the characteristic density and luminosity evolution parameters, respectively, $\Phi(L,z)$ is the observed AGN RLF, and $\Phi_0$ is the local luminosity function. For consistency with the literature, we adopt the local AGN RLF from \citet{Mauch+07}, which is parametrized as a double power law,
% % % % 
\begin{equation}
 \Phi_{0}(L) = \frac{\Phi^{\star}}{(L^{\star}/L)^{\delta_1} + (L^{\star}/L)^{\delta_2}} ~ ,
   \label{eq:lf0}
\end{equation}
where the parameters are the normalization $\Phi^{\star}$=$\frac{1}{0.4}$$\cdot$10$^{-5.5}$~Mpc$^{-3}$~dex$^{-1}$ calculated at the knee position $L^{\star}$=10$^{24.59}$~W~Hz$^{-1}$, and the bright and faint end slopes $\delta_1$=--1.27 and $\delta_2$=--0.49 (with uncertainties of $\pm$0.18 and $\pm$0.04, respectively; see \citealt{Mauch+07}), respectively. These parameters were obtained from 2,661 radio detections at $z$$\sim$0 spanning six decades in $L_{1.4}$, thus ideal to constrain both the faint and bright end of the local RLF. 

Recent AGN RLF studies have also performed a global fit of all redshift bins to retrieve an evolution of $\Phi$(L) with redshift (e.g., \citealt{Novak+18}; \citealt{Kono+22}). In that case, an additional parameter $\beta_L$ (or $\beta_D$) for the redshift evolution is introduced, thus re-formulating the luminosity function of Eq.~\ref{eq:lf_evol} as: 

\begin{equation}
\Phi(L,z) = (1+z)^{\alpha_D + z \cdot \beta_D} \times \Phi_{0} \left[\frac{L}{(1+z)^{\alpha_L + z\cdot\beta_L}}\right] 
\label{eq:lfevol}
\end{equation}

While this formalism is preferable to increase the statistics of input data, it implicitly assumes a simple \textit{linear} trend of the total evolution parameter ($\alpha + z\cdot \beta$) with redshift. This has been proved to be successful in describing the redshift evolution of the \citet{Mauch+07} AGN RLF (e.g., \citealt{Novak+18}; \citealt{Kono+22}). 

We attempt a similar approach by using the Markov chain Monte Carlo (MCMC) algorithm, available in the Python package {\sc emcee} \citep{Foreman-Mackey+13}, to perform a multi-variate fit to these data. Nevertheless, we find that a combined four-parameter fit is loosely constrained by our data. Thus, we investigate two extreme cases of evolution: pure luminosity evolution (PLE, i.e., $\alpha_D$=0) and pure density evolution (PDE, i.e., $\alpha_L$=0). In either cases, we fit each redshift slice independently (i.e., with $\beta_L$=$\beta_D$=0), inferring a best-fit $\alpha_L$ (or $\alpha_D$) in each bin (e.g., \citealt{McAlpine+13}; \citealt{Smolcic+17c}; \citealt{Ceraj+18}; \citeyear{Ceraj+20}; \citealt{Kondapally+22}). We acknowledge that such a simplified method is conceptually similar to fitting L$^{\star}$ (for PLE) or $\Phi^{\star}$ for PDE) at each redshift, assuming a fixed LF shape. However, this formalism enables us to compare our evolving AGN RLF with existing radio-based literature. This approach also allows us to demonstrate that the evolution of radio AGN in SFGs cannot be described by a monotonic trend of the alpha parameters (see below).

Fig.~\ref{fig:rlf_fitting_z} shows the best-fit AGN RLF obtained in each redshift bin through a PLE (black curves) or PDE (grey curves) formalism. The dashed lines around each best-fit RLF are derived by bootstrapping 1000 times over the uncertainties of the evolution parameter $\alpha$ and delimit the uncertainty on the fit in each redshift bin. We list all output parameters and related 1$\sigma$ uncertainties in Table~\ref{tab:fit_z}. In particular, we convert the best-fit evolution parameters into the knee luminosity L$^{\star}$ for PLE and knee normalization $\Phi^{\star}$ for PDE, in each redshift slice. We note that only L$^{\star}$ is free to vary for PLE, whereas only $\Phi^{\star}$ is free to vary for PDE (the other parameter is fixed to its local value). The reduced $\chi$$^2$ (or $\chi$$^2_{\nu}$) highlights that both fitting forms successfully reproduce the observed datapoints (red circles). In agreement with previous studies, we find that either L$^{\star}$ or $\Phi^{\star}$ of radio AGN do increase with redshift, peaking at $z$$\sim$2 and declining towards $z$$\sim$3 (e.g., \citealt{Kondapally+22}). 

\begin{figure}
\centering
     \includegraphics[width=\linewidth]{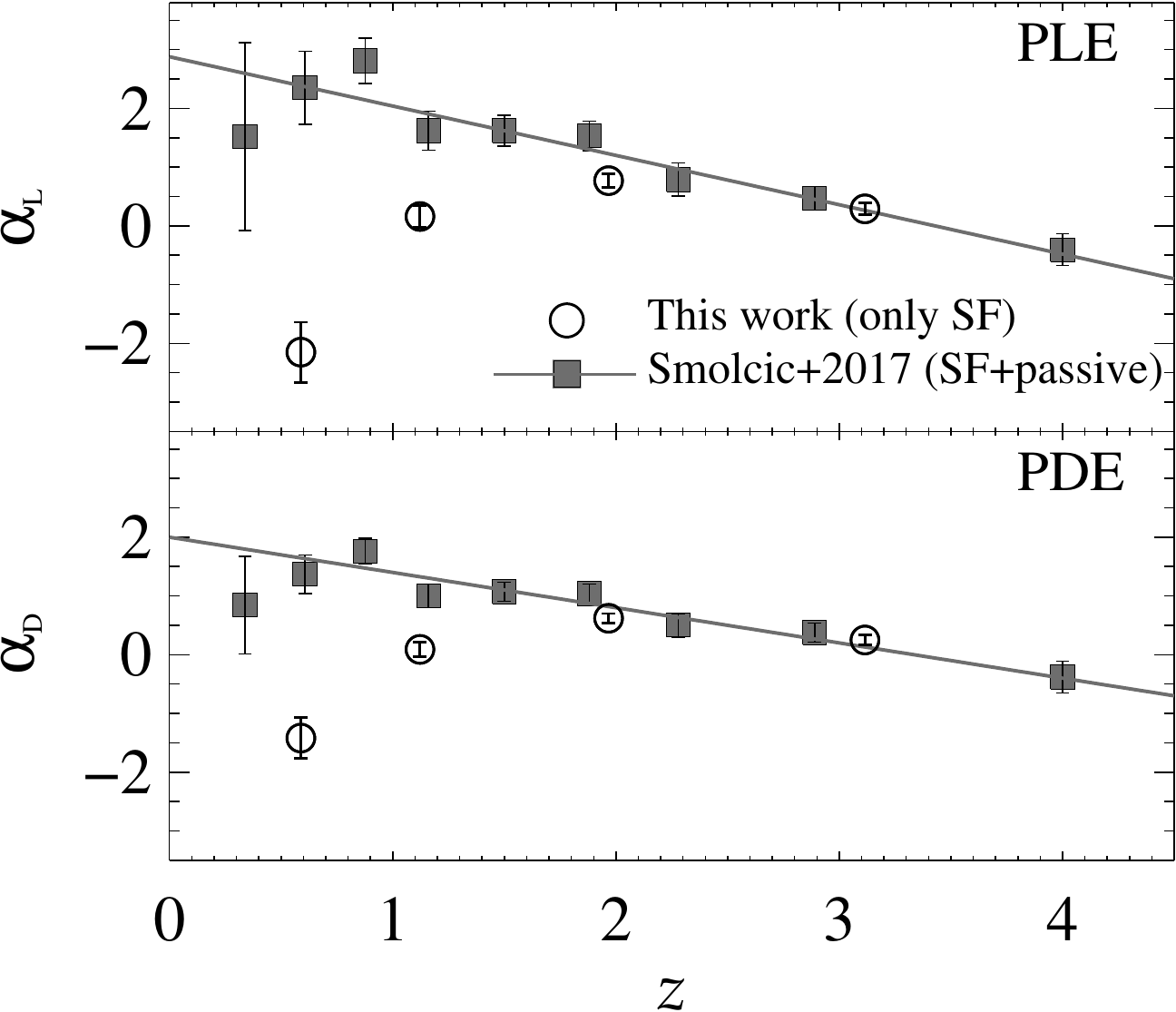}
 \caption{\small Best-fit evolution parameters $\alpha_L$ (upper panel) and $\alpha_D$ (bottom panel) obtained from PLE and PDE fitting forms, respectively. Our estimates (black circles) at each redshift are compared against those by \citeauthor{Smolcic+17c} (\citeyear{Smolcic+17c}, grey squares) that include both SF and passive hosts of radio-excess AGN. Grey lines by \citeauthor{Smolcic+17c} indicate a declining evolution of ($\alpha + z\cdot \beta$) with redshift, while our datapoints display a reversal at $z$$\sim$2 likely ascribed to the lack of passive galaxies that would outnumber SFGs at lower redshifts. See Sect.~\ref{rlf_redshift} for details. 
 }
   \label{fig:rlf_fitting_parms}
\end{figure}

In Fig.~\ref{fig:rlf_fitting_parms} we show our best-fit $\alpha_L$ (upper panel, PLE) $\alpha_D$ (bottom panel, PDE) as a function of redshift. We also display the same parameters obtained from \citet{Smolcic+17c} (grey squares), for the VLA-COSMOS~3~GHz sample, but including both SF and passive hosts. Grey lines by \citeauthor{Smolcic+17c} are obtained by fitting all redshift bins at (0$<$$z$$<$4) and denote a declining evolution of ($\alpha + z\cdot \beta$) with redshift. On the contrary, our datapoints show a reversal below $z$$\sim$2. 

%%%placing figure  ---------------------------
%%%-------------------------------------
  
This difference is likely induced by the lack of passive galaxies in our sample. As mentioned in Sect.~\ref{corr_l14}, about 30\% of radio-excess AGN in the parent VLA-COSMOS~3~GHz sample are hosted in passive galaxies \citep{Smolcic+17c}. However, this fraction strongly drops with increasing redshift, namely: 57\% at 0.1$\leq$$z$$<$0.7; 46\% at 0.7$\leq$$z$$<$1.4; 14\% at 1.4$\leq$$z$$<$2.5 and 7\% at 2.5$\leq z\leq$4.5. Because the relative fraction of passive galaxies strongly varies over redshift, excluding them induces a notable drop in the normalization of the AGN RLF (hence $\alpha_L$ or $\alpha_D$) at $z$$<$2, thus a deviation from the monotonic decline of ($\alpha + z\cdot \beta$) with redshift. Therefore it is no surprise that the best agreement with previous AGN RLFs including all (SF+passive) AGN hosts (e.g., \citealt{Smolcic+17c}; \citealt{Novak+18}; \citealt{Kono+22}; \citealt{Kondapally+22}) is found at the highest redshifts (see Appendix~\ref{Appendix_B}). We motivate our choice of removing passive galaxies in Sect.~\ref{passive}. As a reminder, we are mainly interested in exploring the relationship between radio AGN activity and galaxy growth in SFGs, but disentangling the two populations can elucidate their differential AGN demography and cosmic evolution.

If we were to perform a fitting of RLF datapoints in all redshift bins, we would obtain $\alpha_L$=0.39$\pm$0.22 and $\beta_L$=0.02$\pm$0.09 for PLE ($\chi$$^2_{\nu}$=3.64), while $\alpha_D$=0.11$\pm$0.17 and $\beta_D$=0.11$\pm$0.07 for PDE ($\chi$$^2_{\nu}$=3.55). With either formalisms, the total evolution parameter ($\alpha + z\cdot \beta$) would return a slightly decreasing trend with redshift, that systematically overestimates our AGN RLF at $z$$\lesssim$2, yielding a not very good fit ($\chi$$^2_{\nu}$$>$3; see above). We conclude that a linear evolutionary form is \textit{not} applicable to fitting subsets of radio AGN (e.g., hosted by SFGs), if these follow a different redshift distribution relative to the global radio AGN sample. For this reason, we keep the evolution parameters derived in individual redshift bins (i.e., with $\beta_L$=$\beta_D$=0), as listed in Table~\ref{tab:fit_z}.

We further clarify that our non-monotonic trend of the evolution parameters with redshift (Fig.~\ref{fig:rlf_fitting_parms}) only impacts the assumed \textit{evolution} of the RLF, not its functional \textit{shape} from \citet{Mauch+07}. Nevertheless, we emphasize that our study does not imply that the functional form by \citet{Mauch+07} provides the best possible fit to our data. We simply argue that, at least in the observed $L_{1.4}$ range, the faint- and bright- end slopes seem to fit our datapoints quite well. As long as the faint-end (bright-end) slope does not become steeper (flatter) than unity, the integral will converge (see Sect.~\ref{kld}) and the results will broadly remain unchanged.

\subsection{Removal of $NUVrJ$-passive galaxies}     \label{passive}

As mentioned in the previous sections, we refrain from including $NUVrJ$-\textit{passive} galaxies in our sample, although they dominate the number density of radio-excess AGN at $z$$\lesssim$1 (Sect.~\ref{rlf_redshift}). Our motivations are the following:

\begin{itemize}

\item \ul{\it Same IRRC?} -- Interpreting excess radio emission in passive galaxies implicitly assumes that the same IRRC calibrated for SFGs also applies to passive systems. Nonetheless, this is highly uncertain, since passive galaxies might be affected by contamination in their $L_{IR}$ and $L_{1.4}$ (hence $q_{\mathrm{IR}}$) estimates, which are instead needed in order to disentangle radio emission from star formation or AGN activity. Specifically, ``cirrus'' emission associated with cold dust heated by old ($>$A-type) stellar populations might lower the intrinsic SFR at fixed $L_{IR}$, especialy at low s-SFR (e.g., \citealt{Yun+01}). Moreover, radio emission can suffer from (largely unconstrained) milli-second pulsar contamination, which is negligible for SFGs but it can severely contaminate radio emission from star formation in massive quiescent galaxies (e.g., \citealt{Sudoh+21}), thus complicating a proper IRRC calibration for this class. 

\smallskip

\item \ul{\it Different cosmic evolutions} -- As highlighted in Sect.~\ref{rlf_redshift} and \ref{kld}, passive galaxies host the bulk of radio AGN feedback at $z$$\lesssim$1, and they strongly decline in number density at higher redshifts. As a consequence, mixing passive and star-forming galaxies in the same RLF washes out their differential cosmic evolution, as clearly demonstrated by the non-monotonic redshift trend of ($\alpha + z\cdot \beta$) in SFGs (see  Fig.~\ref{fig:rlf_fitting_parms}). 

\smallskip

\item \ul{\it Different AGN triggering?} -- As discussed in Sect.~\ref{radio_unified}, radio-excess AGN show systematically higher BHARs in SF than in passive galaxies (e.g., \citealt{Delvecchio+18}), supporting a broad link between available cold gas and SMBH growth in radio AGN. Furthermore, as discussed in Sect.~\ref{radio_unified}, recent studies find evidence that radio AGN in SF vs passive hosts have different duty cycle and cosmic evolution (e.g., \citealt{Kondapally+22}). These findings seem to suggest that radio AGN activity in SF vs passive galaxies is intrinsically driven by different mechanisms, hence they should be analysed separately.

\smallskip

\item \ul{\it Comparison with literature} -- We aim to quantify the incidence, power and evolution of radio AGN activity specifically within SF hosts. This is to compare radio AGN activity, star formation (set by the MS, e.g., \citealt{Speagle+14}) and X-ray AGN activity (from X-ray stacking; e.g., \citealt{Carraro+20}), all measured across the SFG population, via a self-consistent framework. Moreover, a recent paper by \citet{Ito+22} already inferred the mean radio-AGN luminosity in a sample of $\mathcal{M_*}$-selected passive galaxies from COSMOS2020 \citep{Weaver+22}, via stacking of 3~GHz images \citep{Smolcic+17a}. This is discussed in Sect.~\ref{radio_x}. 
Unlike for SFGs, stacking of quiescent hosts enables to capture the underlying radio-AGN emission, due to high contrast of AGN-over-SF emission. On the contrary, in SFGs radio stacking is insensitive to radio-AGN emission (see D21), hence we follow a novel approach based on the RLF. 

\end{itemize}

  %%%placing figure  ---------------------------
\begin{figure*}
\centering
     \includegraphics[width=\linewidth]{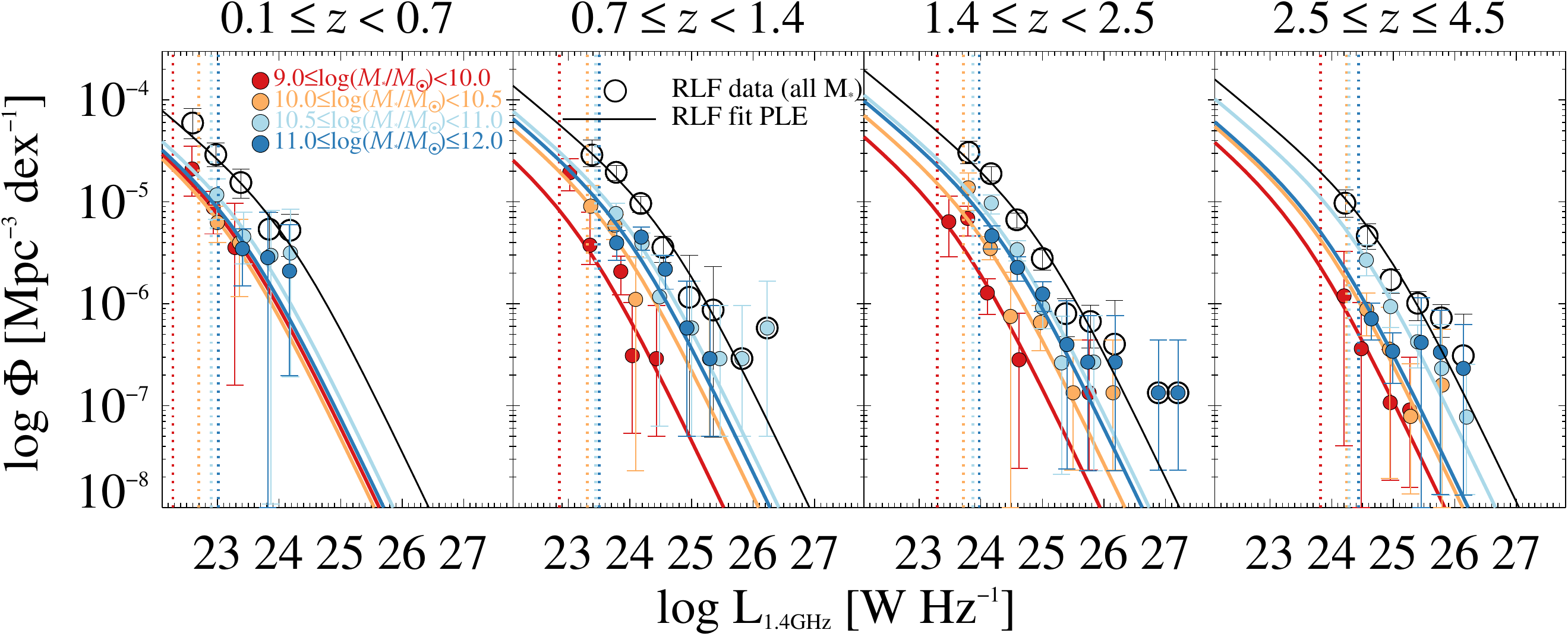}
 \caption{\small AGN RLF split in redshift and $\mathcal{M_{*}}$. Coloured circles are the observed datapoints at each $\mathcal{M_{*}}$, that add up to make the total RLF (black circles) at that redshift. Vertical dotted lines indicate the 1.4~GHz luminosity at +2$\sigma$ above the IRRC at a given ($\mathcal{M_{*}}$,$z$), which sets our $L_{1.4}$ threshold in the same bin to mitigate AGN incompleteness. Solid lines mark the best-fitting RLF obtained with PLE form. See Sect.~\ref{rlf_mass} for details. 
 }
   \label{fig:rlf_mass_ple}
\end{figure*}
%%%-------------------------------------
    
  %%%placing figure  ---------------------------
\begin{figure*}
\centering
     \includegraphics[width=\linewidth]{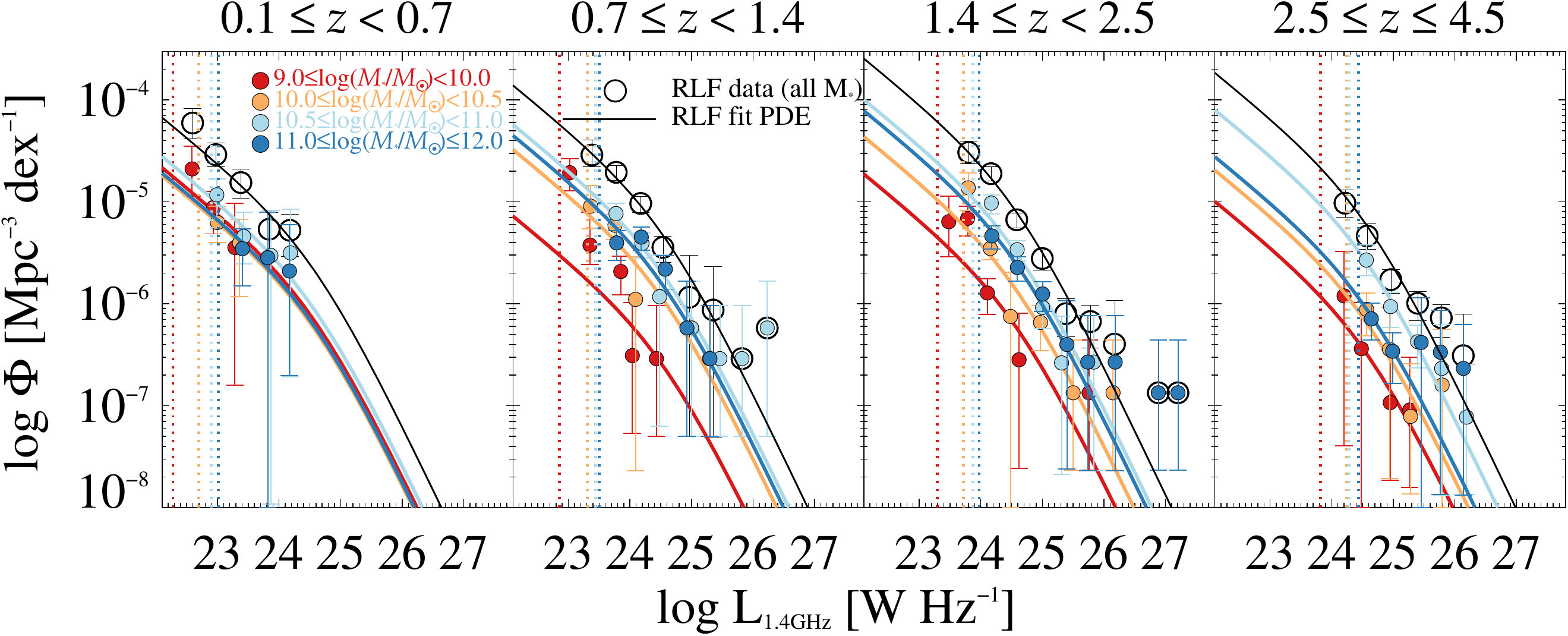}
 \caption{\small Same as in Fig.~\ref{fig:rlf_mass_ple}, but showing the best-fit AGN RLF obtained with PDE form.
 }
   \label{fig:rlf_mass_pde}
\end{figure*}
%%%-------------------------------------

\subsection{Splitting the AGN RLF in $\mathcal{M_{*}}$ bins} \label{rlf_mass}

The main goal of this work is measuring the mean and integrated radio AGN power across SFGs of different $\mathcal{M_{*}}$ and redshifts. To capture the dependence of radio AGN power on $\mathcal{M_{*}}$, we decompose the AGN RLF obtained in the previous sections in various $\mathcal{M_{*}}$ bins, at fixed redshift. To the best of our knowledge, this approach is new in literature, probably since such a distinction requires large statistical samples of radio AGN up to high redshifts, and more importantly a proper treatment of AGN completeness as a function of $\mathcal{M_{*}}$. The $\mathcal{M_{*}}$-dependent IRRC prescription from D21 sets an ideal ground for this method (Sect.~\ref{corr_purity} and Fig.~\ref{fig:histo_class}). Therefore, we split our AGN RLF datapoints (coloured circles) into four different $\mathcal{M_{*}}$ bins, as displayed in Fig.~\ref{fig:rlf_mass_ple}. Empty black circles indicate the global RLF at that redshift bin across the full range 9$<$$\log(\mathcal{M_{*}}$/$\mathcal{M_{\odot}})$$<$12. As done for the RLF in each redshift bin, we again define the 1.4 GHz luminosity corresponding to +2$\sigma$ from the IRRC, this time calculated from the median $\mathcal{M_{*}}$ and redshift of the corresponding bin (vertical dotted lines). Because of the nearly linear relation of $L_{1.4}^{\mathrm{SF}}$ with $\mathcal{M_*}$ set by the IRRC \citep{Delvecchio+21}, this radio luminosity threshold tends to increase with increasing mass, though at low $\mathcal{M_{*}}$ (e.g., middle panels of Fig.~\ref{fig:rlf_mass_pde}) it can reach below the reference $L_{1.4}$ threshold taken at that redshift, since this latter was computed at the median $\mathcal{M_{*}}$ of the full underlying sample (black dotted lines in bottom panel of Fig.~\ref{fig:a1}). All RLF datapoints used in this analysis (i.e., above the corresponding $L_{1.4}$ threshold) are listed in Table~\ref{tab:rlf_observed}, for each $\mathcal{M_{*}}$ and redshift bin.

In order to track the macroscopic evolution of the radio-excess AGN population across different $\mathcal{M_{*}}$, here we explore the same PLE and PDE fitting forms, as used in Sect.~\ref{rlf_redshift}. Hence, for PLE we fix the knee normalization to $\Phi^{\star}$=$\frac{1}{0.4}$$\cdot$10$^{-5.5}$~Mpc$^{-3}$~dex$^{-1}$, and for PDE we fix the knee luminosity $L^{\star}$=10$^{24.59}$~W~Hz$^{-1}$, i.e., their local values \citep{Mauch+07}. Figs.~\ref{fig:rlf_mass_ple} and ~\ref{fig:rlf_mass_pde} show the best-fit RLF split in $\mathcal{M_{*}}$ bins in the case of PLE and PDE fitting, respectively. We stress that the sum of best-fit RLFs over all $\mathcal{M_{*}}$ bins (coloured lines) is fully consistent with the best-fit RLF obtained at that redshift (black solid line), despite having been derived independently from one another. 
As a sanity check, we further verified that in each ($\mathcal{M_{*}}$,$z$) bin, the integrated number density of radio AGN (from either PLE or PDE form) never exceeds the number density of all $\mathcal{M_*}$-selected SFGs.

We retrieve the best-fit $L^{\star}$ or $\Phi^{\star}$ in each bin, and estimate their 1$\sigma$ uncertainties via bootstrapping over the error bars of all datapoints in the same bin, as described in Sect.~\ref{boot}. Output parameters for each ($\mathcal{M_{*}}$,$z$) bin are listed in Table~\ref{tab:fit_mass}. As expected, binning also with $\mathcal{M_{*}}$ adds-on noise in the evolution of the AGN RLF. Nevertheless, we observe a clear stratification in $\mathcal{M_{*}}$ for both PLE and PDE fitting forms. AGN at lower $\mathcal{M_{*}}$ are typically both less common \textit{and} less luminous compared to AGN at higher $\mathcal{M_{*}}$, with a peak being reached at intermediate-to-high stellar masses, i.e., 10.5$<$$\log(\mathcal{M_{*}}$/$\mathcal{M_{\odot}})$$<$11 (cyan lines).

For sake of clarity, hereafter we only show the results obtained from the PLE form. Although the results are fully consistent with one another, we acknowledge that a PLE fitting performs slightly better than PDE in reproducing the RLF at low $\mathcal{M_{*}}$, which is a critical domain to establish the global $\mathcal{M_{*}}$ dependence of the integrated radio AGN power. This choice is also in line with several previous studies of the AGN RLF (e.g., \citealt{Sadler+07}; \citealt{Smolcic+09}; \citealt{McAlpine+13}; \citealt{Padovani+15}; \citealt{Smolcic+17c}).

% % %   UPDATE TABLE
\begin{table*}
\caption{AGN RLF fitting parameters derived in various $\mathcal{M_{*}}$ bins, at each redshift. Output parameter are listed for both PLE and PDE forms, as detailed in Sect.~\ref{rlf_mass}. The best-fitting RLF for each model are shown in Figs.~\ref{fig:rlf_mass_ple} and \ref{fig:rlf_mass_pde}.}
\centering
\begin{tabular}{cccccc}
\hline\hline
$z$-bin &  $\mathcal{M_{*}}$-bin   &   \multicolumn{2}{c}{PLE} & \multicolumn{2}{c}{PDE}   \\
{} & $\log(\mathcal{M_{*}}$/$\mathcal{M_{\odot}})$  &\multicolumn{2}{l}{\phantom{}}\dotfill & \multicolumn{2}{c}{\phantom{}} \dotfill  \\
{} & {}  & $\log L^{\star}(\mathcal{M_{*}},z)$ & $\chi^{2}_{\nu}$ & $\Phi^{\star}(\mathcal{M_{*}},z)\cdot 10^{-6}$ & $\chi^{2}_{\nu}$ \\
{} & {} & [$\log$(W~Hz$^{-1}$)] & {} & [Mpc$^{-3}$~dex$^{-1}$] & {}  \\   
\hline
${0.1} \leq z < {0.7}$  &  9--10    &  $23.35 \pm 0.29$ & 0.15 & $1.32 \pm 0.60$ &  0.37     \\
                        &  10--10.5 &  $23.27 \pm 0.19$ & 0.07 & $1.10 \pm 0.41$ &  $<$0.01  \\
                        &  10.5--11 &  $23.58 \pm 0.15$ & 0.26 & $1.70 \pm 0.44$ &  0.20     \\
 \smallskip             &  11--12   &  $23.43 \pm 0.35$ & 0.43 & $1.16 \pm 0.40$ &  0.16     \\ 

${0.7} \leq z < {1.4}$  &  9--10    &  $23.25 \pm 0.15$ & 1.34 & $0.44 \pm 0.17$ &  1.91     \\
                        &  10--10.5 &  $23.81 \pm 0.14$ & 0.86 & $1.98 \pm 0.48$ &  1.13     \\
                        &  10.5--11 &  $24.14 \pm 0.12$ & 0.48 & $3.48 \pm 0.69$ &  0.49     \\
 \smallskip             &  11--12   &  $24.00 \pm 0.13$ & 1.81 & $2.73 \pm 0.54$ &  1.08     \\

${1.4} \leq z < {2.5}$  &  9--10    &  $23.66 \pm 0.21$ & 1.01 & $1.14 \pm 0.34$ &  1.52     \\
                        &  10--10.5 &  $24.07 \pm 0.10$ & 0.89 & $2.65 \pm 0.51$ &  1.02     \\
                        &  10.5--11 &  $24.46 \pm 0.09$ & 1.07 & $6.13 \pm 1.30$ &  1.30     \\
 \smallskip             &  11--12   &  $24.35 \pm 0.12$ & 0.71 & $4.77 \pm 1.05$ &  0.59     \\

 ${2.5} \leq z \leq {4.5}$  &  9--10   &  $23.56 \pm 0.21$ & 0.13 & $0.62 \pm 0.19$ &  0.16     \\
                        &  10--10.5 &  $23.85 \pm 0.10$ & 0.25 & $1.25 \pm 0.30$ &  0.41     \\
                        &  10.5--11 &  $24.38 \pm 0.11$ & 0.23 & $4.86 \pm 1.25$ &  0.19     \\
                        &  11--12   &  $23.94 \pm 0.14$ & 0.73 & $1.69 \pm 0.57$ &  0.66     \\                        
\hline
\end{tabular} \label{tab:fit_mass}
\end{table*}

% % % -------------------------------------------------------------------------------
\section{Results: radio AGN activity across the SFG population} \label{results}
% % % -------------------------------------------------------------------------------

In this section we explore the integrated power emitted by radio AGN across the $\mathcal{M_{*}}$-selected SFG population. First we calculate the cumulative (kinetic) AGN luminosity exerted as a function of redshift and $\mathcal{M_{*}}$, highlighting its evolution compared to other works (Sect.~\ref{kld}). Then we follow a statistical approach to average the integrated radio AGN power at fixed ($\mathcal{M_{*}}$,$z$) across the entire $\mathcal{M_{*}}$-selected population of SFGs, deriving the ``radio-AGN main sequence'' that links \textit{mean} radio AGN power and SFG stellar mass over cosmic time (Sect.~\ref{rams}).

\subsection{Kinetic AGN luminosity density} \label{kld}

  %%%placing figure  ---------------------------
\begin{figure}
\centering
     \includegraphics[width=\linewidth]{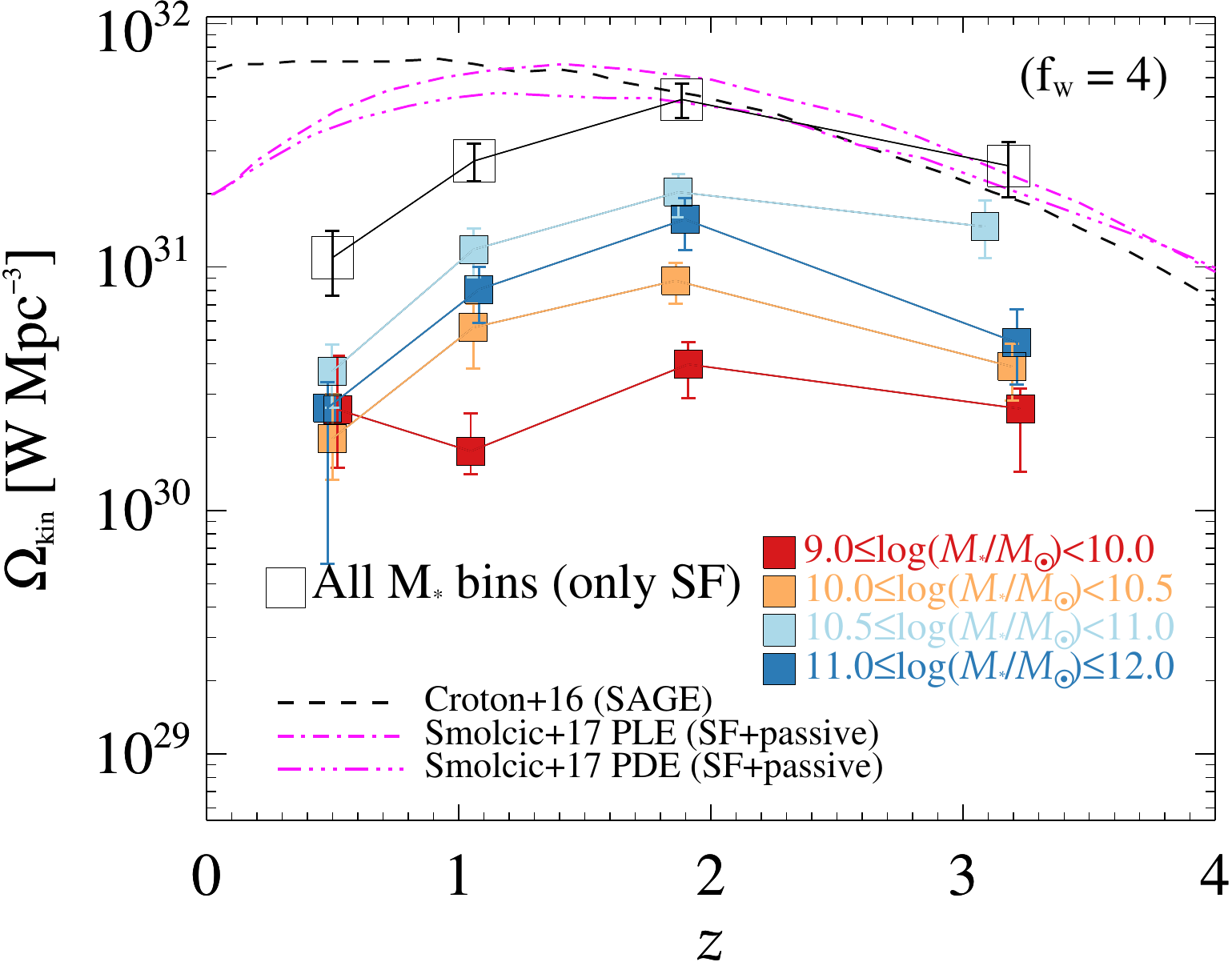}
 \caption{\small Kinetic AGN luminosity density, $\Omega_{\mathrm{kin}}$ as a function of redshift and dissected in $\mathcal{M_{*}}$ bins (coloured squares). The sum across all $\mathcal{M_{*}}$ at each redshift is marked with black open squares. For comparison, we show the integrated values from \citeauthor{Smolcic+17c} (\citeyear{Smolcic+17c}, purple lines), both for PLE (dot-dashed) and for PDE (triple dot-dashed) fitting forms. Model predictions from SAGE \citep{Croton+16} including radio-mode AGN feedback are displayed as a function of redshift (black dashed line). We assume $f_W$=4 as in \citet{Smolcic+17c} for consistency. See Sect.~\ref{kld} for further details.  
 }
   \label{fig:kld}
\end{figure}
%%%-------------------------------------

The integral of the AGN RLF allows us to assess the cumulative luminosity released through radio AGN activity. This originates from mass accretion onto the central SMBH, and is channeled in kinetic form via collimated jets propagating throughout the galaxy (but see \citealt{Panessa+19} for alternative origins of radio emission in radio quiet AGN). Though these jet signatures are widespread in powerful radio AGN at high-$z$ (e.g., \citealt{Nesvadba+17}; \citealt{Collet+16}; \citealt{Spingola+20}; but see \citealt{Radcliffe+18},\citeyear{Radcliffe+21} for high-$z$ radio-faint AGN) or in nearby radio AGN (e.g., \citealt{Jarvis+19}; \citealt{Brienza+21}; \citealt{Venturi+21}; \citealt{Girdhar+22}), only a small fraction of the kinetic energy carried by the jet and deposited in the interstellar medium is observable with monochromatic observations, the rest being dissipated in the environment (see reviews by e.g., \citealt{McNamara+07}; \citealt{Gitti+12}). Several scaling relations have been proposed in the literature to convert monochromatic radio to kinetic luminosity (e.g., \citealt{Willott+99}; \citealt{Birzan+04}; \citeyear{Birzan+08}; \citealt{Merloni+07}; \citealt{Cavagnolo+10}; \citealt{OSullivan+11}; \citealt{Daly+12}; \citealt{Godfrey+16}). A comprehensive overview of these relations and their unknowns is given in \citet{Smolcic+17c} (see their Appendix~A).  

For consistency with \citet{Smolcic+17c}, we compute the kinetic AGN luminosity $L_{\mathrm kin}$ from the following relation formulated by \citet{Willott+99} (scaled to 1.4~GHz),
\begin{equation}
 \log(L_{\mathrm{kin}}) = 0.86 \cdot \log(L_{1.4}^{\mathrm{AGN}}) + 14.08 + 1.5 \cdot \log(f_W) ,
    \label{eq:willott}
 \end{equation}
where $L_{\mathrm kin}$ is given in W, $L_{1.4}^{\mathrm{AGN}}$ is the AGN-related 1.4~GHz luminosity in units of W~Hz$^{-1}$. The parameter $f_W$ encapsulates all uncertainties on the energetics and geometry of the jet, ranging from $f_W$$\approx$1 to $f_W$$\approx$20 (we assume $f_W$=4, see below).

The kinetic AGN luminosity density at a given redshift, $\Omega_{\mathrm{kin}}(z)$, is computed by multiplying $L_{\mathrm kin}$ by the AGN RLF, $\Phi(L_{1.4}^{\mathrm{AGN}})$, and integrating in 1.4~GHz luminosity,
\begin{equation}
\Omega_{\mathrm{kin}}(z) = \int_{L_{1.4}^{\mathrm{SF}}}^{L_{1.4}^{\mathrm{max}}} \Phi(L_{1.4}^{\mathrm{AGN}}) \cdot L_{\mathrm{kin}} \cdot ~ d(\log L_{1.4}^{\mathrm{AGN}})  ~.
   \label{eq:kld}
\end{equation}
We extend this calculation also in different $\mathcal{M_{*}}$ bins, computing $\Omega_{\mathrm{kin}}(\mathcal{M_{*}},z)$ simply from the corresponding best-fit AGN RLF (Sect.~\ref{rlf_mass}). Unlike previous studies that usually assume an arbitrary minimum luminosity, we set the minimum to match $L_{1.4}^{\mathrm{SF}}(\mathcal{M_{*}},z)$, that is the 1.4~GHz luminosity corresponding to the IRRC at a given ($\mathcal{M_{*}}$,$z$). This value is motivated by the need to account for radio-faint AGN that display no excess radio emission (i.e., lying within the scatter of the IRRC). We remind the reader that the RLF datapoints cover a $L_{1.4}^{\mathrm{AGN}}$ range down to +2$\sigma$ above the IRRC, in which we are able to correct for AGN purity, and in which our sample is $\approx$90\% complete in $L_{1.4}^{\mathrm{AGN}}$ (see Appendix~\ref{completeness}). Instead, we now extrapolate the best-fitting RLF down to the corresponding $L_{1.4}^{\mathrm{SF}}$ to factor in extra radio-AGN emission not yet accounted for. The entity of such correction is inherently tied to the assumed shape of the AGN RLF (i.e., \citealt{Mauch+07}) at the faint end regime, and will be discussed in Sect.~\ref{rams}.

Fig.~\ref{fig:kld} displays the kinetic AGN luminosity density $\Omega_{\mathrm{kin}}$ dissected in $\mathcal{M_{*}}$ (coloured squares), that add-up to make the total $\Omega_{\mathrm{kin}}$ at a given redshift (black open squares). Error bars at 1$\sigma$ level are obtained by bootstrapping 1000 times over the uncertainties on the AGN RLF, listed in Tables~\ref{tab:fit_z} and \ref{tab:fit_mass}. 

The global $\Omega_{\mathrm{kin}}(\mathcal{M_{*}},z)$ displays a clear stratification in $\mathcal{M_{*}}$. Specifically, radio AGN in galaxies at 9$<$$\log(\mathcal{M_{*}}$/$\mathcal{M_{\odot}}$)$<$10 display the smallest contribution, while the population at 10.5$<$$\log(\mathcal{M_{*}}$/$\mathcal{M_{\odot}}$)$<$11 dominates the kinetic AGN luminosity density at all redshifts. 

Irrespective of the $\mathcal{M_{*}}$ bin, the integrated kinetic AGN luminosity density in SFGs peaks at $z$$\sim$2 and declines towards $z$$\sim$0, though this drop is more gentle in low-$\mathcal{M_{*}}$ galaxies. We emphasize that the RLF in each $\mathcal{M_{*}}$ bin is fitted independently from the others, without forcing any internal monotonic $\mathcal{M_{*}}$ trend. Hence, the strong reported $\mathcal{M_{*}}$ stratification is genuine. We further note that the total $\Omega_{\mathrm{kin}}(\mathcal{M_{*}},z)$ summed over all $\mathcal{M_{*}}$ bins (black open squares) is in good agreement with the $\Omega_{\mathrm{kin}}(z)$ inferred from the integrated RLF at every redshift.

For comparison, we also show the global $\Omega_{\mathrm{kin}}(z)$ inferred by \citet{Smolcic+17c} for PLE (dot-dashed line) and PDE (triple dot-dashed line) forms. For consistency with their work and for illustrative purposes, we also assume $f_W$=4, though we note that any constant value in the range $f_W$=1--20 would rigidly scale the data without affecting our main conclusions. However, we acknowledge that a $\mathcal{M_{*}}$ and/or $z$-dependent $f_W$ could alter $\Omega_{\mathrm{kin}}$, but this possibility has been so far unexplored. Following this naive assumption ($f_W$=4), we find fully consistent $\Omega_{\mathrm{kin}}(z)$ measurements at z$\gtrsim$2 with \citet{Smolcic+17c}, while our data are a factor 2--3$\times$ lower at lower redshifts. This is, again, most likely caused by the lack of passive galaxies in our sample. This apparent discrepancy closely resembles the offset seen in the evolution parameters $\alpha_L$ and $\alpha_D$ in Fig.~\ref{fig:rlf_fitting_parms}. This further strengthens that radio-AGN activity taking place in passive galaxies dominates the kinetic luminosity density at $z$$\lesssim$1.

The black dashed line in Fig.~\ref{fig:kld} marks the prediction from the semi-analytical galaxy evolution ({\sc sage}, \citealt{Croton+16}) model. Compared to the former version by \citet{Croton+06}, this updated model incorporates a realistic coupling between gas cooling and radio-mode (or ``jet-mode'') AGN heating, which is a desirable refinement to predict the long-term impact of radio AGN feedback on the surrounding gas. This cooling-heating cycle is modulated through the so-called ``radio mode efficiency'' parameter ($k_R$=0.08; see Eq. (16) and Sect. 9.1 in \citealt{Croton+16}). In {\sc sage}, this parameter is used to modulate the BHAR (or $\dot{m}_{\mathcal{BH}}$), hence the AGN accretion luminosity $L_{\mathrm{AGN}}$~=~$\eta$$\dot{m}_{\mathcal{BH}}$$c^2$, where $\eta$=0.1 is the standard radiative efficiency, and $c$ is the speed of light. Assuming that a given fraction of the accretion energy is channeled in kinetic - rather than radiative - form, we can re-scale $L_{\mathrm{AGN}}$ to our derived $L_{\mathrm kin}$, and compare their volume-averaged luminosity density across cosmic time (Fig.~\ref{fig:kld}). We find a good agreement at z$\gtrsim$2 with the shape and normalization of the {\sc sage} model, while at lower redshifts our estimated $\Omega_{\mathrm{kin}}(z)$ lies 3--6$\times$ lower. This is not surprising, since the missing population of passive (i.e., quenched) galaxies in our study is the one undergoing most likely radio mode AGN feedback in the model, as a means to permanently turn off gas cooling and star formation. Taking $f_W$=15, as suggested by observations of radio lobes inflating X-ray cavities in local galaxy clusters (e.g., \citealt{Birzan+04}; \citeyear{Birzan+08}; \citealt{Merloni+07}; \citealt{Cavagnolo+10}; \citealt{OSullivan+11}), would match {\sc sage} predictions at $z$$\sim$0.5, albeit overboosting our data at z$\gtrsim$1 to 3--4$\times$ above the model. It is evident how the large uncertainties on $f_W$ can esily accommodate an agreement, although we stress that alternative scaling relations to Eq.~\ref{eq:willott} would yield $\Omega_{\mathrm{kin}}(z)$ systematically above {\sc sage} predictions (see Appendix A in \citealt{Smolcic+17c}).

  %%%placing figure  ---------------------------
\begin{figure*}
\centering
     \includegraphics[width=3.6in]{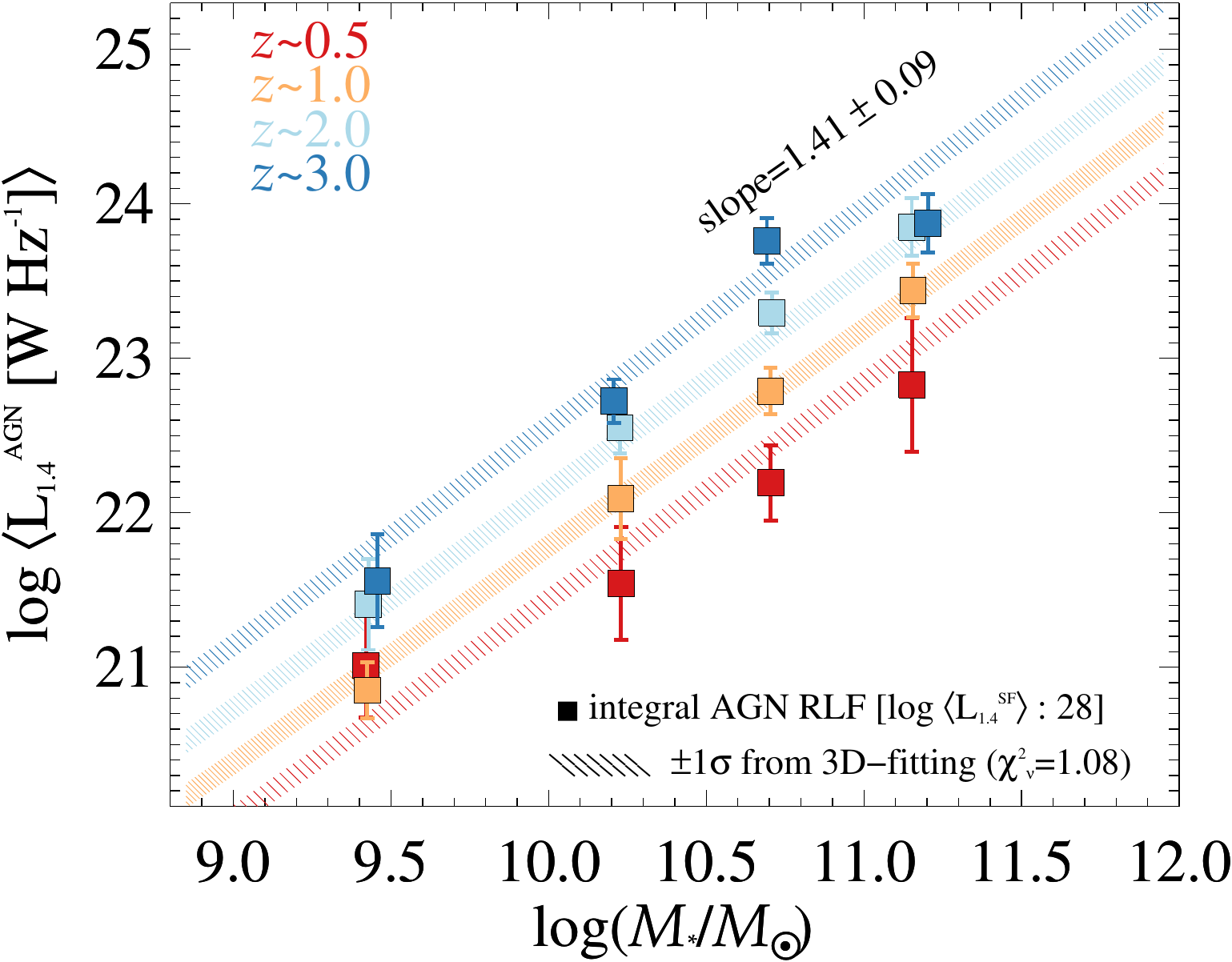}
     \includegraphics[width=3.58in]{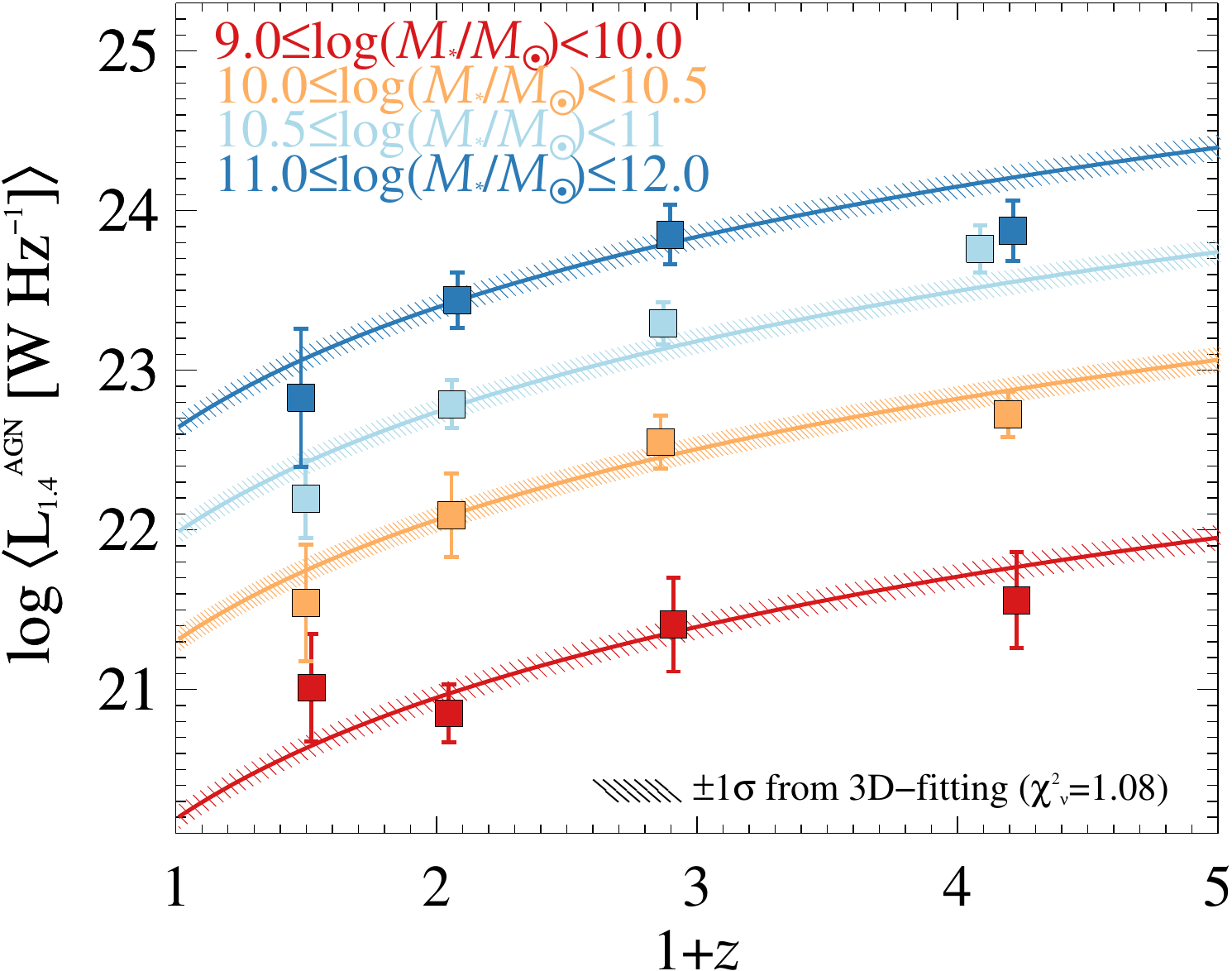}

 \caption{\small \textit{Left}: Radio-AGN main sequence (RAMS), relating \textit{mean} radio AGN power $\langle{L_{1.4}^{\mathrm{AGN}}}\rangle$ averaged across all ($NUVrJ$-based) SFGs and galaxy $\mathcal{M_{*}}$, at different redshifts. Individual points are inferred from the integral of the AGN RLF at each ($\mathcal{M_{*}}$,$z$), following Eq.~\ref{eq:mean} and coloured by $\mathcal{M_{*}}$. We fit all points with a three-dimensional function (Eq.~\ref{eq:rams}). Dashed areas encompass the $\pm$1$\sigma$ scatter around the best-fit values obtained by bootstrapping over the uncertainties. \textit{Right}: redshift evolution of the same same $\langle{L_{1.4}^{\mathrm{AGN}}}\rangle$ measurements, coloured by $\mathcal{M_{*}}$ and fitted as a function of (1+$z$)$^{P_2}$ (see Eq.~\ref{eq:rams}), with $P_2$=2.51$\pm$0.34. 
 }
   \label{fig:rams}
\end{figure*}
%%%-------------------------------------

% % %   UPDATE TABLE
\begin{table}
\caption{Parameters used to compute the radio-AGN main sequence (RAMS) in Sect.~\ref{rams}: the median redshift ($\langle z \rangle$), stellar mass ($\langle M_* \rangle$) and 1.4~GHz luminosity from SF ($\langle$$L_{1.4}^{\mathrm{SF}}$$\rangle$) averaged over all $\mathcal{M_{*}}$-selected SFGs in each bin; the number of all $NUVrJ$-selected SFGs ($N_{\mathrm{SFG}}$); the mean radio AGN power averaged across $N_{\mathrm{SFG}}$ ($\langle L_{1.4}^{\mathrm{AGN}} \rangle$). }
\centering
\begin{tabular}{ccccc}
\hline\hline
  \multicolumn{5}{c}{RAMS parameters (Sect.~\ref{rams})} \\
\multicolumn{5}{c}{\phantom{}}\dotfill  \\
$\langle z \rangle$ & $\langle \log(\mathcal{M_{*}}) \rangle$  & $\log \langle (L_{1.4}^{\mathrm{SF}}) \rangle$ & $N_{\mathrm{SFG}}$ & $\log \langle (L_{1.4}^{\mathrm{AGN}}) \rangle$ \\
{} & $\log$(M$_{\odot})$  &  [$\log$(W~Hz$^{-1}$)] &    &   [$\log$(W~Hz$^{-1}$)]  \\   
\hline
0.52            &   9.42    & 21.67   & 15463 &  $21.01 \pm 0.34$  \\
0.50            &   10.23   & 22.32   & 3610  &  $21.54 \pm 0.38$  \\
0.50            &   10.70   & 22.54   & 1708  &  $22.19 \pm 0.25$  \\
0.48 \smallskip &   11.15   & 22.65   & 251   &  $22.83 \pm 0.43$  \\ 

1.05            &   9.42    & 22.02   & 49264 &  $20.85 \pm 0.19$  \\
1.06            &   10.23   & 22.74   & 10593 &  $22.09 \pm 0.26$  \\
1.06            &   10.70   & 23.02   & 5166  &  $22.79 \pm 0.15$  \\
1.08 \smallskip &   11.16   & 23.19   & 813   &  $23.44 \pm 0.17$  \\

1.91            &   9.43    & 22.23   & 63448 &  $21.41 \pm 0.30$  \\
1.86            &   10.23   & 23.05   & 13265 &  $22.55 \pm 0.16$  \\
1.87            &   10.71   & 23.45   & 6677  &  $23.30 \pm 0.13$  \\
1.90 \smallskip &   11.15   & 23.71   & 1315  &  $23.85 \pm 0.19$  \\

3.23             &  9.46    & 22.51   & 52301 &  $21.56 \pm 0.30$  \\
3.19             &  10.21   & 23.33   & 9365  &  $22.72 \pm 0.14$  \\
3.09             &  10.69   & 23.78   & 2902  &  $23.76 \pm 0.14$  \\
3.21             &  11.20   & 24.17   & 622   &  $23.87 \pm 0.19$  \\                        
\hline
\end{tabular} \label{tab:rams}
\end{table}

\subsection{The radio-AGN main sequence (RAMS)} \label{rams}

Each AGN RLF fit obtained in Sect.~\ref{rlf_mass} is calibrated on a carefully-selected sample of radio-excess AGN (at $>$2$\sigma$ from the IRRC) hosted in SFGs, taking into account flux incompleteness, classification purity and $\mathcal{M_{*}}$-dependent radio emission from star formation at each redshift. Therefore, we are well-placed to explore the intrinsic relationship between \textit{mean} radio AGN power and galaxy $\mathcal{M_{*}}$ in SFGs, factoring in the contribution of radio-faint AGN located within the IRRC.

We thus proceed in two steps: firstly, we compute the cumulative power produced by the radio AGN population in SFGs at a given ($\mathcal{M_{*}},z$); secondly, we divide this integrated value by the number of all ($NUVrJ$-selected) SFGs contained in each bin to infer a representative sample-averaged radio-AGN luminosity.

Similarly to the calculation of the kinetic AGN luminosity density (Sect.~\ref{kld}), we integrate the best-fit AGN RLF above the radio luminosity set by the IRRC, at each ($\mathcal{M_{*}}$,$z$). However, a key difference now is that we want to assess the integrated power released by radio AGN across the \textit{full} SFG population. The sample of SFGs is selected by $\mathcal{M_{*}}$ \citep{Jin+18} and counts 236,763 galaxies identified via $NUVrJ$-colours at 9$\leq$$\log$($\mathcal{M_{*}}$/$\mathcal{M_{\odot}}$)$\leq$12 (from D21). These galaxies lie on the ``star-forming main sequence'' (MS), while the subset of radio-detections stands slightly above MS (i.e., at higher $L_{IR}$ at fixed $\mathcal{M_{*}}$), especially at low $\mathcal{M_{*}}$, since the radio flux limit sets a roughly horizontal cut in SFR. Based on the IRRC, lower $L_{IR}$ (or SFR) imply also lower $L_{1.4}$ in $\mathcal{M_{*}}$-selected SFGs than in radio detections. Thus, we adjust the new radio luminosity to match the $L_{IR}$/$L_{1.4}$ ratio of a typical MS galaxy at a given ($\mathcal{M_{*}}$,$z$). Specifically, we read $L_{IR}$ from the MS fitting form presented in \citeauthor{Daddi+22b} (\citeyear{Daddi+22b}, see their Table 1). They used median $L_{IR}$ measurements obtained in D21 via IR-mm stacking for the same sample of $\mathcal{M_{*}}$-selected SFGs. From $L_{IR}$ we compute the mean 1.4~GHz luminosity of the IRRC for MS galaxies ($\langle$$L_{1.4}^{\mathrm{SF}} \rangle$), at each $\mathcal{M_{*}}$ and redshift. These values set the minimum of the integral in Eq.~\ref{eq:cumulative} and are reported in Table~\ref{tab:rams}. The integrated radio-AGN luminosity can be therefore expressed as:
\begin{equation}
\sum{L_{1.4}^{\mathrm{AGN}}} = \int_{\langle L_{1.4}^{\mathrm{SF}} \rangle}^{L_{1.4}^{\mathrm{max}}} \Phi(L_{1.4}^{\mathrm{AGN}}) \cdot \langle{V_{\mathrm{max}}}\rangle \cdot L_{1.4}^{\mathrm{AGN}} \cdot d(\log L_{1.4}^{\mathrm{AGN}}) ~ .
  \label{eq:cumulative}
\end{equation}
Differently from the $\Omega_{\mathrm{kin}}$ calculation, our $\sum{L_{1.4}^{\mathrm{AGN}}}$ is dimensionally a luminosity, not a luminosity density. Hence, in Eq.~\ref{eq:cumulative} we multiply the best-fit $\Phi(L_{1.4}^{\mathrm{AGN}})$ by a \textit{characteristic} $\langle V_{\mathrm{max}} \rangle$, taken as the median value across the underlying population in the same ($\mathcal{M_{*}}$,$z$) bin. We propagate the dispersion around $\langle V_{\mathrm{max}} \rangle$ when assessing the uncertainty on $\sum{L_{1.4}^{\mathrm{AGN}}}$. By integrating above $\langle$$L_{1.4}^{\mathrm{SF}} \rangle$, we are implicitly assuming that no radio AGN is ``active'' below the value set by the IRRC. This limit is empirically-motivated by a ($\mathcal{M_{*}}$,$z$) dependent IRRC prescription (D21)\footnote{We note that the contribution of ``quiescent SMBHs'' (here assumed to have $L_{1.4}^{\mathrm{AGN}}$=0) will be factored in by applying Eq.~\ref{eq:mean}.}. This approach ensures a fully self-consistent treatment of radio emission from SF and AGN activity.

As a sanity check, we quantify the effect of changing integration limits and extrapolating the LF in the faint end:

\noindent
\textbullet{} We find that the extra portion of the integral counted in the faint-end extrapolation (i.e., between $L_{1.4}^{\mathrm{SF}}$ and the faintest observed $L_{1.4}$-bin) is only about 20\%, while the total number of AGN included in the extrapolation increases by a factor of 3--4. Therefore, this extrapolation does not substantially alter the \textit{integrated} AGN luminosity density, while it is necessary to account for the global \textit{incidence} of radio AGN.

\noindent
\textbullet{} We stress that setting the lower integration bound to a fixed canonical value (e.g., 10$^{22}$~W~Hz$^{-1}$, \citealt{Ceraj+18}) overestimates $\sum{L_{1.4}^{\mathrm{AGN}}}$ by up to 50\% at the highest $\mathcal{M_{*}}$. Thus, we argue that accounting for the evolving $\langle$$L_{1.4}^{\mathrm{SF}} \rangle$ with ($\mathcal{M_*}$,$z$) is necessary for minimizing AGN-vs-SF cross-contamination. Instead, increasing the \textit{upper} integration limit $L_{1.4}^{\mathrm{max}}$ from 10$^{28}$ (as in \citealt{Novak+18}; \citealt{Ceraj+20}) to infinity would boost $\sum{L_{1.4}^{\mathrm{AGN}}}$ by only $\lesssim$10\%. 
  
\noindent
\textbullet{} Over 50\% of the cumulative AGN luminosity density is produced by AGN within $\pm$1~dex from the corresponding L$^{\star}$ ($\mathcal{M_*}$,$z$), although these sources are a relatively small fraction of the galaxy population.

 \smallskip

By dividing the cumulative radio-AGN luminosity $\sum{L_{1.4}^{\mathrm{AGN}}}$ by the number of all $\mathcal{M_{*}}$-selected SFGs in the same bin ($N_{\mathrm{SFG}}$), we can compute the \textit{mean} radio-AGN luminosity:
\begin{equation}
\langle{L_{1.4}^{\mathrm{AGN}}}\rangle = \frac{\sum{L_{1.4}^{\mathrm{AGN}}}}{N_{\mathrm{SFG}}}  ~.
  \label{eq:mean}
\end{equation}
This method follows the same logic of stacking, in which a representative sample-averaged measurement is inferred by combining detections and non-detections. The main difference is that stacked luminosities are corrected for SF contamination \textit{a-posteriori}; instead, our prescription set by the IRRC allows us to quantify and remove galaxy contamination \textit{a-priori}, by integrating down to $L_{1.4}^{\mathrm{SF}}$. Moreover, stacking is broadly sensitive to the average signal from the \textit{dominant} underlying population. Since radio emission from 3~GHz-undetected SFGs in COSMOS is primarily originated from SF (D21), radio stacking reveals a notable radio-excess only for passive galaxies \citep{Ito+22}. Instead, our statistical approach, backed-up with a detailed assessment of the evolving AGN RLF, can also account for hidden radio AGN in SFGs. Some caveats and limitations inherent to our approach are discussed in Sect.~\ref{caveats}.    

The ``radio-AGN main sequence'' (RAMS hereafter) between $\langle{L_{1.4}^{\mathrm{AGN}}}\rangle$ and galaxy $\mathcal{M_{*}}$ (in $\log$-space) is presented in Fig.~\ref{fig:rams} (left). Mean $\langle{L_{1.4}^{\mathrm{AGN}}}\rangle$ measurements (squares) are coloured by redshift to highlight the evolution of this relationship. We use the IDL routine {\sc mpfit2dfun.pro} to perform a three-dimensional fitting in the $\langle{L_{1.4}^{\mathrm{AGN}}}\rangle$--$\mathcal{M_*}$--$(1+z)$ $\log$-space, assuming the following analytical expression,

\begin{equation}
\log \langle L_{1.4}^{\mathrm{AGN}} / \mathrm{W~Hz^{-1}} \rangle = P_1 + P_2 \cdot \log(1+z) + (P_3) \cdot \left(\log  \frac{\mathcal{M_*}}{\mathcal{M_{\odot}}} - 10 \right) ~.
  \label{eq:rams}
\end{equation}

The three best-fit parameters are: the intercept $P_1$=(20.97$\pm$0.16), the $\log$(1+z) slope $P_2$=(2.51$\pm$0.34), and the $\log$($\mathcal{M_{*}}$) slope $P_3$=(1.41$\pm$0.09). Our three-dimensional fitting yields $\chi^2_{\nu}$=1.08. Error bars on the best-fit parameters are given at 1$\sigma$ level. By bootstrapping over the uncertainties on ($P_1,P_2,P_3$) at the mean redshift of each bin, the error on $\log \langle L_{1.4}^{\mathrm{AGN}} \rangle$ is about 0.08~dex, at fixed ($\mathcal{M_{*}}$,$z$). For visual purposes, we show this error as a function of $\mathcal{M_{*}}$ at the mean redshift of each bin (coloured dashed areas) in Fig.~\ref{fig:rams} (left). All measurements are listed in Table~\ref{tab:rams}.

With a significantly super-linear slope of 1.41$\pm$0.09 (i.e., roughly 4$\sigma$ steeper than unity) between $\langle L_{1.4}^{\mathrm{AGN}} \rangle$ and galaxy $\mathcal{M_{*}}$, this trend suggests that more massive SFGs contain, on average, brighter radio AGN. 

Moreover, as shown in Fig.~\ref{fig:rams} (right), the RAMS normalization at fixed $\mathcal{M_{*}}$ clearly evolves with redshift ($\propto$(1+z)$^{2.51\pm0.34}$), mimicking the evolution of the MS relation (e.g., \citealt{Daddi+22b}) and the evolution of the molecular gas fraction in galaxies (roughly $\propto$(1+z)$^{2.5}$; e.g., \citealt{Saintonge+17}; \citealt{Tacconi+18}; \citealt{Liu+19}; \citealt{Tacconi+20}; \citealt{Decarli+20}; \citealt{Walter+20}, \citealt{Wang+22}). This trend suggests that the RAMS is in place at least since $z$$\sim$3.

% % % -------------------------------------------------------------------------------
\section{Discussion} \label{discussion}
% % % -------------------------------------------------------------------------------

In this section, we caution the reader about some caveats and limitations related to the RAMS (Sect.~\ref{caveats}). Then we further discuss the main implications of the existence of a RAMS in the framework of AGN-galaxy co-evolution: the relative contribution of AGN vs SF-driven radio emission (Sect.~\ref{agn_vs_sf}), the triggering of radio-AGN activity in SFGs (Sect.~\ref{enhancement}), and the long-term imprinting of AGN feedback on galaxy star formation (Sect.~\ref{feedback}).

\subsection{Possible caveats and limitations of the RAMS}  \label{caveats}

\begin{itemize}

\item The naming ``radio-AGN main sequence'' intentionally echoes both the ``star-forming main sequence'' (MS; e.g., \citealt{Speagle+14}) as well as the ``AGN main sequence'' (e.g., \citealt{Mullaney+12}) obtained from X-ray data. However, we emphasize that, unlike the MS of star-forming galaxies and similar to that of X-ray AGN, our RAMS is not visible for individual galaxies as it is ``hidden'' by intrinsic (radio) AGN variability.

\smallskip

\item Our $\langle L_{1.4}^{\mathrm{AGN}} \rangle$ measurements should \textit{not} be interpreted as the ``typical'' (i.e., most likely) radio-AGN luminosity observed in a MS galaxy at ($\mathcal{M_{*}}$,$z$). To address this, we would need to compute the probability distribution that a galaxy of a given ($\mathcal{M_{*}}$,$z$) hosts a radio AGN with that luminosity. Assuming that radio AGN triggering is a stochastic process, a sample-averaged AGN emission should match a time-averaged emission. Therefore, each $\langle L_{1.4}^{\mathrm{AGN}} \rangle$ estimate could be interpreted as a time-averaged luminosity of the AGN over the entire galaxy's lifecycle. This is further discussed in Sect.~\ref{feedback}.

\smallskip
 
\item The strong $\mathcal{M_{*}}$ dependence reported in Eq.~\ref{eq:rams} is not artificially induced by the $\mathcal{M_{*}}$-evolving 1.4~GHz luminosity limit from star formation ($\langle L_{1.4}^{\mathrm{SF}} \rangle$). On the contrary, we do find enhanced radio AGN activity in more massive galaxies \textit{on top} of increasing SF-driven radio emission. As a consequence, taking a fixed $q_{\mathrm{IRRC}}$ (e.g., 2.64 from the local Universe, \citealt{Bell03}) would have led to a higher number of radio-excess AGN in high-$\mathcal{M_{*}}$ galaxies than in this work, hence steepening the correlation between $\langle L_{1.4}^{\mathrm{AGN}} \rangle$ and galaxy $\mathcal{M_{*}}$.

\smallskip
 
\item We assume a fixed LF shape from \citet{Mauch+07} throughout the full ($\mathcal{M_*}$,$z$) range studied in this work. As detailed in Sect.~\ref{rlf_redshift}, any functional form with a faint (bright)-end slope flatter (steeper) than unity would yield a converged integral, hence the cumulative and average radio-AGN luminosity (Sect.~\ref{kld} and \ref{rams}) would remain stable. Since we do not find systematic deviations from \citeauthor{Mauch+07}'s LF across the observed $L_{1.4}$ range, we do not explore alternative functional forms.

\smallskip

\item It is perhaps confusing that the mean radio AGN power, at each $\mathcal{M_{*}}$ and redshift, is much lower than the observed $L_{1.4}^{\mathrm{AGN}}$ range covered by our RLF datapoints (e.g., Fig.~\ref{fig:rlf_mass_ple}). We clarify that this is purely the result of averaging the cumulative radio-AGN luminosity across the entire sample of $\mathcal{M_{*}}$-selected galaxies in each bin, which outnumber radio detections by $>$100 at the lowest $\mathcal{M_{*}}$ (see Table~\ref{tab:rams}). However, an important remark is that the majority of radio AGN feedback is originated from a small fraction of galaxies close to L$^{\star}$, implying relatively \textit{short} AGN phases at high radio power and much longer periods at low AGN power.

\smallskip

\item Stellar mass incompleteness might affect the real $N_{\mathrm{SFG}}$ at low $\mathcal{M_{*}}$. However, we note that correcting for missing low-$\mathcal{M_{*}}$ galaxies would \textit{decrease} the resulting $\langle L_{1.4}^{\mathrm{AGN}} \rangle$, strengthening the observed super-linear trend with $\mathcal{M_{*}}$.

\smallskip

\item Missing radio AGN might alter the RAMS shape and normalization. It is, indeed, possible that we are underestimating the identified number of radio-faint AGN within the IRRC, since they do not feature a radio excess (Sect.~\ref{histo}). Although constraining their demography is critical for obtaining a full AGN census, in our analysis we have conservatively restricted our $L_{1.4}$-range to above +2$\sigma$ from the luminosity at the IRRC, as in this regime we reach the highest degree of AGN purity and completeness. Then, by extrapolating the best-fit AGN RLF down to $\langle L_{1.4}^{\mathrm{SF}} \rangle$, we are implicitly factoring in the statistical contribution of radio-faint AGN in SF-dominated sources, even if formally undetected at 3~GHz. 

\smallskip

\item We acknowledge that the most massive galaxies ($\mathcal{M_*}$$>$10$^{11}$~$\mathcal{M_{\odot}}$) at $z$$\sim$3 could be under-represented, especially if extremely dusty and, thus, not fully captured by an optical/NIR counterpart catalogue \citep{Laigle+16}. This might partly explain the slightly offset datapoint in Fig.~\ref{fig:rams} at the highest ($\mathcal{M_*}$,$z$). 

\end{itemize}

  %%%placing figure  ---------------------------
\begin{figure}
\centering
     \includegraphics[width=\linewidth]{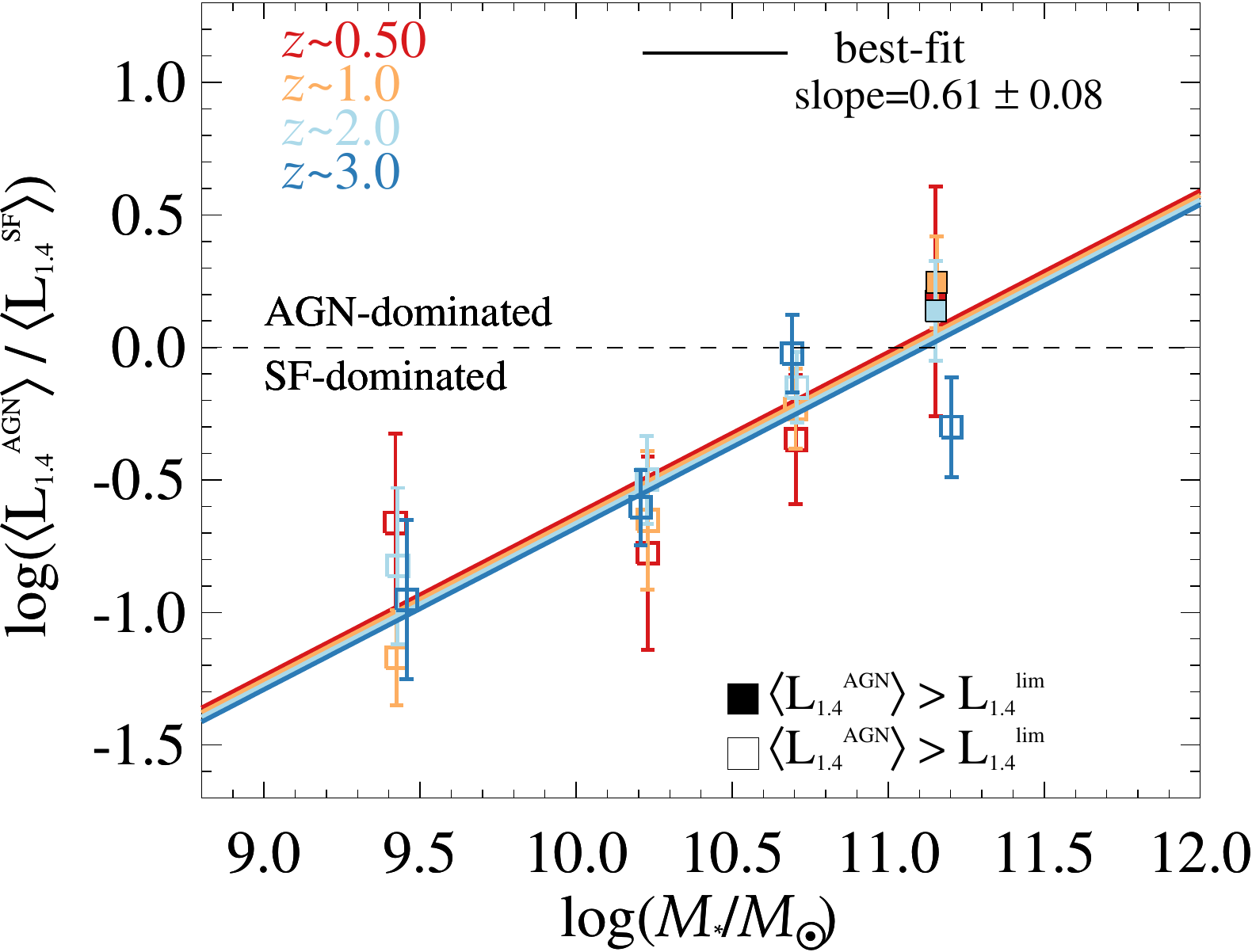}
 \caption{\small Logarithmic ratio between AGN-related and SF-related radio emission ($\langle L_{1.4}^{\mathrm{AGN}} \rangle$ and $\langle L_{1.4}^{\mathrm{SF}} \rangle$, respectively), as a function of $\mathcal{M_{*}}$, coloured by redshift. Each datapoint (square) represents the ratio between average AGN and SF luminosities for $\mathcal{M_{*}}$-selected SFGs in the same bin. The dividing threshold between AGN and SF-dominated regions (black dashed line) is crossed at $\mathcal{M_{*}}$$\sim$10$^{11}$~$\mathcal{M_{\odot}}$, above which galaxy radio emission is mainly powered by AGN jets. Filled squares mark bins in which the mean $\langle L_{1.4}^{\mathrm{AGN}} \rangle$ is above the 5$\sigma$ VLA~3~GHz luminosity limit ($L_{1.4}^{\mathrm{lim}}$, scaled to 1.4~GHz). Coloured lines indicate the best-fit ratio (solid) from Eq.~\ref{eq:lratio}.
 }
   \label{fig:lratio}
\end{figure}
%%%-------------------------------------

\subsection{Radio AGN emission is usually sub-dominant compared to SF} \label{agn_vs_sf}

Having determined a representative (i.e., time-averaged) mean radio-AGN luminosity across a wide $\mathcal{M_{*}}$ and redshift range, we can compare our RAMS with the shape and evolution of radio emission arising from SF processes. The latter is taken from D21 and calibrated for the same $\mathcal{M_{*}}$-selected ($NUVrJ$-based) SFGs. Relating the average power of both phenomena enables us to assess the dominant process of radio emission in SFGs. Eq.~\ref{eq:rams} describes the mean radio AGN power at fixed ($\mathcal{M_*}$,$z$), while the mean SF-related emission $\langle L_{1.4}^{\mathrm{SF}} \rangle$ is implicitly given in Eq.~\ref{eq:bestq}, though in the form of $q_{\mathrm{IRRC}}$. For consistency with the formalism adopted in the AGN part, we use the same analytical function of Eq.~\ref{eq:rams} to also fit the AGN-to-SF luminosity ratio, 
\begin{equation}
\log \left(\frac{\langle L_{1.4}^{\mathrm{AGN}} \rangle}{\langle L_{1.4}^{\mathrm{SF}} \rangle} \right) = Q_1 + Q_2 \cdot \log(1+z) + (Q_3) \cdot \left(\log  \frac{\mathcal{M_*}}{\mathcal{M_{\odot}}} - 10 \right) ~.
  \label{eq:lratio}
\end{equation}
We find the following best-fit parameters: $Q_1$=(--0.63$\pm$0.15), $Q_2$=(--0.05$\pm$0.31), $Q_3$=(0.61$\pm$0.08). The best-fit yields $\chi^{2}_{\nu}$=0.71. Fig.~\ref{fig:lratio} shows the average AGN-to-SF luminosity ratio (squares), as a function of $\mathcal{M_{*}}$ and coloured by redshift. The dashed horizontal line indicates equal contributions from AGN and SF. Coloured lines indicate the best-fit ratio (solid) expressed in Eq.~\ref{eq:lratio}. 

The redshift evolution of $\langle L_{1.4}^{\mathrm{AGN}} \rangle / \langle L_{1.4}^{\mathrm{SF}} \rangle$ is quite weak and consistent with a null slope (Eq.~\ref{eq:lratio}). This is clearly illustrated in Fig.~\ref{fig:lratio} by the nearly redshift-invariant behaviour. Our findings seem to suggest that radio AGN activity and galaxy star formation, at fixed $\mathcal{M_{*}}$, broadly evolve over time with a similar pace. This is probably only part of the full story, since we are mapping radio AGN activity solely inside SFGs. Indeed, we have seen in Sect.~\ref{kld} that the cumulative energy exerted by radio AGN in SF+passive galaxies \citep{Smolcic+17c} or via radio-mode feedback \citep{Croton+16} is notably dominant over that produced in SFGs alone. However, at $z$$>$1 ($NUVrJ$-based) SFGs vastly outnumber passive galaxies, and contain the bulk of radio-excess AGN. Therefore our findings should be providing a representative radio view of SMBH-galaxy growth at the cosmic noon. 

The positive $\mathcal{M_{*}}$ slope obtained in Eq.~\ref{eq:lratio} suggests a steeper $\mathcal{M_{*}}$ dependence of $\langle L_{1.4}^{\mathrm{AGN}} \rangle$ than $\langle L_{1.4}^{\mathrm{SF}} \rangle$. This is not surprising, given the \textit{super-linear} relationship of the RAMS (Eq.~\ref{eq:rams}) and the typically \textit{sub-linear} $\mathcal{M_{*}}$ dependence for the (radio) star-forming MS (D21).\footnote{The consensus range for the MS slope is about 0.5--0.9 (\citealt{Speagle+14} for a review; see also \citealt{Leslie+20} from VLA-COSMOS~3~GHz data), albeit this is affected by the bending at the highest $\mathcal{M_{*}}$, possibly linked to the transition from cold-to-hot gas accretion (e.g., \citealt{Dekel+13}; \citealt{Daddi+22b},\citeyear{Daddi+22}; \citealt{Popesso+22}).} This enhancement of AGN activity with $\mathcal{M_{*}}$ is further discussed in Sect.~\ref{feedback}.

Another take-away message from Fig.~\ref{fig:lratio} is that, at any redshift, radio AGN emission is usually sub-dominant relative to that from star formation. The only exception comes from the most massive galaxies at $\mathcal{M_{*}}$$>$10$^{11}$~$\mathcal{M_{\odot}}$ (at $z$$\lesssim$2), whose radio emission is mainly ($\approx$65\% on average) powered by AGN jets. This result is in line with previous studies (e.g., \citealt{Sabater+19}) finding widespread radio AGN activity in these massive galaxies, albeit without separating between passive and star-forming systems.

However, as mentioned in Sect.~\ref{rams}, the bulk of radio AGN activity originates from relatively short phases in which the AGN emission \textit{is dominant over SF}. This is what still allows us to calculate the mean $\langle L_{1.4}^{\mathrm{AGN}} \rangle$ despite this being on average (i.e., across the galaxy's lifetime) sub-dominant compared to $\langle L_{1.4}^{\mathrm{SF}} \rangle$.

We emphasize that the regions labeled as ``AGN-dominated'' or ``SF-dominated'' do not reflect the probability of a radio AGN being turned on in a given galaxy, but rather a luminosity-weighted probability. In other words, our average measurements carry a degeneracy between timescales and intensity of each episode of AGN activity. This distinction is supported by multiple observations of intermittant jet activity in radio galaxies (e.g., \citealt{Jurlin+20}; \citealt{Brienza+21}), which can occur over timescales comparable to those of galaxy star formation (10$^{7-8}$~yr, see e.g., \citealt{Konar+13}). Breaking such a degeneracy in observations is challenging at high-redshift, since jet-driven emission appears intrinsically more compact ($<$1~kpc; e.g., \citealt{Costa+18}) than in luminosity-matched AGN at $z$$\sim$0 (e.g., \citealt{Bondi+18} based on the same VLA-COSMOS~3~GHz data), and also because radio-faint jets can be easily washed out by stellar-driven radio emission, that is enhanced at high redshift (\citealt{Padovani16}; \citealt{Magliocchetti+18}; \citealt{Delvecchio+18}). Nonetheless, our statistical approach tied to the AGN RLF allows us to dissect the role of radio-AGN duty cycle and intensity of AGN activity, as discussed in Sect.~\ref{duty_cycle} and ~\ref{radio_bright}.

To provide a rough idea of how incomplete is our current picture of radio AGN activity in SFGs, in Fig.~\ref{fig:lratio} we highlight as filled squares those bins in which the mean $\langle L_{1.4}^{\mathrm{AGN}} \rangle$ is formally above the 5$\sigma$ VLA~3~GHz luminosity limit ($L_{1.4}^{\mathrm{lim}}$, corresponding to 20$\mu$Jy/beam at 1.4~GHz). This check allows us to test the detectability of the \textit{mean} radio AGN activity across the galaxy's lifecycle by current deep radio surveys. Unsurprisingly, among AGN-dominated bins, only those at $\mathcal{M_{*}}$$>$10$^{11}$~$\mathcal{M_{\odot}}$ (and $z$$\lesssim$2) would be formally detectable. Even pushing radio sensitivity to below $\mu$Jy levels (e.g., with SKA1-MID or ngVLA), isolating the sub-dominant ($\approx$10--25\% at $\mathcal{M_{*}}$$\sim$10$^{10}$~$\mathcal{M_{\odot}}$, based on Fig.~\ref{fig:lratio}) AGN-related emission at 1.4~GHz will be out of reach also at sub-arcsec angular resolution ($\gtrsim$0.1'' at 0.95-1.76 GHz in SKA1-MID Band-2, e.g., \citealt{Braun+19}; see also \citealt{Sweijen+22} for LOFAR imaging). Thus, a promising alternative comes from Very Long Baseline Interferometry (VLBI) techniques, that is crucial to pin down radio-AGN emission even in the SF-dominated regime (\citealt{Maini+16}; \citealt{HerreraRuiz+18}; \citealt{Muxlow+20}; \citealt{Radcliffe+21}, \citeyear{Radcliffe+21b}). In this context, the added VLBI capability to the SKA will be critical \citep{Paragi+15} to reach a full radio-AGN census.

  %%%placing figure  ---------------------------
\begin{figure}
\centering
     \includegraphics[width=\linewidth]{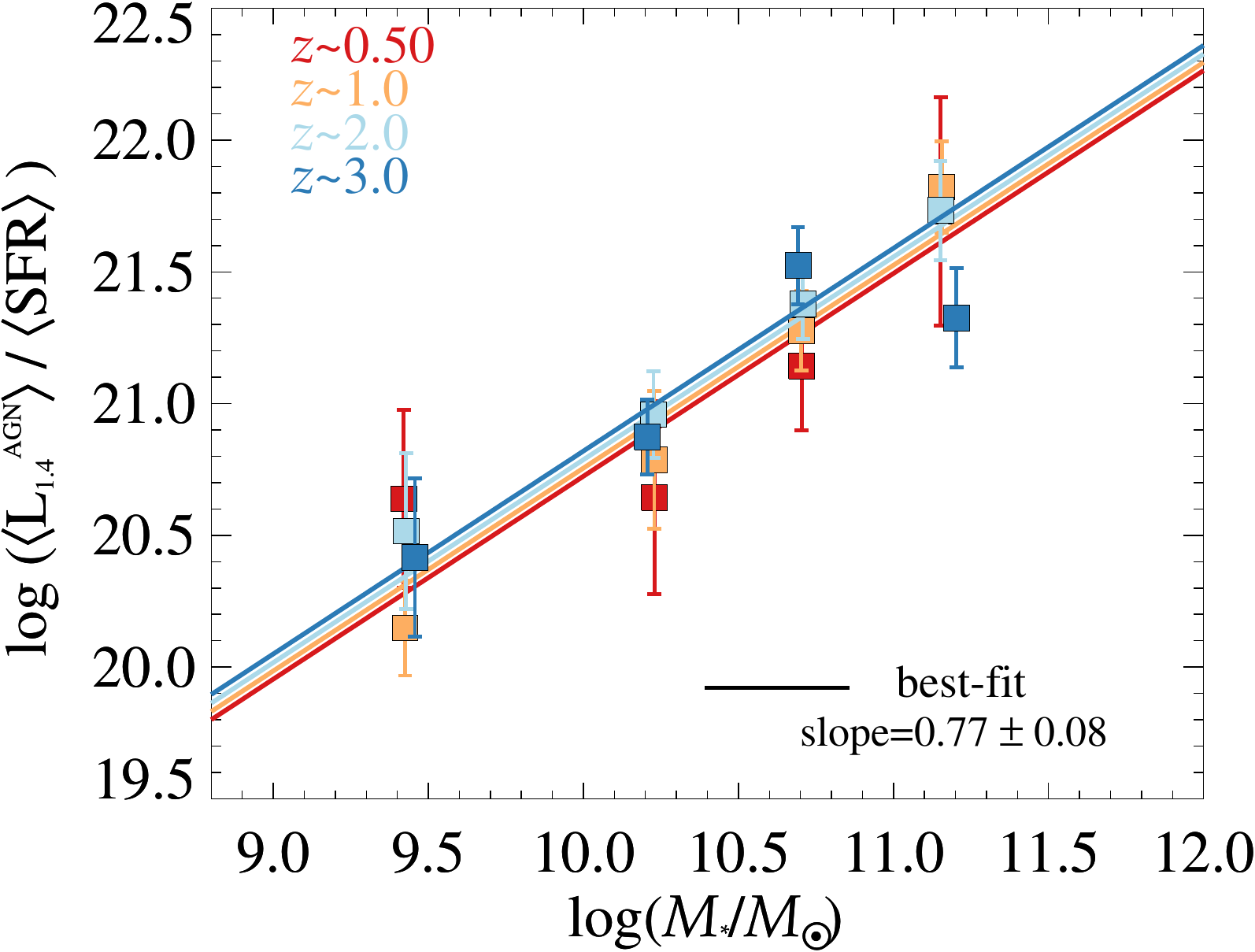}
 \caption{\small Logarithmic ratio between $\langle L_{1.4}^{\mathrm{AGN}} \rangle$ and $\langle \mathrm{SFR} \rangle$, as a function of $\mathcal{M_{*}}$, coloured by redshift. As in Fig.~\ref{fig:lratio}, all parameters are averaged across the entire sample of $\mathcal{M_{*}}$-selected SFGs in the same bin. Coloured lines indicate the best-fit ratio (solid) obtained from Eq.~\ref{eq:lratio_sfr}.
 }
   \label{fig:lratio_sfr}
\end{figure}
%%%-------------------------------------

  %%%placing figure  ---------------------------
\begin{figure*}
\centering
     \includegraphics[width=3.61in]{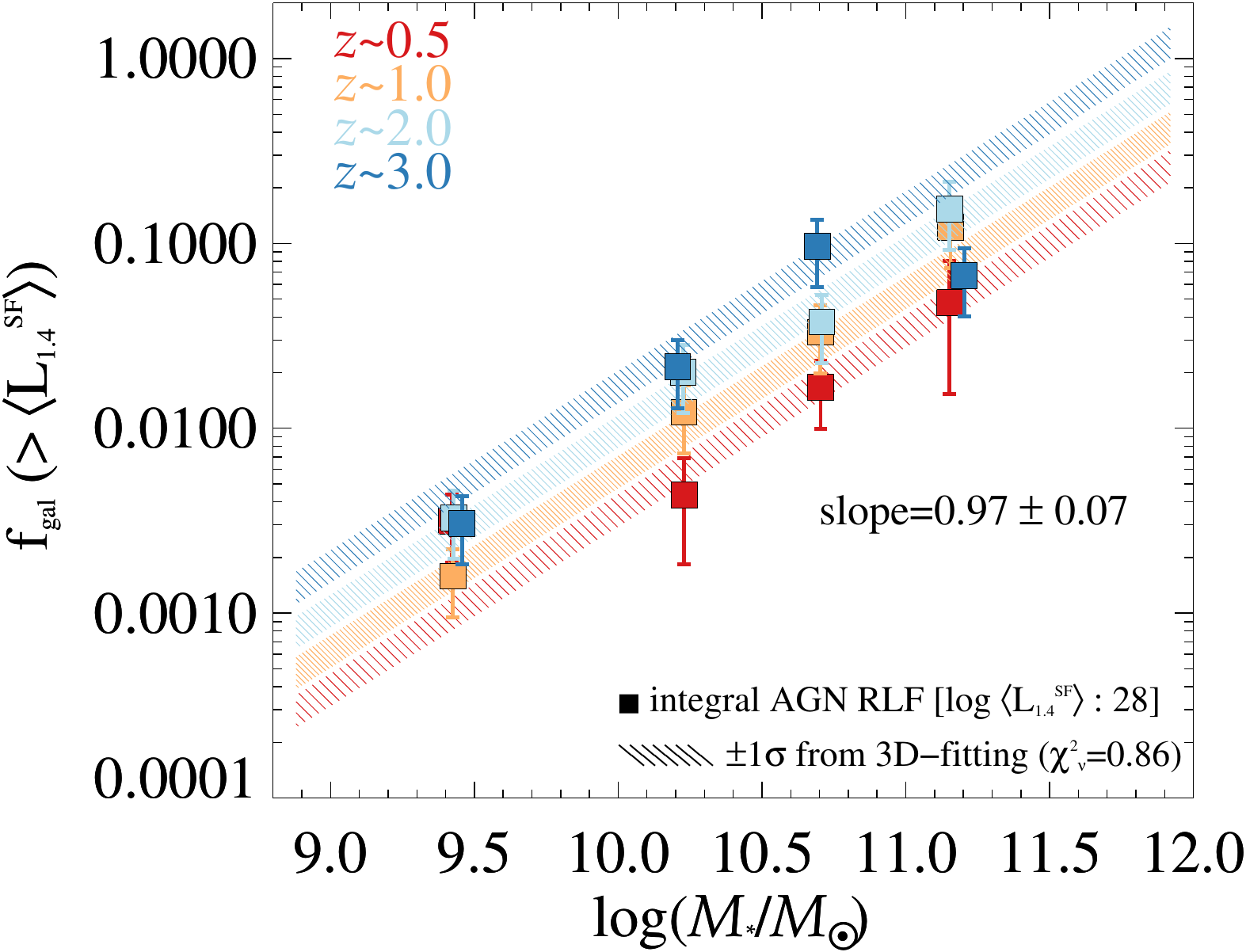}
     \includegraphics[width=3.56in]{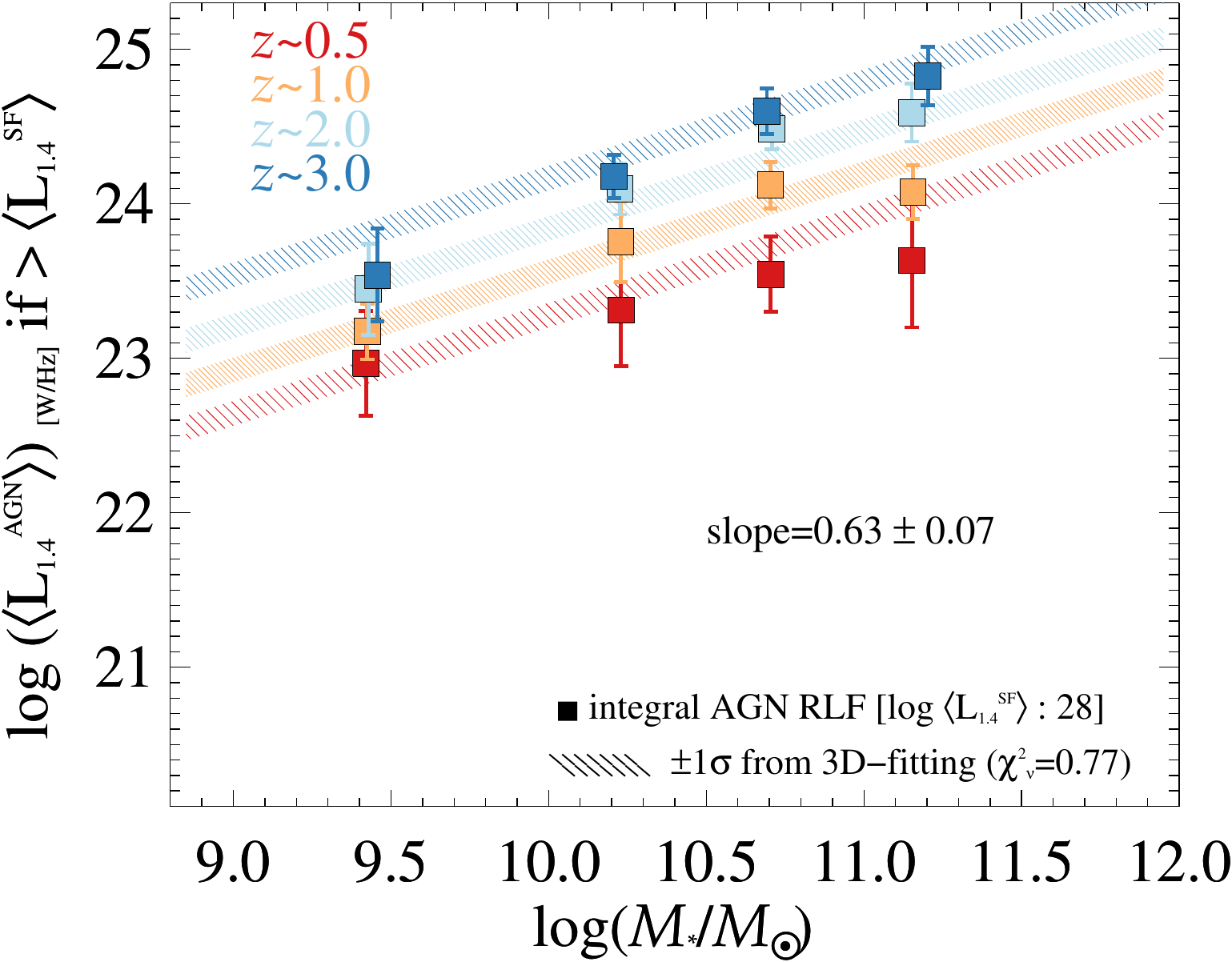}

 \caption{\small \textit{Left}: Fraction of \textit{all SFGs} in ``radio-AGN phase'' ($f_{\mathrm{gal}}$), i.e., having radio luminosity above $\langle L_{1.4}^{\mathrm{SF}} \rangle$, at a given ($\mathcal{M_*}$,$z$). This fraction is equivalent to the radio-AGN duty cycle, as written in Eq.~\ref{eq:duty_cycle}. Symbols and colours are as in Fig.~\ref{fig:rams}. \textit{Right}: mean radio luminosity in ``radio-AGN phase'' ($\langle L_{1.4}^{\mathrm{active}} \rangle$) as a function of ($\mathcal{M_*}$,$z$). 
 }
   \label{fig:duty_cycle}
\end{figure*}
%%%-------------------------------------

\subsection{What drives the super-linear RAMS with $\mathcal{M_{*}}$?} \label{enhancement}

The super-linear (1.41$\pm$0.09) $\log$~$\mathcal{M_{*}}$ dependence of the RAMS suggests enhanced (radio) AGN activity compared to star formation in massive SFGs. To make this trend explicit in terms of SFR, we re-write Eq.~\ref{eq:lratio} by converting $\langle L_{1.4}^{\mathrm{SF}} \rangle$ into $\langle \mathrm{SFR} \rangle$. To this end, we employ the $z$-evolving SFR--$\mathcal{M_{*}}$ fitting form proposed by \citet{Lee+15} but applied to the data from D21 (see Table~1 of \citealt{Daddi+22b}). The resulting expression reads as follows: 
\begin{equation}
\log \left(\frac{\langle L_{1.4}^{\mathrm{AGN}} / \mathrm{W~Hz^{-1}} \rangle}{\langle \mathrm{SFR} / \mathcal{M_{\odot}}~yr^{-1} \rangle} \right) = R_1 + R_2 \cdot \log(1+z) + (R_3) \cdot \left(\log  \frac{\mathcal{M_*}}{\mathcal{M_{\odot}}} - 10 \right) ~.
  \label{eq:lratio_sfr}
\end{equation}
With a reduced chi-square of $\chi^{2}_{\nu}$=0.70, the best-fit parameters are: $R_1$=(20.73$\pm$0.15), $R_2$=(0.08$\pm$0.32), $R_3$=(0.77$\pm$0.08). Because the SFR--$\mathcal{M_{*}}$ relation is flatter than the $L_{1.4}^{\mathrm{SF}}$--$\mathcal{M_{*}}$ relation\footnote{The difference in slope is $\approx$0.15, that is the net $\mathcal{M_{*}}$ dependence of the $L_{IR}$/$L_{1.4}^{\mathrm{SF}}$ ratio (or $q_{\mathrm{IRRC}}$) found in D21.}, the above ratio $\langle L_{1.4}^{\mathrm{AGN}} \rangle / \langle \mathrm{SFR} \rangle $ exhibits a slightly steeper $\mathcal{M_{*}}$ dependence than in Eq.~\ref{eq:lratio}. This is also clearly shown in Fig.~\ref{fig:lratio_sfr}.

\subsubsection{The radio-AGN duty cycle across SFGs}    \label{duty_cycle}

A super-linear trend of $\langle L_{1.4}^{\mathrm{AGN}} \rangle$ with $\mathcal{M_{*}}$ might be attributed to either a higher radio-AGN duty cycle in more massive galaxies, or to an intrinsically brighter radio-AGN phase in more massive galaxies, or to both effects. Breaking such a degeneracy is important to assess the role of the host-galaxy $\mathcal{M_*}$ in driving radio AGN activity. 

We leverage the best-fit RLF obtained at each ($\mathcal{M_*}$,$z$) in order to calculate the fraction of \textit{all SFGs} hosting a radio AGN. For consistency with the formalism used so far, we set $\langle$$L_{1.4}^{\mathrm{SF}}$$\rangle$ as the lower radio-luminosity limit for a galaxy to be in ``AGN phase'', at each ($\mathcal{M_*}$,$z$). Under this assumption, the fraction of galaxy's lifetime spent as AGN, i.e., the radio-AGN duty cycle, corresponds to the fraction of all SFGs above this limit ($f_{gal}~({>\langle L_{1.4}^{\mathrm{SF}} \rangle})$),

\begin{equation}
f_{gal}~({> \langle L_{1.4}^{\mathrm{SF}} \rangle}) = \frac{\int_{L_{1.4}^{\mathrm{SF}}}^{L_{1.4}^{\mathrm{max}}} \Phi(L_{1.4}^{\mathrm{AGN}}) \cdot \langle V_{\mathrm{max}} \rangle \cdot ~ d(\log L_{1.4}^{\mathrm{AGN}})}{N_{\mathrm{SFG}}} ~, 
   \label{eq:duty_cycle}
\end{equation}
where the numerator is the total number of galaxies in radio-AGN phase, scaled from the AGN RLF in a given ($\mathcal{M_*}$,$z$) bin. 
Fig.~\ref{fig:duty_cycle} (left) shows the radio-AGN duty cycle as a function of $\mathcal{M_*}$, colour-coded by redshift. We find a close-to-linear best-fit relation ($\propto$$\mathcal{M_{*}}^{0.97\pm0.07}$). A significant evolution is also observed, although weaker than that of the RAMS ($\propto$(1+$z$)$^{1.40\pm0.27}$). This result suggests that more massive and/or more distant star-forming galaxies undergo a progressively higher radio-AGN duty cycle. The duty cycle reaches close to 10\% at $\mathcal{M_*}$$\gtrsim$10$^{11}$~$\mathcal{M_{\odot}}$.
Other studies have also reported similar results. For instance, \citealt{Best+05} found that the radio-AGN duty cycle in local passive galaxies strongly rises as $\propto$M${_*}^{2.5}$ (see also \citealt{Sabater+19}). When selecting only star-forming hosts (based on different cuts in specific-SFR; sSFR=SFR/$\mathcal{M_{*}}$), \citet{Kondapally+22} found a slightly shallower behaviour, with the radio-AGN fraction scaling as $\propto$$\mathcal{M_{*}}^{1.37\pm0.57}$ at redshift 0.3$<$$z$$\leq$1.5 (see their Figure 9). Consistently with our work, \citet{Kondapally+22} found that the fraction of SFGs hosting a radio AGN reaches $\approx$10\% at 10$^{11.5}$~$\mathcal{M_{\odot}}$, albeit using a different AGN nomenclature (i.e., ``SF-LERGs'', see details in Sect.~\ref{radio_unified}). Instead, the marginally steeper dependence of the duty cycle on $\mathcal{M_{*}}$ from \citeauthor{Kondapally+22}, can be partly explained by their different integration limit: the authors computed the duty cycle as the fraction of star-forming galaxies hosting a radio AGN with $\log$[$L_{\mathrm{150~MHz}}$ / W~Hz$^{-1}$]$\geq$24, which corresponds to $\log$[L$_{\mathrm{1.4~GHz}}$ / W~Hz$^{-1}$]$\geq$23.3 (if assuming $\gamma$=--0.75). This threshold is roughly consistent with our lower integration limit, $L_{1.4}^{\mathrm{SF}}$, at $\mathcal{M_*}$$\gtrsim$10$^{11}$~$\mathcal{M_{\odot}}$ and $z$$\sim$1, whereas it is systematically higher (by 1~dex) than $L_{1.4}^{\mathrm{SF}}$ at $\mathcal{M_*}$$\sim$10$^{10}$~$\mathcal{M_{\odot}}$ and/or at lower redshifts (see Table~\ref{tab:rams}). Hence, we stress that adopting a ($\mathcal{M_*}$,$z$)-dependent integration limit enables us to reach a globally \textit{higher} incidence of \textit{radio-faint} AGN, especially in lower-$\mathcal{M_*}$ galaxies, which a more conservative luminosity cut would likely miss. This explains our flatter $\mathcal{M_*}$ trend compared to that found in \citet{Kondapally+22}, despite obtaining consistent numbers for the most massive galaxies.

\subsubsection{The AGN luminosity in ``radio-bright phase''}    \label{radio_bright}

Given the \textit{linear} correlation between radio-AGN duty cycle and $\mathcal{M_{*}}$ (Fig.~\ref{fig:duty_cycle}, left), we conclude that the \textit{super-linear} $\mathcal{M_{*}}$ trend of the RAMS (Fig.~\ref{fig:rams}, $1.41\pm0.09$) cannot be explained by a larger incidence of radio AGN alone, but it must be steepened by the fact that the typical AGN luminosity during a ``radio-bright phase'' is also higher in more massive than in less massive galaxies. 

As a sanity check, we calculate the mean AGN luminosity during a ``radio-AGN phase'' ($\langle L_{1.4}^{\mathrm{active}} \rangle$), again based on the best-fit RLF:
\begin{equation}
\langle L_{1.4}^{\mathrm{active}} \rangle = \frac{\int_{L_{1.4}^{\mathrm{SF}}}^{L_{1.4}^{\mathrm{max}}} \Phi(L_{1.4}^{\mathrm{AGN}}) \cdot L_{1.4}^{\mathrm{AGN}} \cdot ~ d(\log L_{1.4}^{\mathrm{AGN}})}{\int_{L_{1.4}^{\mathrm{SF}}}^{L_{1.4}^{\mathrm{max}}} \Phi(L_{1.4}^{\mathrm{AGN}}) \cdot ~ d(\log L_{1.4}^{\mathrm{AGN}})}  ~ .
   \label{eq:active}
\end{equation}
This mean AGN luminosity is obtained as a weighted-average over the RLF. A similar formalism has been used in \citet{Aird+22} for X-ray AGN. Fig.~\ref{fig:duty_cycle} (right) displays $\langle L_{1.4}^{\mathrm{active}} \rangle$ across all ($\mathcal{M_{*}}$,$z$) bins. Not surprisingly, we again observe a positive correlation with $\mathcal{M_{*}}$ (as $\propto$$\mathcal{M_{*}}$$^{0.63\pm0.07}$), though weaker than that seen for the duty cycle, but a more significant evolution with redshift (as $\propto$(1+$z$)$^{1.96\pm0.27}$). 

We acknowledge that part of the above $\mathcal{M_{*}}$ dependence could be explained by a roughly constant (kinetic) Eddington ratio and a fixed $\mathcal{M_{BH}}$/$\mathcal{M_*}$ ratio. In this assumption, more massive galaxies would simply \textit{appear} more (radio) luminous because they host more massive BHs, while the Eddington ratio is, in fact, constant. A more detailed analysis of the Eddington ratio distributions in radio AGN is postponed to a future work.

Based on these results, we argue that our empirical RAMS can be explained only by a combination of both a higher duty cycle, and a brighter radio-AGN phase in more massive and/or higher-$z$ galaxies.

\subsection{Mapping integrated AGN feedback across the galaxy population} \label{feedback}

The strong $\mathcal{M_{*}}$ stratification seen in the best-fit AGN RLF (Fig.~\ref{fig:rlf_mass_ple}) is reflected into the super-linear trend reported in our RAMS (Fig.~\ref{fig:rams}). This behaviour corroborates the idea that radio AGN activity is enhanced relative to star formation in the most massive SFGs, at least since $z$$\sim$3. 
In this context, we compare our findings with recent radio and X-ray studies to discuss the role of star-forming host galaxies in triggering different types of AGN feedback.

\subsubsection{Radio AGN at the crossroad: towards a single AGN population in SFGs?} \label{radio_unified}

It has been argued that more massive galaxies undergo a higher radio-AGN duty cycle (e.g., \citealt{Sabater+19}, at $z$$<$0.3), i.e., the fraction of the galaxy's lifetime in which a radio AGN is switched on, on average, strongly increases with $\mathcal{M_{*}}$. At higher redshift ($z<$2.5), a similar analysis has been presented in \citet{Kondapally+22}, who exploited 150~MHz data from the first data release of the LOFAR Two-meter Sky Survey Deep Fields (LoTSS-Deep) survey (\citealt{Tasse+21}; \citealt{Sabater+21}; \citealt{Kondapally+21}; \citealt{Duncan+21}). 

Following the historical dichotomy observed in the local Universe between radiatively efficient and inefficient radio AGN (e.g., \citealt{Best+12}), \citet{Kondapally+22} separate AGN between low-excitation and high-excitation radio galaxies (LERGs and HERGs, respectively), studying their space density and incidence as a function of galaxy $\mathcal{M_{*}}$. The authors classify LERGs based on the presence of ($>$0.7~dex, i.e., 3$\sigma$) radio-excess from a ridgeline set by Best et al. (in prep.) and the absence of optical-MIR signatures of AGN activity from SED-fitting. On the contrary, HERGs in their study consist of both radio-excess and SED-based AGN. The population of LERGs is further split among quiescent and star-forming LERGs, based on the sSFR of the host galaxy. Therefore, our radio-excess AGN in SF hosts are broadly consistent with the combined (SF-)HERG and SF-LERG population in their study. Interestingly, \citet{Kondapally+22} find that quiescent LERGs show systematically different properties and evolution compared to SF-LERGs, which instead resemble more closely the HERG population. For instance, they observe a flatter $\mathcal{M_{*}}$ dependence of the fraction of SF-LERGs (slope $\sim$1.37$\pm$0.57), compared to that of quiescent LERGs (slope $\sim$2.5 at all redshifts). The incidence of SF-LERGs increases with redshift following the evolution of the cold gas fraction ($\propto$(1+$z$)$^{2.5}$, e.g., \citealt{Tacconi+18}), while quiescent LERGs do not show hints of evolution. This behaviour suggests that a different fueling mechanism, likely associated with the availability of cold gas supply, is responsible for triggering SF-LERGs. The link with the HERG population is further reinforced by the self-similar fractions of SF-LERGs and HERGs at all redshifts.

Therefore, a plausible scenario is that HERGs and SF-LERGs are both triggered by cold gas accretion, that is more largely available in star-forming than in passive systems. Unsurprisingly, HERGs are mostly hosted in SFGs \citep{Delvecchio+17}, and X-ray stacking of radio-excess AGN reveals systematically ($>3$$\times$) higher BHARs in star-forming than in quiescent hosts (split by $NUVrJ$ colours), at fixed radio power and redshift (\citealt{Delvecchio+18}; see also \citealt{Magliocchetti+18}). While it is true that local LERGs can also be fueled by cold gas accretion (e.g., \citealt{Ruffa+19}), this is possibly driven by sporadic cold gas filaments due to radiative cooling of the hot halo (\citealt{Hardcastle18}, see \citealt{Hardcastle+20} for a review) or chaotic accretion (\citealt{Gaspari+15},\citeyear{Gaspari+17}), that anyway do not switch LERGs to ``radiative mode'' AGN.

We reiterate that the main difference between (SF-)HERGs and SF-LERGs in the literature is the presence of excess emission in their X-ray/MIR/optical data relative to pure star formation. However, SMBH accretion is a stochastic process and can potentially vary over $\lesssim$Myr timescales (e.g., \citealt{Schawinski+15}), biasing these criteria to \textit{instantaneous} rather than \textit{long-term} AGN activity. Additionally, the above criteria are sensitive to the contrast of AGN-vs-host light, hence they can potentially miss relatively faint AGN in highly star-forming systems. In this framework, splitting AGN by host-galaxy type (as well as $\mathcal{M_*}$,$z$) might be more effective in capturing the historical SMBH fueling due to cold gas accretion in the host, as compared to using conventional ``single-epoch'' AGN diagnostics. In other words, radio-excess AGN in SF hosts (both SF-LERGs and SF-HERGs) and radio-excess AGN in quiescent hosts (both Q-LERGs and Q-HERGs) might behave broadly as two distinct AGN populations, irrespective of their on-going SMBH growth.

This simple picture, in which the long-term AGN activity is primarily modulated by the host-galaxy type (but internally varying with $\mathcal{M_*}$,$z$) is further corroborated by the comparison between average radio and X-ray AGN activity in SFGs (Sect.~\ref{radio_x}).

\subsubsection{Radio and X-ray AGN in SF hosts: a common fueling scenario?} \label{radio_x}

Our analysis has demonstrated that the incidence, evolution, integrated and mean luminosity of radio-excess AGN follow a strong trend with galaxy $\mathcal{M_{*}}$. Here we discuss the quantitatively similar $\mathcal{M_{*}}$ dependence seen in the average X-ray properties of $\mathcal{M_{*}}$-selected samples of SFGs. 

A well-known study by \citet{Mullaney+12} put forward the idea that the average BHAR/$\mathrm{SFR}$ ratio in MS galaxies is both redshift and $\mathcal{M_{*}}$-invariant at $\mathcal{M_*}>$10$^{10}$~$\mathcal{M_{\odot}}$ and 0.5$<$$z$$<$2.5. This ``hidden AGN main sequence'' lies at BHAR/$\mathrm{SFR}$$\sim$10$^{-3}$, consistent with a constant $M_{\mathcal{BH}}$/$\mathcal{M_{*}}$ ratio, thus in line with the empirical BH-bulge mass scaling relations at $z$$\sim$0 \citep{Kormendy+13}. In the past decade, similar studies on larger samples and wider $\mathcal{M_{*}}$ ranges have been favouring an increasing BHAR/$\mathrm{SFR}$ ratio with $\mathcal{M_{*}}$ (\citealt{Rodighiero+15}; \citealt{Yang+18}; \citealt{Aird+19}; \citealt{Bernhard+19}; \citealt{Carraro+20}; \citealt{Delvecchio+20}), suggesting that BHAR is enhanced relative to SFR in more massive galaxies. The BHAR/$\mathrm{SFR}$ ratio evolves as $\propto$M${_*}^{0.5-0.7}$ and, when adopting the same bending MS, is consistent with a redshift-invariant ratio. Pulling out the SFR term, the effective BHAR scales as $\propto$$\mathcal{M_*}^{1.5}$, that is remarkably similar to that observed for the mean $\langle{L_{1.4}^{\mathrm{AGN}}}\rangle$ (Sect.~\ref{rams}). 

We note that X-ray based BHAR estimates are usually scaled from the mean X-ray luminosity $\langle{L_{X}}\rangle$ obtained from either X-ray stacking (e.g., \citealt{Rodighiero+15}; \citealt{Yang+18}; \citealt{Carraro+20}), or via Bayesian modeling of X-ray detections and non-detections (e.g., \citealt{Aird+19}; \citealt{Bernhard+19}). These techniques can smooth over short-term fluctuations due to AGN flickering ($\lesssim$Myrs; e.g., \citealt{Chen+13}; \citealt{Hickox+14}). Recurrent AGN activity is also seen in the radio (e.g., \citealt{Jurlin+20}; \citealt{Brienza+21}), albeit over longer timescales (10$^{7-8}$~yr, see e.g., \citealt{Konar+13}). Therefore, our sample-average $\langle{L_{1.4}^{\mathrm{AGN}}}\rangle$ measurements should be, if anything, less affected by AGN variability than from the X-rays. Nonetheless, radio-faint AGN emission can suffer more from ``host galaxy dilution'' \citep{Padovani17}, which we have addressed by adopting a ($\mathcal{M_{*}}$,$z$)-dependent IRRC (D21).

\begin{figure}
\centering
     \includegraphics[width=\linewidth]{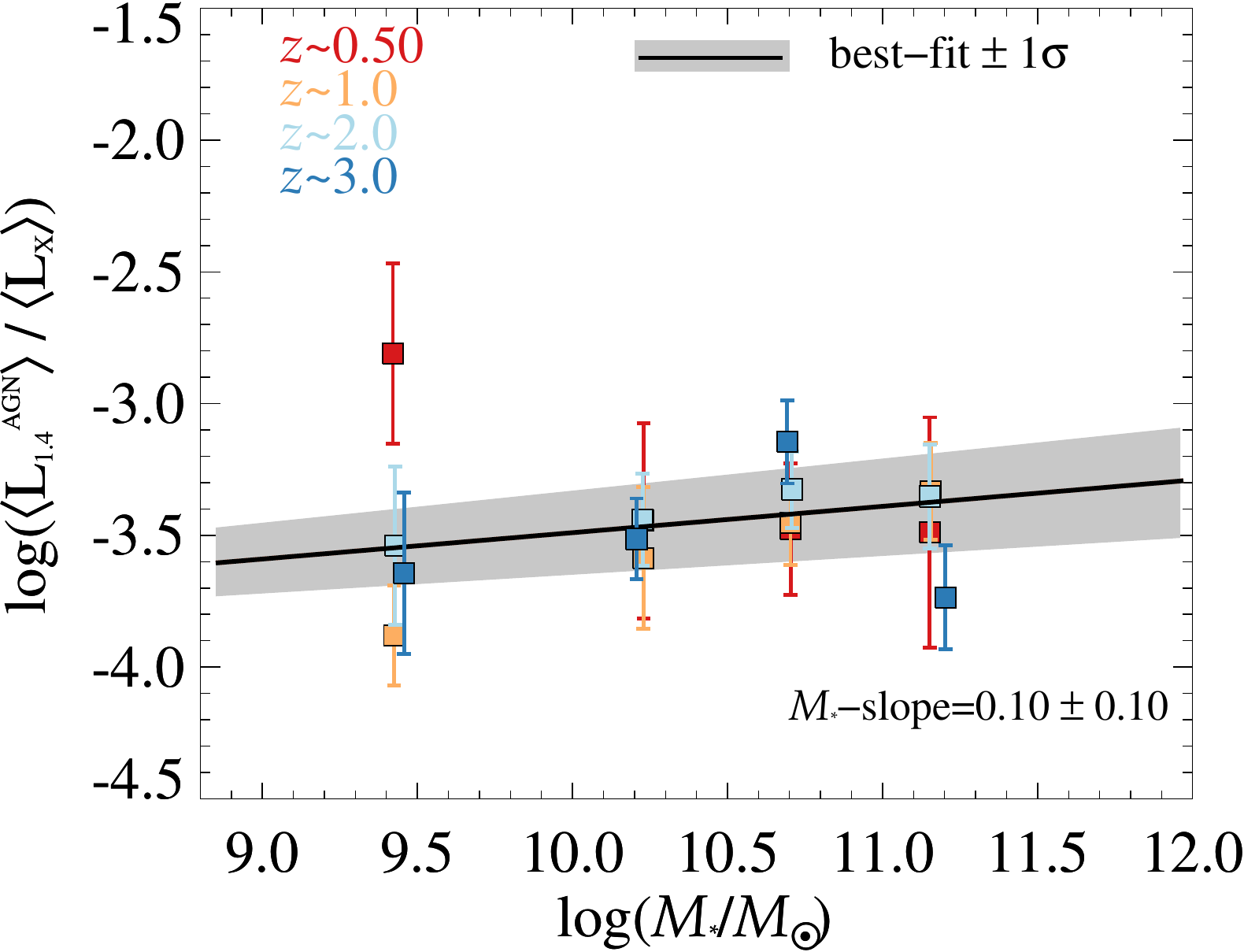}
 \caption{\small Logarithmic ratio between our $\langle L_{1.4}^{\mathrm{AGN}} \rangle$ measurements and $\langle L_{X} \rangle$ obtained from X-ray stacking \citep{Carraro+20}, as a function of $\mathcal{M_{*}}$, coloured by redshift. The black line indicates the best-fit ratio with $\mathcal{M_{*}}$, by imposing a $z$-invariant trend, which returns a $\mathcal{M_{*}}$-slope of 0.10$\pm$0.10. The grey shaded area marks the $\pm$1$\sigma$ confidence interval after propagating the parameter uncertainties. See Sect.~\ref{radio_x} for details.
 }
   \label{fig:lratio_agn}
\end{figure}

Fig.~\ref{fig:lratio_agn} displays the logarithmic ratio between $\langle{L_{1.4}^{\mathrm{AGN}}}\rangle$ and $\langle{L_{X}}\rangle$ (scaled to the same units) as a function of $\mathcal{M_*}$, colour-coded by redshift. X-ray measurements are taken from the SFG sample of \citet{Carraro+20}, for consistency with this work. Indeed, \citet{Carraro+20} stacked X-ray images of a $\mathcal{M_*}$-selected sample, and identified SFGs via $NUVrJ$ colour criteria as in this work. Other similar studies adopted different galaxy classifications based on sSFR (e.g., \citealt{Aird+19}; \citealt{Ito+22}), and thus the results are not fully comparable to those of \citet{Carraro+20}, albeit qualitatively consistent with each other. A 3-D fitting of all $\langle{L_{1.4}^{\mathrm{AGN}}}\rangle$/$\langle{L_{X}}\rangle$ datapoints with $\mathcal{M_{*}}$ and redshift yields a weak, poorly significant dependence on both $\mathcal{M_{*}}$ (0.14$\pm$0.10) and redshift (0.00$\pm$0.38). Imposing for simplicity a $z$-invariant relationship leads to an even weaker $\mathcal{M_{*}}$ trend (0.10$\pm$0.10), and consistent with a constant ratio of $\log$($\langle{L_{1.4}^{\mathrm{AGN}}}\rangle$/$\langle{L_{X}}\rangle$)$\approx$--3.5 (black line in Fig.~\ref{fig:lratio_agn}). The best-fit expression yields $\chi^2_{\mathrm{red}}$=1.48. Intriguingly, this ($\langle{L_{1.4}^{\mathrm{AGN}}}\rangle$/$\langle{L_{X}}\rangle$) ratio resembles the ``radio loudness'' parameter $R_X$ typically used to separate ``radio quiet'' from ``radio loud'' AGN (\citealt{Terashima+03}, see also \citealt{Lambrides+20}).

Therefore, we conclude that mean X-ray and radio AGN luminosities in SFGs seem to evolve in a strikingly similar fashion with both $\mathcal{M_{*}}$ and redshift, suggesting that similar mechanisms trigger and sustain long-term X-ray and radio AGN activity in SFGs.

  %%%placing figure  ---------------------------
%%%-------------------------------------

Such a similar behaviour with $\mathcal{M_{*}}$ extends beyond average measurements, encompassing the global evolution of the AGN luminosity function. Indeed, the characteristic knee luminosity L$^{\star}$ of the X-ray AGN luminosity function (XLF), at fixed redshift, has been reported to increase with $\mathcal{M_{*}}$ in a qualitatively similar fashion to this work, although partly induced by a $\mathcal{M_{*}}$-invariant characteristic Eddington ratio ($\lambda_{Edd}$, or sBHAR$\propto$$L_{X}$/$\mathcal{M_{*}}$; \citealt{Aird+12}) assumed in the XLF modeling \citep{Delvecchio+20}. Moreover, the integrated AGN luminosity density in the X-rays is dominated by AGN in MS galaxies with 10.5$<$$\log$($\mathcal{M_{*}}$/$\mathcal{M_{\odot}}$)$<$11, showing a peak at $z$$\sim$2 (e.g., \citealt{Delvecchio+20}). This is fully consistent with our reported evolutionary trend of the kinetic AGN luminosity density in SFGs (Sect.~\ref{kld}). Similarly, the average $\langle{L_{X}}\rangle$/$\mathrm{SFR}$ ratio, at fixed $\mathcal{M_{*}}$, evolves with redshift following the MS relation, closely resembling the evolution of $\langle{L_{1.4}^{\mathrm{AGN}}}\rangle$/$\mathrm{SFR}$ (Sect.~\ref{enhancement}).

Such a degree of consistency in the evolution of the X-ray and radio AGN population in SFGs is surprising given the completely different methodologies adopted in each of the above studies. However, these results add to the evidence discussed earlier (Sect.~\ref{radio_unified}) that the availability of cold gas supply modulates the long-term fueling onto the SMBH, broadly independent of the (instantaneous) AGN diagnostics at a given wavelength. The small overlap (10--15\%) previously reported between X-ray and radio AGN populations (e.g., \citealt{Goulding+14}; \citealt{Azadi+15}; \citealt{Delvecchio+17}; \citealt{Ji+22}), even for the same galaxy type, could be attributed to intrinsic AGN variability, which globally reaches a duty cycle of 10\% in the most massive ($\mathcal{M_*}$$\gtrsim$10$^{11}$~$\mathcal{M_{\odot}}$) SFGs (see Fig.~\ref{fig:duty_cycle}, left), consistently with the X-ray/radio AGN overlap in $\mathcal{M_*}$-matched galaxies.

We reiterate that both $\langle{L_{1.4}^{\mathrm{AGN}}}\rangle$ and $\langle{L_{X}}\rangle$ are averaged across the entire SFG population, at each $\mathcal{M_{*}}$ and redshift, thus smoothing over the entire AGN duty-cycle. We note that studying galaxies at fixed redshift, stellar mass and galaxy type enables us to roughly lock the expected SFR of the host, hence the baryonic accretion rate modulated by the dark matter halo \citep{Daddi+22b}. At fixed redshift, more massive SFGs are embedded in more massive halos, hence it is plausible to expect that stochastic gas accretion triggers more frequent and/or more luminous AGN activity. 

In support of our arguments, widespread X-ray and radio AGN activity in massive galaxies ($\mathcal{M_{*}}$$>$10$^{10}$~$\mathcal{M_{\odot}}$) has been recently reported in \citet{Ito+22}. They exploited the most recent optical-to-MIR photometry from the COSMOS2020 catalogue \citep{Weaver+22}, stacking radio (VLA-3~GHz; \citealt{Smolcic+17a}) and X-ray (\textit{Chandra}; \citealt{Civano+16}; \citealt{Marchesi+16}) images of the underlying $\mathcal{M_{*}}$-selected sample in bins of $\mathcal{M_{*}}$ and redshift. Though the strongest $\mathcal{M_{*}}$ dependence was found for quiescent galaxies, also in SFGs \citet{Ito+22} found a rising trend of both $\langle{L_{X}}\rangle$ and $\langle{L_{1.4}^{\mathrm{AGN}}}\rangle$ with $\mathcal{M_{*}}$, out to $z$$\sim$3. 

Hence, in this simple framework the average radio and X-ray nuclear activity might broadly trace each other because they are fuelled via similar mechanisms (i.e., stochastic cold gas accretion). This ``hidden'' connection, however, does not imply that radio and X-ray AGN are switched on at the same time. Clearly, the small overlap between \textit{detected} X-ray and radio AGN suggests that these phases of AGN feedback are broadly \textit{unsynchronized}, and only smoothing over the SMBH duty cycle reveals their long-term connection.

As a final caveat, by ``cold gas accretion'' we are neither implying that gas fueling of SF and BH growth happens at the same time, nor that the physical mechanisms funnelling the gas inward are necessarily the same. We simply mean that the long-term growth of both BHs and galaxies is possibly modulated by the amount of usable gas already in the host \citep{Harrison17}, irrespective of the internal/external mechanisms that did channel the gas inward. Because more massive and more distant SFGs have more cold gas available than the rest of the population (\citealt{Tacconi+18}; \citeyear{Tacconi+20}; \citealt{Liu+19}; \citealt{Wang+22}), we interpret the higher average (radio and X-ray) AGN activity in these galaxies as broadly driven by larger supply of cold gas in the host, on statistical basis. However, we also acknowledge the fact that SMBH and star formation fueling could be further promoted by other aspects, such as the compactness or geometry of SF regions, and the morphology of the host galaxy (e.g., \citealt{GomezGuijarro+22}; \citealt{Aird+22}), which might affect the efficiency of channelling gas towards the smallest ($<<$kpc) scales, and thus the effective gas accretion rate. For this reason, we caution that understanding the higher AGN-to-SF luminosity ratios seen in more massive SFGs (Figs.~\ref{fig:lratio}, ~\ref{fig:lratio_sfr}) might require additional, possibly non-linear, processes happening during cold gas accretion.

\section{Summary and conclusions} \label{summary}

We have presented a novel approach to assess the relationship between radio AGN activity and galaxy stellar mass across $\mathcal{M_{*}}$-selected SFGs. In particular, we performed this analysis on radio-excess AGN, factoring in the statistical contribution of radio-faint AGN within the IRRC, which is critical to compute a representative sample-averaged radio AGN power across the galaxy population. To achieve this goal, we have exploited deep VLA-COSMOS 3-GHz data to identify bona-fide ``radio-excess'' AGN, i.e., showing ($>$2$\sigma$) excess radio emission relative to that expected from pure star formation (i.e. the IRRC, see Sect.~\ref{intro}), at each ($\mathcal{M_*}$,$z$). From this, we have built the 1.4 GHz luminosity function (AGN RLF) of radio-excess AGN in SFGs at 0.1$\leq$$z$$\leq$4.5, following previous works (e.g., \citealt{Smolcic+17c}; \citealt{Novak+18}; \citealt{Ceraj+18}; \citealt{Butler+19}; \citealt{Kono+22}). Then, for the first time, we decomposed the AGN RLF at each redshift in different $\mathcal{M_{*}}$ bins over the range 9$<$$\log$($\mathcal{M_{*}}$/$\mathcal{M_{\odot}}$)$<$12, fitting and integrating each luminosity function down to the minimum $L_{1.4}^{\mathrm{SF}}$ set by the IRRC at the same ($\mathcal{M_*}$,$z$). 

Our main results can be summarized as follows:
\begin{enumerate}

\item The integrated radio-AGN luminosity density across SF galaxies is mostly driven by massive galaxies with 10.5$<$$\log$($\mathcal{M_{*}}$/$\mathcal{M_{\odot}}$)$<$11 and peaks at $z$$\sim$2 (Sect.~\ref{kld}). 
 
\item When averaging this cumulative radio AGN power across \textit{all} $\mathcal{M_{*}}$-selected galaxies at each ($\mathcal{M_*}$,$z$), we obtain a \textit{super-linear} (slope of 1.41$\pm$0.09) ``radio-AGN main sequence'' (RAMS) that links the \textit{mean} (i.e., time-averaged across the galaxy's lifecycle) radio AGN power ($\langle{L_{1.4}^{\mathrm{AGN}}}\rangle$) and galaxy $\mathcal{M_{*}}$ since $z$$\sim$3 (Sect.~\ref{rams}).
 
\item The mean radio AGN power $\langle{L_{1.4}^{\mathrm{AGN}}}\rangle$ at fixed $\mathcal{M_{*}}$ evolves with redshift in a similar fashion to the MS relation ($\propto$(1+z)$^{2.5}$; see e.g., \citealt{Speagle+14}), suggesting that long-term radio AGN activity and galaxy star formation proceed at a similar pace through cosmic time, at least since $z$$\sim$3 (Sect.~\ref{rams}).
 
\item The comparison between radio emission from AGN vs SF reveals that AGN emission is typically dominant over SF only at $\mathcal{M_{*}}$$>$10$^{11}$~$\mathcal{M_{\odot}}$ (Sect.~\ref{agn_vs_sf}). This is also the only $\mathcal{M_{*}}$ range in which current deep radio-continuum surveys (e.g., VLA-COSMOS~3~GHz, \citealt{Smolcic+17a}) are able to formally detect individual radio AGN with luminosity $\langle{L_{1.4}^{\mathrm{AGN}}}\rangle$. Probing typical radio AGN activity in less massive galaxies requires deeper, new generation radio surveys (e.g., SKA1-MID, ngVLA), ideally complemented with VLBI techniques.
 
\item We find that the bulk of radio AGN activity originates from relatively short phases in which the AGN emission \textit{is dominant over SF} (Sect.~\ref{rams}). This is what still allows us to calculate the mean $\langle L_{1.4}^{\mathrm{AGN}} \rangle$ despite this being on average (i.e., across the galaxy's lifetime) sub-dominant compared to $\langle L_{1.4}^{\mathrm{SF}} \rangle$.

\item The super-linear dependence of $\langle{L_{1.4}^{\mathrm{AGN}}}\rangle$ on $\mathcal{M_{*}}$ suggests that radio AGN activity is strongly enhanced in more massive SFGs, as compared to the shallower evolution of galaxy star formation along the ``star-forming main sequence'' (e.g., \citealt{Schreiber+15}). 

\item We dissect the effects of radio-AGN duty cycle (Sect.~\ref{duty_cycle}) and AGN luminosity in radio-bright phase (Sect.~\ref{radio_bright}) for explaining the shape and evolution of the RAMS. Our analysis suggests that more massive (and higher-$z$) galaxies have higher mean $\langle L_{1.4}^{\mathrm{AGN}} \rangle$ due to a combination of both a higher duty cycle \textit{and} an intrinsically brighter radio-AGN phase.

\item Intriguingly, the super-linear trend of $\langle{L_{1.4}^{\mathrm{AGN}}}\rangle$ with $\mathcal{M_{*}}$ closely resembles the evolution of X-ray AGN activity with $\mathcal{M_{*}}$ (e.g., \citealt{Carraro+20}; \citealt{Ito+22}), possibly suggesting a common fueling scenario (Sect.~\ref{radio_x}). While our analysis favours a long-term X-ray/radio connection, these AGN phases are likely \textit{unsynchronized} due to a relatively short X-ray/radio AGN duty cycle -- $\sim$10\% even in the most massive galaxies ($\mathcal{M_*}$$\gtrsim$10$^{11}$~$\mathcal{M_{\odot}}$) -- which is consistent with the small X-ray/radio AGN overlap (10--15\%) seen in $\mathcal{M_*}$-matched SFGs.

\end{enumerate}

%________________________________________________________________

\begin{acknowledgements}
We thank the anonymous referee for a positive report. ID thanks K. Ito for useful discussions. The Cosmic Dawn Center (DAWN) is funded by the Danish National Research Foundation under grant No. 140. SJ acknowledges the Villum Fonden research grants 37440, 13160 and the financial support from European Union's Horizon 2021 research and innovation program under the Marie Sk\l{}odowska-Curie grant agreement No. 101060888. CS acknowledges financial support from the Italian Ministry of University and Research - Project Proposal CIR01\_00010.
% % % The authors are grateful to...

\end{acknowledgements}

 \bibliographystyle{aa} % style aa.bst
 \bibliography{biblio} % your references Yourfile.bib

\begin{appendix}

\section{Extras on sample selection} \label{Appendix_ref}

\begin{figure}
\centering
     \includegraphics[width=\linewidth]{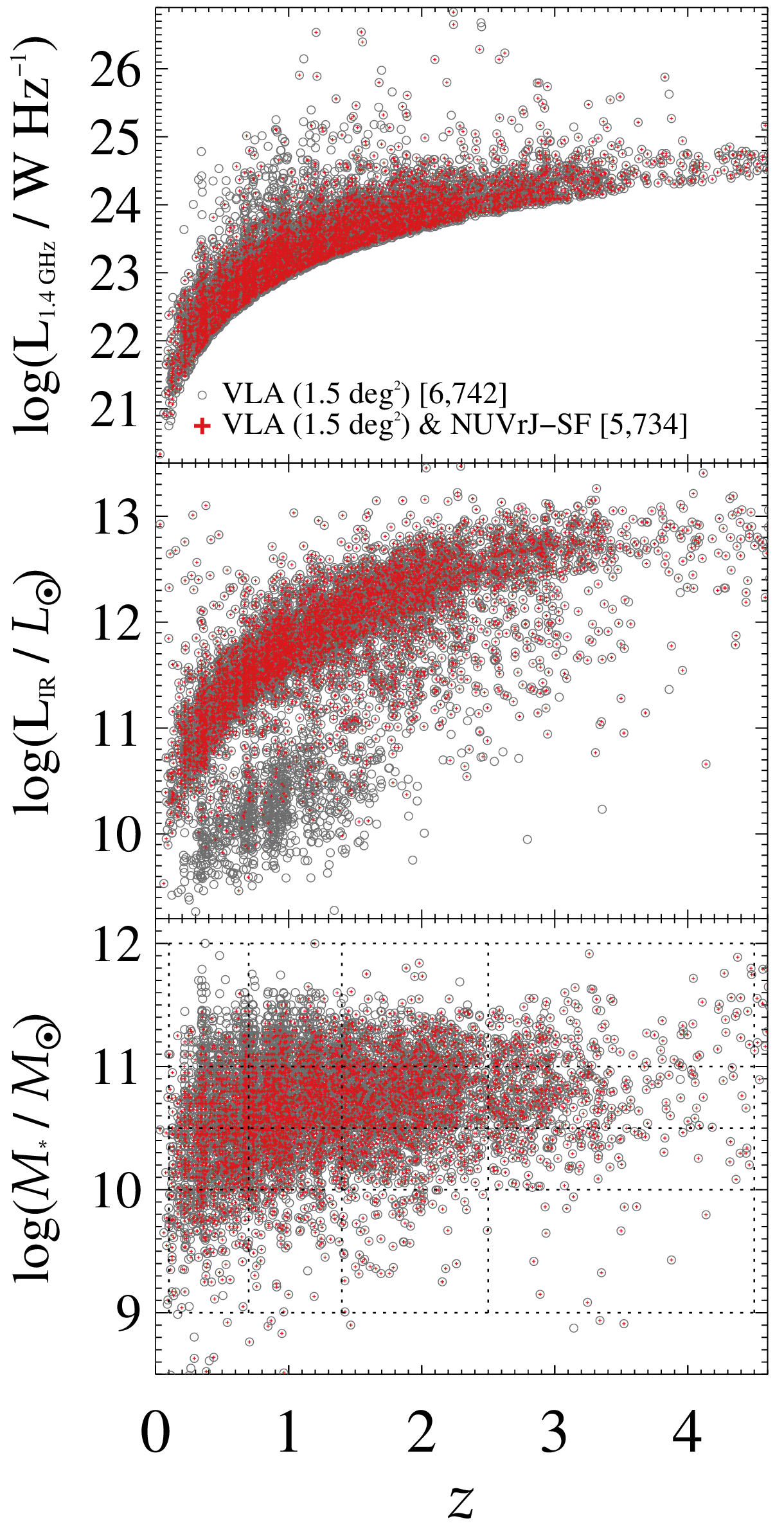}
 \caption{\small Distributions of L$_{1.4}$ (top), L$_{IR}$ (middle) and galaxy $\mathcal{M_{*}}$ (bottom) of our sample, as a function of redshift. Grey empty circles are VLA sources with optical/NIR counterpart \citep{Laigle+16} within the UltraVISTA 1.5~deg$^2$ area, while red crosses mark the subset classified as ``star-forming'' based on $NUVrJ$ criteria. Dotted lines in the bottom panel enclose the $\mathcal{M_{*}}$--$z$ space of the final sample (5,658 sources).
 }
   \label{fig:ref2}
\end{figure}

We display some relevant distributions for our VLA 3~GHz detections, before and after applying the selection cuts outlined in Table~\ref{tab:sample}. Particularly, Fig.~\ref{fig:ref2} shows VLA sources with optical/NIR counterpart \citep{Laigle+16} within the UltraVISTA 1.5~deg$^2$ area (empty circles), as well as its subset classified as ``star-forming'' (red crosses). From top to bottom, the panels show the distribution with redshift of L$_{1.4}$, L$_{IR}$ and galaxy $\mathcal{M_{*}}$, respectively. As discussed in Appendix~\ref{Appendix_C}, the removal of $NUVrJ$-passive galaxies reduces our sample by a factor of 2--3$\times$ at $z$$\lesssim$$1$. At these redshifts, such a cut leaves us with higher L$_{IR}$ (middle panel), and slightly lower L$_{1.4}$ (top panel) and $\mathcal{M_{*}}$ (bottom panel) than the parent VLA sample. We refer the reader to Sect.~\ref{passive} for a detail explanation on why $NUVrJ$-passive galaxies were removed.

%%placing figure  ---------------------------
%%%-------------------------------------

\begin{figure*}
\centering
     \includegraphics[width=\linewidth]{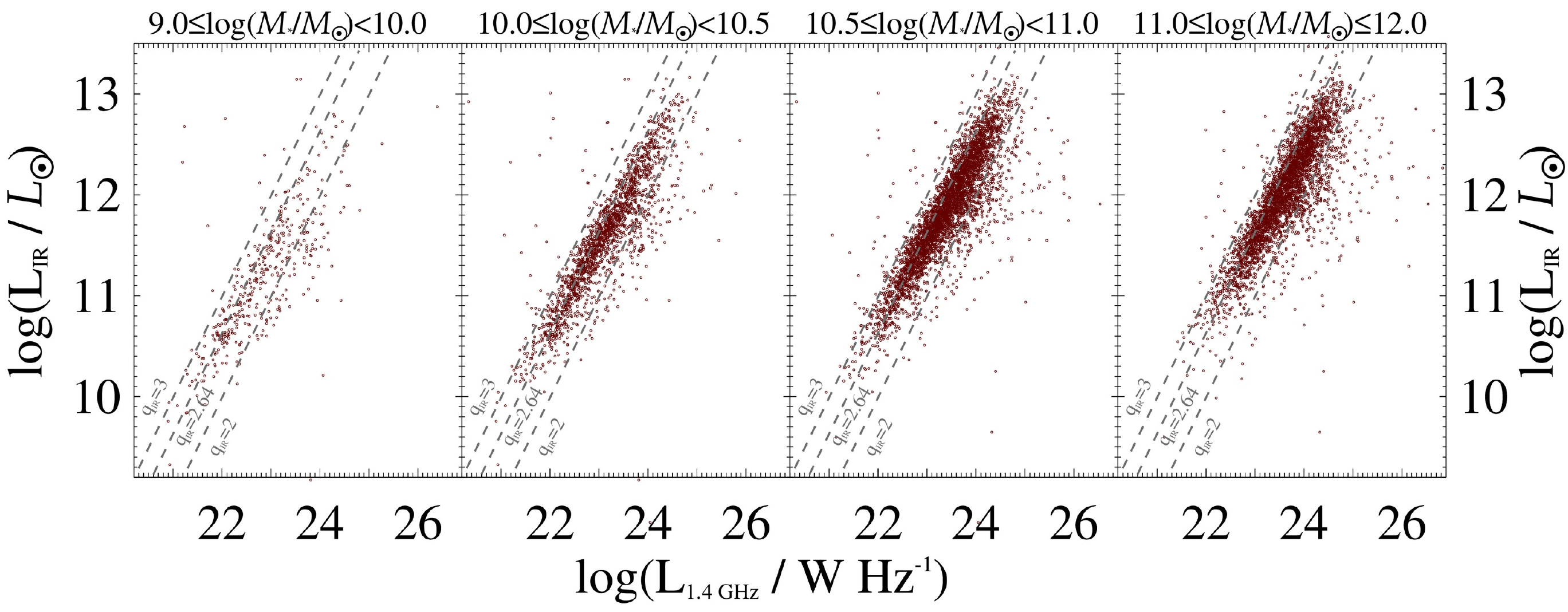}
 \caption{\small Distribution of our final sample of all 3~GHz detection classified as ``star-forming'' (5,658, see Table~\ref{tab:sample}), as a function of L$_{IR}$ and L$_{1.4}$ and split in different $\mathcal{M_{*}}$ bins (increasing from left to right). Dashed lines mark constant q$_{IR}$ lines as a reference.
 }
   \label{fig:ref3}
\end{figure*}

Fig.~\ref{fig:ref3} shows the L$_{IR}$-vs-L$_{1.4}$ distributions of our final sample of 5,658 radio-detected (S/N$>$5 at 3~GHz) star-forming galaxies across 1.5~deg$^2$, and within the range 0.1$\le$$z$$\le$4.5 and 9$\le$$\log(\mathcal{M_{*}}$/$\mathcal{M_{\odot}}$)$\le$12 (see Table~\ref{tab:sample}).
Dashed lines mark constant q$_{IR}$ lines as a reference, including the local value (q$_{IR}$=2.64; \citealt{Bell03}). At high $\mathcal{M_{\odot}}$, both L$_{IR}$ and L$_{1.4}$ increase, due to galaxies becoming more star-forming. Moreover, at high luminosities the distributions appear to shift to lower q$_{IR}$: this is caused by a redshift trend within our flux-limited VLA sample (i.e. q$_{IR}$ apparently decreasing with redshift, hence with luminosity, see D21). Controlling for such internal redshift dependence is needed to coherently compare the q$_{IR}$ distributions in different $\mathcal{M_{*}}$ bins. This is why in Sect.~\ref{histo} (and in Fig.~\ref{fig:histo_class}) we have re-scaled each observed q$_{IR}$ by the corresponding q$_{IRRC}$ at the same ($\mathcal{M_{*}}$,$z$).

% 
%%%placing figure  ---------------------------
%%%-------------------------------------

% 

\section{Statistical corrections to the AGN RLF} \label{Appendix_A}

%%%placing figure  ---------------------------
\begin{figure*}
\centering
     \includegraphics[width=\linewidth]{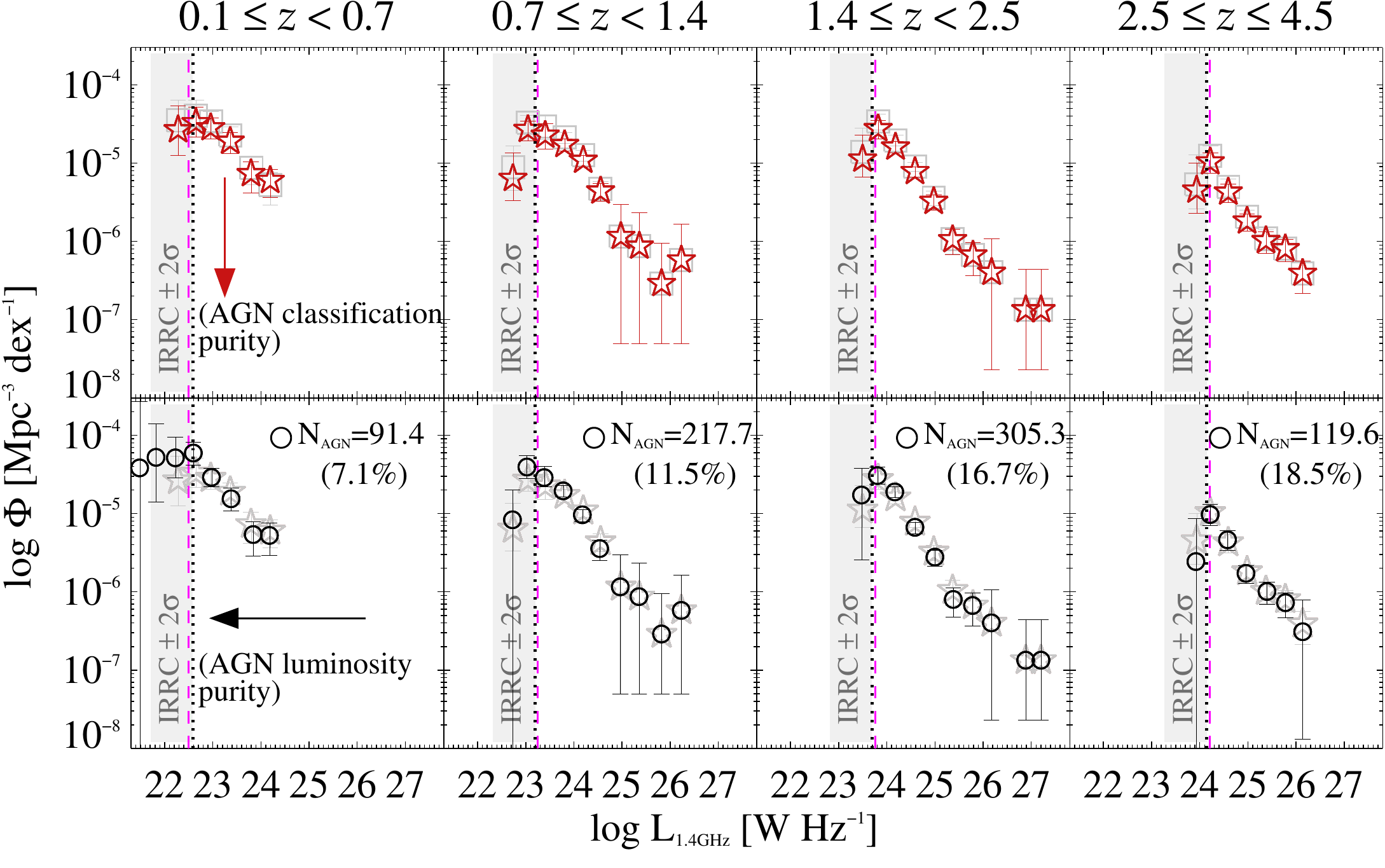}
 \caption{\small 1.4~GHz luminosity function of radio-excess AGN in ($NUVrJ$) SFGs (AGN RLF) at different redshifts (increasing from left ro right). The grey shaded area encloses the $\pm$2$\sigma$ locus (i.e., $\pm$0.44~dex) around the luminosity corresponding to the IRRC. All datapoints below the +2$\sigma$ boundary (black dotted line) are conservatively disregarded to minimize incompleteness. \textit{(Top panels)}: the former AGN RLF (grey squares) is corrected for AGN classification purity according to $f_{N}^{\mathrm{AGN}}$ (see Sect.~\ref{corrections} and Fig.~\ref{fig:histo_class}). This step lowers $\Phi$(L) at the faint end (red stars). \textit{(Bottom panels)}: The $f_{N}^{\mathrm{AGN}}$-corrected AGN RLF (grey stars) is further corrected for AGN luminosity purity at 1.4~GHz, based on $f_{1.4}^{\mathrm{AGN}}$. This step mildly steepens $\Phi$(L) at the faint end (black circles), while the brightest bins remain unchanged. The magenta dashed line indicates the luminosity corresponding to 90\% AGN completeness (L$_{90\%}$) in each $z$-bin. After applying the above corrections, the final numbers and fractions relative to the parent VLA-COSMOS~3GHz sample are reported in each $z$-bin. See text for details.
 }
   \label{fig:a1}
\end{figure*}
%%%-------------------------------------

\subsection{Correction for classification purity} \label{corr_purity}

The comoving number density $\Phi$(L) obtained in four redshift bins is shown in the upper panels of Fig.~\ref{fig:a1} (grey open squares). As highlighted previously, the 2$\sigma$ threshold at $\Delta q_{\mathrm{IRRC}}$=--0.44~dex was designed to minimize AGN contamination within ``normal'' SFGs (D21). Reversing this approach for identifying radio-excess AGN is, however, at the expense of AGN purity (i.e., higher galaxy contamination). Correcting for mis-classified AGN can be done on statistical basis as explained in Sect.~\ref{corrections} and Fig.~\ref{fig:histo_class}. Briefly, instead of counting each radio-excess AGN as unity, we weigh it as the corresponding $f_{N}^{\mathrm{AGN}}$ in the 1/V$_{\mathrm max}$ calculation. This approach enables us to statistically correct for source mis-classifications by assigning each object a purity 0$\leq$$f_{N}^{\mathrm{AGN}}$$\leq$1 of being a radio-excess AGN. 

Corrected values of $\Phi$(L) are marked in Fig.~\ref{fig:a1} as red starred symbols. As expected, the number density remains unchanged at the bright-end, while at low $L_{1.4}$ it drops by an amount proportional to 1-$f_{N}^{\mathrm{AGN}}$, that is $\approx$30--40\%. Therefore, this correction is quite important to clean up our AGN sample. We clarify that each object is treated as both an AGN and a galaxy: in the \textit{AGN} RLF each object is counted as $f_{N}^{\mathrm{AGN}}$, while in the complementary \textit{galaxy RLF} (not shown here), the same object is counted as (1--$f_{N}^{\mathrm{AGN}}$). As a reference, the vertical dotted line in each $z$-bin indicates the 1.4~GHz luminosity threshold corresponding to +2$\sigma$ from the IRRC. This $L_{1.4}$ threshold is converted from the $\Delta q_{\mathrm{IRRC}}$ space, by taking for simplicity the median $L_{IR}$ and the median $\mathcal{M_*}$ of the sample in that $z$-bin. This is for visual purposes in Fig~\ref{fig:a1}. Instead, in practice the correction for AGN classification purity is applied to each object based on its \textit{observed} $\Delta q_{\mathrm{IRRC}}$ value, irrespective of its $L_{1.4}$.

It is likely that radio-excess AGN are under-represented (or at least their fraction is highly uncertain) at radio luminosities below our +2$\sigma $$L_{1.4}$ threshold, since we assumed that the peak of the $\Delta q_{\mathrm{IRRC}}$ distribution is entirely made by SFGs (see also Fig.~\ref{fig:histo_class}), while radio AGN fill the residual part of the distribution. For this reason, throughout this manuscript we disregard all datapoints placed at $\Delta q_{\mathrm{IRRC}}$$>$--0.44~dex (i.e., the --2$\sigma$ boundary from the IRRC). This interval is indicated as the shaded area in Fig.~\ref{fig:a1}. 
After applying the above corrections, the final numbers and fractions relative to the parent VLA-COSMOS~3GHz sample are reported in each $z$-bin. See text for details.

\subsection{Correction for $L_{1.4}$ purity} \label{corr_l14}

The correction for AGN classification purity acts solely on the counting, not on the luminosity of each object. Thus, following the motivation presented in Sect.~\ref{corrections}, we attempt at decomposing the total 1.4~GHz luminosity of each source between AGN- and galaxy-driven contributions. 

As done in \citet{Ceraj+18} and Fig.~\ref{fig:histo_class} (grey dashed lines), we compute the AGN-related fraction at 1.4~GHz, $f_{1.4}^{\mathrm{AGN}}$. Then we multiply each $L_{1.4}$ by the corresponding $f_{1.4}^{\mathrm{AGN}}$ to obtain the AGN-related luminosity, i.e., $L_{1.4}^{\mathrm{AGN}}$. 
The global effect on the AGN RLF is shown in Fig.~\ref{fig:a1} (bottom panels). This correction for $L_{1.4}$ re-distributes the former dataset (grey stars) to lower luminosities. Specifically, except for the brightest $L_{1.4}$ bins that are unchanged, the (pure) AGN-related $\Phi$(L) steepens in the faint end at the expense of bins at moderate $L_{1.4}$. Of course, in the faintest sources, this step shifts $L_{1.4}^{\mathrm{AGN}}$ formally below the luminosity limit of the survey (or within the $\pm$2$\sigma$ locus of the IRRC, grey shaded area), in which case the source is removed from our final sample. 

In this study, we refrain from re-calculating the V$_{\mathrm max}$ of each source using the new AGN-scaled flux density, since the detectability of each object remains bounded to its combined AGN+galaxy flux density. Uncertainties on $\Phi$(L) after applying the afore-mentioned corrections are discussed in Sect.~\ref{boot}.

Finally, we double-check that correcting only for AGN $L_{1.4}$ purity (i.e., without applying an AGN classification purity), but extending this to all radio-detected galaxies, returns a consistent LF, although about 20\% more AGN just above the +2$\sigma$ $L_{1.4}$ threshold (black dotted line in Fig.~\ref{fig:a1}). This is because a correction for AGN $L_{1.4}$ purity alone does not distinguish between a radio-excess AGN and a $>$2$\sigma$ outlier SFG, hence we go ahead with this two-step correction.

\subsection{AGN completeness and final numbers} \label{completeness}

A potential caveat of our approach is related to the fact that we correct for AGN classification and $L_{1.4}$ purity based on the offset of each object from the IRRC (i.e., $\Delta q_{\mathrm{IRRC}}$), but re-scaled to $L_{1.4}$ space based on the median $L_{IR}$ of the sample. Thus, in principle, even a moderately-bright radio AGN hosted by a starbursting SFG (i.e., well above average in $L_{IR}$) would display no radio excess, hence it might not be counted in our final AGN RLF. 

In order to quantify such incompleteness in $L_{1.4}^{\mathrm{AGN}}$, we proceed as follows. We consider the subset of all radio-detected galaxies (black histogram in Fig.~\ref{fig:histo_class}) with $\Delta q_{\mathrm{IRRC}}$$>$--0.44~dex, i.e., not showing radio excess. Then we look at their cumulative distribution of $L_{1.4}^{\mathrm{AGN}}$ (i.e., $L_{1.4}$$\times$$f_{1.4}^{\mathrm{AGN}}$), and we set the luminosity corresponding to the 90$^{th}$ percentile as our completeness limit at 90\% level (L$_{90\%}$, see magenta dashed line in Fig.~\ref{fig:a1}). We redo this check in each $z$-bin. This L$_{90\%}$ nearly matches our +2$\sigma$ $L_{1.4}^{\mathrm{AGN}}$ threshold\footnote{We note that the 100\% completeness limit would be placed only 0.1--0.15~dex above L$_{90\%}$.}. This is because, when moving closer to $\Delta q_{\mathrm{IRRC}}$$\sim$0 (i.e., the IRRC), the distribution is populated by SFGs and progressively radio-fainter AGN, so the brightest ``missing'' radio AGN will hardly be at above our +2$\sigma$ $L_{1.4}^{\mathrm{AGN}}$ threshold (black dotted line in Fig.~\ref{fig:a1}). However, this check demonstrates that our approach delivers a $\approx$90\% complete sample of radio AGN.

\smallskip

The final numbers and fractions relative to the parent VLA-COSMOS~3GHz sample (5,658 sources) are reported in each $z$-bin for convenience (non-integer numbers reflect the sum over $f_{N}^{\mathrm{AGN}}$). The radio-excess AGN fraction is significantly redshift dependent. This is mostly a selection effect induced by the redshift-dependent luminosity cut corresponding to the limiting flux of the survey. As a consequence, at higher redshift only the brightest radio sources (i.e., more likely AGN) are detectable. 

However, we acknowledge that our global fraction of radio-excess AGN (734/5,658$\sim$13\%) is significantly smaller than that found in \citet{Smolcic+17b} (1,814/7,729$\sim$23\%). This difference is partly introduced by our rather conservative corrections for classification and $L_{1.4}$ purity. Accounting for the radio AGN within $\pm$2$\sigma$ of the IRRC (i.e., without radio excess) - in turn likely underestimated - would bring the radio AGN fraction to $\sim$19\%. Another concomitant effect is played by the absence of passive galaxies in our sample. Despite being only $\approx$15\% of the parent VLA-3~GHz sample, about a third of those ($NUVrJ$) passive systems would be classified as ``radio-excess AGN'' according to our criteria, increasing the global AGN fraction by an additional $\sim$5\%, thus in line with \citet{Smolcic+17b}. The observed prevalence of radio AGN in passive systems (e.g., \citealt{Gobat+18}; \citealt{Magdis+21}; \citealt{Kondapally+22}; \citealt{Ito+22}) might be interpreted as both a quenching effect due AGN-driven jets hampering star formation (e.g., \citealt{Heckman+14}), but also partly as a selection effect, due to passive galaxies simply showing higher constrast between AGN-vs-host radio emission, at fixed $L_{1.4}$. We test the impact of source classification and sample selection (i.e., SF vs passive) on the AGN RLF in Appendix~\ref{Appendix_B}. The take-away message from this test is that source classification methods do not significantly alter the shape and normalization of the AGN RLF. Instead, the lack of passive galaxies in our sample induces a systematically 2--3$\times$ lower AGN RLF at $z$$\lesssim$1 (nearly constant with $L_{1.4}$), as seen in Fig.~\ref{fig:b1}.

\section{Impact of input assumptions on the AGN RLF}   \label{Appendix_B}

We test the impact of our sample selection and radio-excess criterion on the evolving AGN RLF since $z$$\sim$3. To this end, we directly compare our observed RLF against that derived by \citet{Smolcic+17c} from the same VLA-COSMOS~3~GHz data. For sake of consistency with their study, here we show our un-corrected datapoints (i.e., before applying AGN purity and luminosity corrections, see Sect.~\ref{corrections}).

\begin{figure*}
\centering
     \includegraphics[width=6.5in]{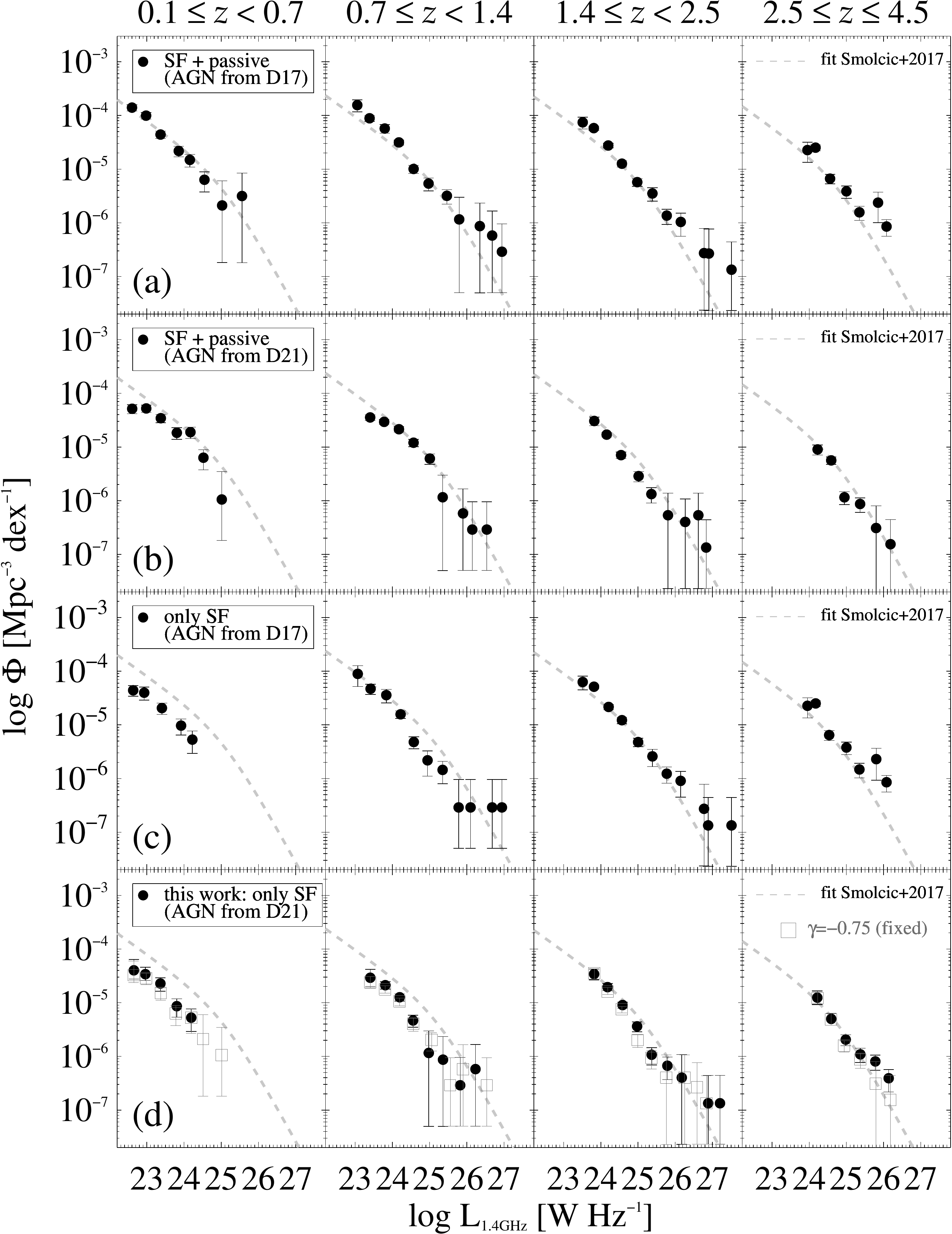}
 \caption{\small Observed radio-excess 1.4~GHz luminosity function divided in four redshift bins (increasing from left to right). Datapoints (black circles) are binned with $\Delta$($\log L_{1.4}$)=0.4~dex. Correction for flux completeness is equally applied to all realizations, following previous VLA-COSMOS based studies (e.g., \citealt{Smolcic+17c}; \citealt{Novak+18}; \citealt{Ceraj+18}). The best-fit PLE model from \citet{Smolcic+17c} is overlaid to each panel. This Figure is split in four rows, as follows. \textit{(a)}: Dataset from \citeauthor{Smolcic+17c} (\citeyear{Smolcic+17c}, including SF and passive galaxies) obtained by applying the radio-excess criterion from \citet{Delvecchio+17}. \textit{(b)}: Dataset from \citet{Smolcic+17c} but identifying radio-excess AGN according to D21. \textit{(c)}: Radio-excess AGN from this work (i.e only within $NUVrJ$-based SFGs) identified based on \citet{Delvecchio+17}. \textit{(d)}: This work following D21 (same as the pre-corrected RLF shown in Fig.~\ref{fig:a1}, top panel). The equivalent RLF by assuming a fixed spectral index ($\gamma$=--0.75) is shown for comparison (empty squares). See text for details.
 }
   \label{fig:b1}
\end{figure*}

Fig.~\ref{fig:b1} displays four realizations of the same AGN RLF, by changing either sample selection or radio-excess criterion, or both. Individual datapoints (black circles) are shown in the same four redshift intervals. In each panel, the best-fit RLF from \citeauthor{Smolcic+17c} (\citeyear{Smolcic+17c}, using PLE fitting model) is overlaid for comparison. Fig.~\ref{fig:b1} is split in four rows, as follows.

\textit{(a)}: Observed AGN RLF re-computed following \citet{Smolcic+17c} in each redshift bin. The sample includes both SF and passive galaxies, in which radio-excess AGN are identified based on the redshift-dependent criterion of \citet{Delvecchio+17}. As expected the datapoints are in very good agreement with the best-fit solution obtained in \citet{Smolcic+17c}. 

\textit{(b)}: Same as \textit{(a)}, but applying the radio-excess criterion from D21, based on a 2$\sigma$ offset from the IRRC at the $\mathcal{M_{*}}$ and redshift of each source. The choice of a different radio-excess threshold does not strongly affect the RLF, except at the faint end at $z$$<$1.4, where the criterion from D21 rejects some additional AGN.

\textit{(c)}: Same as (a) but changing sample selection. Here only radio detections classified as $NUVrJ$-based SFGs are displayed. As further discussed in Sect.~\ref{rlf_redshift}, the lack of radio AGN in passive galaxies induces a 2--3$\times$ drop at $z$$<$1 that is roughly independent of $L_{1.4}$, while at higher redshifts the RLF remaines identical. 

\textit{(d)}: Dataset from this work, i.e., including only SF radio detections, with radio-excess AGN being identified from D21 (same as top panel of Fig.~\ref{fig:a1}). As hinted at from \textit{(c)}, removing passive galaxies still results in a drop at $z$$<$1, while at higher redshifts the RLF is in unchanged. The equivalent AGN RLF obtained by taking a fixed 1.4--3~GHz spectral index ($\gamma$=--0.75) is shown for comparison (empty squares). The global consistency among RLFs indicates that our spectral index assumption does not affect the shape and evolution of the RLF.
 
In summary, these comparisons suggest that our different radio-excess AGN criterion method does not significantly alter the shape and normalization of the AGN RLF, at any redshift. Instead, the lack of passive galaxies in our sample induces a systematically 2--3$\times$ lower normalization, especially at $z$$\lesssim$1. This offset is roughly independent of $L_{1.4}$, hence in this work we can still adopt the faint- and bright-end slopes inferred by \citet{Mauch+07} on a local sample of radio AGN.

%%%placing figure  ---------------------------
%%%-------------------------------------

\section{AGN RLF datapoints}   \label{Appendix_C}
Table~\ref{tab:rlf_observed} summarizes all AGN RLF datapoints used in this work, split in redshift and $\mathcal{M_{*}}$. 

% % %   UPDATE TABLE
\begin{table*}
\caption{Luminosity function of radio-excess AGN in SFGs, split in redshift (Fig.~\ref{fig:rlf_fitting_z}) and $\mathcal{M_{*}}$ (Figs.~\ref{fig:rlf_mass_ple},~\ref{fig:rlf_mass_pde}) bins. Only luminosity bins above the corresponding $>$2$\sigma$ cut from the IRRC (i.e., those used in the fitting) are shown. Error bars are given at 1$\sigma$ level. }
\centering
\begin{tabular}{ccc|ccc}
\hline\hline
$z$-bin &  $L_{1.4}^{\mathrm{AGN}}$ ($z$)&   $\Phi(L,z)$$\cdot$10$^{-6}$  &      $\mathcal{M_{*}}$-bin  &   $L_{1.4}^{\mathrm{AGN}}$ ($\mathcal{M_*}$)  &   $\Phi(L,z,\mathcal{M_{*}})$$\cdot$10$^{-6}$   \\
        & [$\log$(W~Hz$^{-1}$)]   &   [Mpc$^{-3}$~dex$^{-1}$]    &    [$\log(\mathcal{M_{*}}$/$\mathcal{M_{\odot}})$]  &  [$\log$(W~Hz$^{-1}$)]   &   [Mpc$^{-3}$~dex$^{-1}$]  \\
\hline
  \smallskip
$0.1 \leq z < 0.7$  &  22.60   &  59.38$^{+36.13}_{-41.96}$  &       9.0--10.0  &   22.59    &  21.03$^{+6.76}_{-11.43}$  \\ 
 \smallskip  &    22.97        &  29.03$^{+20.20}_{-22.14}$  &                  &   22.93    &  8.80$^{+4.85}_{-4.85}$  \\  
 \smallskip  &    23.38        &  15.46$^{+9.70}_{-10.94}$   &                  &   23.28    &  3.56$^{+2.52}_{-0.16}$  \\  
 \smallskip  &    23.84        &  5.39$^{+2.86}_{-2.86}$     &      10.0--10.5  &   23.00    &  6.27$^{+3.46}_{-4.00}$  \\  
 \smallskip  &    24.18        &  5.26$^{+2.92}_{-2.92}$     &                  &   23.36    &  3.96$^{+1.18}_{-1.18}$  \\   
 \smallskip  &  &   &                                               10.5--11.0  &   22.99    &  11.73$^{+6.78}_{-6.78}$  \\ 
  \smallskip &  &   &                                                           &   23.43    &  4.57$^{+1.21}_{-2.45}$  \\  
  \smallskip &  &   &                                                           &   23.87    &  2.98$^{+2.32}_{-0.01}$  \\  
  \smallskip &  &   &                                                           &   24.19    &  3.16$^{+2.14}_{-0.19}$  \\  
  \smallskip &  &   &                                               11.0--12.0  &   23.40    &  3.47$^{+1.50}_{-1.50}$  \\  
  \smallskip &  &   &                                                           &   23.82    &  2.84$^{+2.29}_{-0.03}$  \\  
  \smallskip &  &   &                                                           &   24.17    &  2.10$^{+1.78}_{-0.20}$  \\  
\hline

  \smallskip
$0.7 \leq z < 1.4$  &  23.38    &  28.81$^{+17.09}_{-22.11}$  &       9.0--10.0  &    23.02  &  19.32$^{+12.14}_{-12.76}$  \\  
   \smallskip       &    23.78  &  19.50$^{+16.11}_{-16.36}$  &                  &    23.35  &  3.76$^{+0.44}_{-2.44}$  \\  
  \smallskip        &    24.18  &  9.65$^{+7.94}_{-7.94}$     &                  &    23.85  &  2.07$^{+1.23}_{-1.23}$  \\  
   \smallskip       &    24.54  &  3.57$^{+2.55}_{-2.55}$     &                  &    24.04  &  0.31$^{+0.40}_{-0.05}$  \\  
  \smallskip        &    24.96  &  1.16$^{+0.67}_{-0.05}$     &                  &    24.43  &  0.29$^{+0.38}_{-0.05}$  \\  
   \smallskip       &    25.35  &  0.87$^{+0.59}_{-0.05}$     &      10.0--10.5  &    23.36  &  9.07$^{+3.66}_{-5.47}$  \\  
   \smallskip       &    25.82  &  0.29$^{+0.38}_{-0.05}$     &                  &    23.76  &  5.89$^{+4.27}_{-4.27}$  \\  
   \smallskip       &    26.23  &  0.58$^{+0.50}_{-0.05}$     &                  &    24.09  &  1.10$^{+0.68}_{-0.02}$  \\  
    \smallskip &  &   &                                                          &    25.29  &  0.29$^{+0.38}_{-0.05}$  \\  
  \smallskip &  &   &                                                10.5--11.0  &    23.77  &  7.67$^{+5.63}_{-5.84}$  \\  
  \smallskip &  &   &                                                            &    24.20  &  3.91$^{+2.83}_{-2.83}$  \\  
   \smallskip &  &   &                                                           &    24.48  &  1.17$^{+0.66}_{-0.06}$  \\  
   \smallskip &  &   &                                                           &    25.00  &  0.58$^{+0.50}_{-0.05}$  \\  
  \smallskip &  &   &                                                            &    25.46  &  0.29$^{+0.38}_{-0.05}$  \\  
   \smallskip &  &   &                                                           &    25.82  &  0.29$^{+0.38}_{-0.05}$  \\  
     \smallskip &  &   &                                                         &    26.23  &  0.58$^{+0.50}_{-0.05}$  \\  
     \smallskip &  &   &                                             11.0--12.0  &    23.78  &  3.96$^{+2.68}_{-2.68}$  \\  
  \smallskip &  &   &                                                            &    24.18  &  4.51$^{+3.35}_{-3.35}$  \\  
    \smallskip &  &   &                                                          &    24.57  &  2.19$^{+1.39}_{-1.39}$  \\  
   \smallskip &  &   &                                                           &    24.93  &  0.58$^{+0.50}_{-0.05}$  \\  
    \smallskip &  &   &                                                          &    25.30  &  0.29$^{+0.38}_{-0.05}$  \\  
\hline
\hline
\end{tabular} \label{tab:rlf_observed}
\end{table*}

\begin{table*}
\caption*{Continued from Table~\ref{tab:rlf_observed}.}
\centering
\begin{tabular}{ccc|ccc}
\hline\hline
$z$-bin &  $L_{1.4}^{\mathrm{AGN}}$ ($z$)&   $\Phi(L,z)$$\cdot$10$^{-6}$  &      $\mathcal{M_{*}}$-bin  &   $L_{1.4}^{\mathrm{AGN}}$ ($\mathcal{M_*}$)  &   $\Phi(L,z,\mathcal{M_{*}})$$\cdot$10$^{-6}$   \\
        & [$\log$(W~Hz$^{-1}$)]   &   [Mpc$^{-3}$~dex$^{-1}$]    &    [$\log(\mathcal{M_{*}}$/$\mathcal{M_{\odot}})$]  &  [$\log$(W~Hz$^{-1}$)]   &   [Mpc$^{-3}$~dex$^{-1}$]  \\
\hline
  \smallskip
  $1.4 \leq z < 2.5$  &  23.80  &  30.55$^{+21.40}_{-23.80}$  &             9.0--10.0   &   23.48  &  6.38$^{+1.32}_{-2.90}$  \\  
     \smallskip  &    24.17     &  18.90$^{+15.53}_{-16.03}$  &                         &   23.79  &  6.88$^{+4.66}_{-4.63}$  \\  
    \smallskip   &    24.58     &  6.67$^{+5.57}_{-5.58}$  &                            &   24.11  &  1.28$^{+0.80}_{-0.79}$  \\  
    \smallskip   &    25.00     &  2.78$^{+2.16}_{-2.16}$  &                            &   24.62  &  0.28$^{+0.24}_{-0.02}$  \\  
    \smallskip   &    25.38     &  0.80$^{+0.48}_{-0.48}$  &                            &   25.76  &  0.13$^{+0.17}_{-0.02}$  \\  
    \smallskip   &    25.78     &  0.67$^{+0.37}_{-0.37}$  &                10.0--10.5  &   23.80  &  13.73$^{+8.24}_{-8.24}$  \\  
    \smallskip   &    26.17     &  0.40$^{+0.28}_{-0.02}$  &                            &   24.15  &  3.49$^{+2.42}_{-2.72}$  \\  
    \smallskip   &    26.88     &  0.13$^{+0.17}_{-0.02}$  &                            &   24.48  &  0.75$^{+0.53}_{-0.02}$  \\  
    \smallskip   &    27.20     &  0.13$^{+0.17}_{-0.02}$  &                            &   24.97  &  0.65$^{+0.34}_{-0.34}$  \\  
  \smallskip &  &   &                                                                   &   25.50  &  0.13$^{+0.17}_{-0.02}$  \\  
    \smallskip &  &   &                                                                 &   26.15  &  0.13$^{+0.17}_{-0.02}$  \\  
   \smallskip &  &   &                                                      10.5--11.0  &   24.17  &  9.72$^{+7.79}_{-7.60}$  \\  
   \smallskip &  &   &                                                                  &   24.59  &  3.39$^{+2.62}_{-2.62}$  \\  
   \smallskip &  &   &                                                                  &   25.00  &  0.92$^{+0.57}_{-0.57}$  \\  
    \smallskip &  &   &                                                                 &   25.31  &  0.26$^{+0.23}_{-0.02}$  \\  
    \smallskip &  &   &                                                                 &   25.83  &  0.27$^{+0.23}_{-0.02}$  \\  
     \smallskip &  &   &                                                    11.0--12.0  &   24.18  &  4.63$^{+3.42}_{-3.44}$  \\  
     \smallskip &  &   &                                                                &   24.59  &  2.27$^{+1.66}_{-1.66}$  \\  
    \smallskip &  &   &                                                                 &   25.00  &  1.25$^{+0.84}_{-0.84}$  \\  
    \smallskip &  &   &                                                                 &   25.40  &  0.40$^{+0.27}_{-0.02}$  \\  
     \smallskip &  &   &                                                                &   25.74  &  0.27$^{+0.23}_{-0.02}$  \\  
     \smallskip &  &   &                                                                &   26.18  &  0.27$^{+0.23}_{-0.02}$  \\  
        \smallskip &  &   &                                                             &   26.88  &  0.13$^{+0.17}_{-0.02}$  \\  
        \smallskip &  &   &                                                             &   27.20  &  0.13$^{+0.17}_{-0.02}$  \\  
 \hline  
  \smallskip
  $2.5 \leq z \leq 4.5$   &    24.22  &  9.71$^{+6.18}_{-7.15}$  &     9.0--10.0           &    24.20  &  1.19$^{+0.87}_{-0.04}$  \\  
    \smallskip            &    24.58  &  4.63$^{+3.10}_{-3.45}$  &                         &    24.48  &  0.36$^{+0.31}_{-0.04}$  \\  
   \smallskip             &    24.96  &  1.72$^{+1.28}_{-1.28}$  &                         &    24.95  &  0.11$^{+0.14}_{-0.02}$  \\  
    \smallskip            &    25.40  &  1.01$^{+0.69}_{-0.69}$  &                         &    25.27  &  0.09$^{+0.12}_{-0.02}$  \\  
     \smallskip           &    25.78  &  0.73$^{+0.47}_{-0.47}$  &     10.0--10.5          &    24.57  &  0.88$^{+0.49}_{-0.49}$  \\  
     \smallskip           &    26.14  &  0.31$^{+0.18}_{-0.01}$  &                         &    24.94  &  0.36$^{+0.20}_{-0.02}$  \\  
          \smallskip &  &   &                                                              &    25.28  &  0.08$^{+0.10}_{-0.01}$  \\  
       \smallskip &  &   &                                                                 &    25.79  &  0.16$^{+0.24}_{-0.07}$  \\  
       \smallskip &  &   &                                             10.5--11.0          &    24.56  &  2.68$^{+1.58}_{-1.88}$  \\  
       \smallskip &  &   &                                                                 &    24.96  &  0.94$^{+0.59}_{-0.59}$  \\   
       \smallskip &  &   &                                                                 &    25.40  &  0.42$^{+0.23}_{-0.23}$  \\  
       \smallskip &  &   &                                                                 &    25.79  &  0.23$^{+0.16}_{-0.01}$  \\  
       \smallskip &  &   &                                                                 &    26.19  &  0.08$^{+0.10}_{-0.01}$  \\  
        \smallskip &  &   &                                            11.0--12.0          &    24.64  &  0.71$^{+0.41}_{-0.44}$  \\  
        \smallskip &  &   &                                                                &    24.99  &  0.34$^{+0.16}_{-0.16}$  \\   
        \smallskip &  &   &                                                                &    25.45  &  0.42$^{+0.31}_{-0.02}$  \\  
        \smallskip &  &   &                                                                &    25.77  &  0.33$^{+0.20}_{-0.01}$  \\  
        \smallskip &  &   &                                                                &    26.14  &  0.23$^{+0.16}_{-0.01}$  \\  
   \hline
   \hline
   \end{tabular}
\end{table*}

\end{appendix}

\end{document}